\documentclass[letterpaper,english]{article}
\usepackage[T1]{fontenc}
\usepackage{color}
\usepackage{array}
\usepackage{multirow}
\usepackage{amsmath}
\usepackage{amssymb}
\usepackage{graphicx}
\usepackage{setspace}
\makeatletter

\providecommand{\tabularnewline}{\\}

\newenvironment{lyxcode}
	{\par\begin{list}{}{
		\setlength{\rightmargin}{\leftmargin}
		\setlength{\listparindent}{0pt}
		\raggedright
		\setlength{\itemsep}{0pt}
		\setlength{\parsep}{0pt}
		\normalfont\ttfamily}%
	 \item[]}
	{\end{list}}

\date{}
\usepackage[margin=1in]{geometry}

\usepackage{babel}

\makeatother

\usepackage{babel}
\begin{document}
\title{\textbf{Space Occupancy in Low-Earth Orbit}}
\maketitle
\noindent \begin{center}
Claudio Bombardelli\footnote{Associate Professor, Space Dynamics Group}
\par\end{center}

\noindent \begin{center}
\textit{Technical University of Madrid (UPM), Madrid, 28040,}{} Spain,
\par\end{center}

\noindent \begin{center}
Gabriele Falco\footnote{Graduate Student, Department of Industrial Engineering}
\par\end{center}

\noindent \begin{center}
\textit{University of Naples Federico II, Naples, 80125,}{} Italy,
\par\end{center}

\noindent \begin{center}
Davide Amato\footnote{Postdoctoral Research Associate, Smead Aerospace Engineering Sciences} 
\par\end{center}

\noindent \begin{center}
\textsl{University of Colorado, Boulder, CO 80309,} 
\par\end{center}

\noindent \begin{center}
and
\par\end{center}

\noindent \begin{center}
Aaron J. Rosengren\footnote{Assistant Professor, Department of Mechanical and Aerospace Engineering} 
\par\end{center}

\noindent \begin{center}
\textsl{UC San Diego, La Jolla, CA 92093 }
\par\end{center}

\noindent \begin{center}
\textbf{abstract}
\par\end{center}

With the upcoming launch of large constellations of satellites in
the low-Earth orbit  (LEO) region it will become important to organize
the physical space occupied by the different operating satellites
in order to minimize critical conjunctions and avoid collisions. Here,
we introduce the definition of space occupancy as the domain occupied
by an individual satellite as it moves along its nominal orbit under
the effects of environmental perturbations throughout a given interval
of time. After showing that space occupancy for the zonal problem
is intimately linked to the concept of frozen orbits and proper eccentricity,
we provide frozen-orbit initial conditions in osculating element space
and obtain the frozen-orbit polar equation to describe the space occupancy
region in closed analytical form. We then analyze the problem of minimizing
space occupancy in a realistic model including tesseral harmonics,
third-body perturbations, solar radiation pressure, and drag. The
corresponding initial conditions, leading to what we call minimum
space occupancy (MiSO) orbits, are obtained numerically for a set
of representative configurations in LEO. The implications for the
use of MiSO orbits to optimize the design of mega-constellations are
discussed.

\section*{Introduction}

Preserving and sustaining the Low Earth Orbit (LEO) environment as
a valuable resource for future space users has motivated space actors
to consider mechanisms to control the growth of man-made debris. These
prevention, mitigation, and remediation actions will become more and
more urgent following the launch of upcoming mega-constellations of
satellites to provide high-bandwidth, space-based internet access.
Envisioned mega-constellation designs involve the deployment of thousands
of satellite at nominally equal altitude and inclination and distributed
over a number of orbital planes for optimized ground coverage. The
concentration of such a high number of satellites in a relatively
small orbital region can lead to a high risk of in-orbit collisions
and an escalation of required collision avoidance maneuvers\cite{virgili2016risk,radtke2017interactions,le2018space}.
In this scenario, any design solution that can limit potential collisions
and required maneuvers as much as possible would be highly welcomed.

A possible collision mitigation action that can be implemented at
a negligible cost for space operators is to minimize the potential
interference of a satellite with the rest of its constellation members
by a judicious orbit design within the limits imposed by mission requirements.
Ideally, if each individual satellite could be confined to within
a region of space with zero overlap between the rest of the constellation
members, the endogenous collision risk and frequency of collision
avoidance maneuvers of a constellation of satellites would be reduced
to zero.

In a perturbation-free environment, the obvious solution would be
to adopt a sequence of orbits of equal eccentricity, but slightly
different semi-major axes. Considering a more accurate model that
includes zonal harmonic perturbations, non-intersecting orbits can
still be achieved by placing the individual satellites in non-overlapping
frozen orbits of slightly different semi-major axes. Frozen orbits
(see \cite{aorpimai2003analysis} and references therein) show the
remarkable property of having constant altitude at equal latitude\footnote{Note that here, and in the rest of the article, we employ the terms
``altitude'' and ``latitude'' to refer to ``geocentric altitude''
and ``geocentric''latitude''} which is a consequence of the fact that their singly-averaged eccentricity,
argument of pericenter, and inclination, are constant.

When tesseral harmonics, third-body effects, and non-gravitational
perturbations are accounted for, perfectly frozen orbits cease to
exist, which makes it impossible to achieve control-free, constant-altitude
orbital motion at equal latitude. Accordingly, one can attempt to
minimize residual altitude oscillations by adopting initial conditions
near to the ones corresponding to a frozen orbit in the zonal problem
and to approach the absolute minimum by a slight variation of the
initial state vector. To our knowledge, there has been no effort in
the available literature to obtain initial conditions leading to an
absolute minimum of the altitude variations of a LEO orbiting spacecraft
in a given time span\footnote{Note that the main requirement for many Earth observations missions
that are flying, or have flown, in near-frozen orbits (like TOPEX-Poseidon,
Jason and Sentinel) is to minimize ground track error over the repeat
pattern rather than altitude oscillations \cite{bhat1998topex}.}. Note that quasi-frozen orbits including tesseral harmonics, third-body
perturbations and solar radiation pressure have been obtained in the
literature using a double-averaging approach (\cite{nie2018lunar,abad2009analytical,condoleo2016frozen,tresaco2016frozen}),
which certainly provides an increase in orbit lifetime and stability
but does not necessarily lead to a minimization of altitude oscillations
at equal latitude. 

In this article, we employ analytical and numerical methods to study
what we call ``space occupancy range'' (SOR), ``space occupancy
area'' (SOA), and ``space occupancy volume'' (SOV) of a satellite
in LEO. The first quantity corresponds to the extent of the equal-latitude
radial displacement of the satellite in a given time span, while the
second and third represent, respectively, the total surface area and
volume swept by the satellite throughout a given time span as it moves
in its osculating orbital plane (SOA) or in the orbital space (SOV).
Moreover, we employ a high-fidelity numerical algorithm to determine
MiSO initial conditions for an orbit with a given semi-major axis
and inclination. Once these initial conditions are established and
the dynamical behavior of these orbits is well understood, we propose
to organize the orbital space of future mega-constellations by distributing
the different satellites in non-overlapping MiSO shells thus minimizing
the number of critical conjunctions between satellites of different
orbital planes, and, consequently, the frequency of collision avoidance
maneuvers.

The structure of the article is the following. First, we provide a
definition of space occupancy and review frozen-orbit theory for the
zonal problem starting from the seminal 1966 article by Cook \cite{cook1966perturbations}.
Next, we show how space occupancy can be directly related to the concept
of proper eccentricity. We then derive simple analytical formulas
to obtain near-frozen initial conditions in osculating element space
based on the Kozai-Brower-Lyddane mean-to-osculating element transformations
and obtain a compact and accurate analytical expression for the polar
equation of a frozen orbit in the zonal problem.

In the last section of the article, we investigate space occupancy
considering a high-fidelity model including high-order tesseral harmonics,
lunisolar perturbations and non-gravitational perturbations (solar
radiation pressure and drag). It is important to underline that an
accurate modeling of the Earth attitude and rotation (including precession,
nutation and polar motion) is taken into account when computing high-order
tesseral harmonics.

Numerical simulations are conducted in order to obtain minimum space
occupancy initial conditions and map the minimum achievable space
occupancy for different altitudes and inclinations in LEO. The time
evolution of the space occupancy of MiSO orbits under the effect of
environmental perturbations is also investigated in detail. Finally,
the implications of these results on the design of minimum-conjunction
mega-constellations of satellites for future space-based internet
applications are discussed.

\section*{Space Occupancy: Definition}

We define the \textit{space occupancy range} of an orbiting body,
of negligible size compared to its orbital radius, over the interval
$\left[t_{0},\,t_{0}+\Delta t\right]$, as the maximum altitude variation
for fixed latitude experienced by the body throughout that time interval:

\[
\mathrm{SOR}\left(t_{0},\Delta t\right)=\max\left\{ \Delta r\left(\phi\right),\phi\in\left[0,\phi_{max}\right],t\in\left[t_{0},\,t_{0}+\Delta t\right]\right\} 
\]

where the maximum reachable latitude $\phi_{max}$ can be taken, with
good approximation, as the mean orbital inclination $\hat{i}$. 

Based on the preceding definition, there are two ways of following
the \textit{time evolution} of the SOR, depending whether $t_{0}$
or $\Delta t$ is held constant, which leads to the definition of
a cumulative vs. fixed-timespan SOR function. The \textit{cumulative
SOR} is a monotonic function of the time span $\Delta t$ that describes
how a spacecraft, starting from a fixed epoch $t_{0}$, occupies an
increasing range of radii as its orbit evolves in time under the effect
of the different perturbation forces. Conversely, the \textit{fixed-timespan
SOR} is a function that measures how the SOR changes as the initial
epoch of the measurement interval moves forward in time while the
timespan $\Delta t$ is held fixed. 

We define the \textit{space occupancy area} over the interval $\left[t_{0},\,t_{0}+\Delta t\right]$,
as the smallest two-dimensional region in the mean orbital plane containing
the motion of the orbiting body as its orbit evolves throughout that
time interval.

Finally, we define the \textsl{space occupancy volume} over the interval
$\left[t_{0},\,t_{0}+\Delta t\right]$, as the volume swept by the
SOA when the orbital plane precesses around the polar axis of the
primary body.

When the most important perturbation terms are those stemming from
the zonal harmonic potential with a dominant second order ($J_{2}$)
term, as it is in the case of LEO, the SOA is an annulus of approximately
constant thickness and whose shape will be shown, in this article,
to correspond to an offset ellipse. Under the same hypothesis the
SOV takes the shape of a barrel whose characteristics will also be
studied.

\section*{Frozen Orbits for the Zonal Problem}

The theory of frozen orbits was pioneered by Graham E. Cook in his
seminal 1966 paper \cite{cook1966perturbations}. Here, we summarize
Cook's equations and their implications for the space occupancy concept.
In line with Cook, the dynamical model we refer to in this section
accounts for the effect of $J_{2}$ plus an arbitrary number of odd
zonal harmonics. 

Let us employ dimensionless units of length and time, taking the Earth
radius $R_{\oplus}$ as the reference length and $1/n_{\oplus}$ as
the reference time with $n_{\oplus}$ indicating the mean motion of
a Keplerian circular orbit of radius $R_{\oplus}$. Let us indicate
with $\hat{e}$, $\hat{\omega}$, $\hat{a}$, $\hat{n}$ and $\hat{i}$
the mean value (i.e., averaged over the mean anomaly) of the eccentricity,
argument of periapsis, semi-major axis, mean motion and inclination,
respectively, where the latter is considered constant after neglecting
its small-amplitude long-periodic oscillations.

The differential equations describing the evolution of the mean eccentricity
vector perifocal components, $\xi=\hat{e}\cos\hat{\omega}$ and $\eta=\hat{e}\sin\hat{\omega}$,
are \cite{cook1966perturbations}:

\begin{equation}
\left\{ \begin{array}{c}
\dot{\xi}=-k\left(\eta+e_{f}\right),\\
\\
\dot{\eta}=k\xi,
\end{array}\right.\label{eq:ecc_vector_sec=000026LP}
\end{equation}
where:

\[
k=\frac{3\hat{n}J_{2}}{\hat{a}^{3}}\left(1-\frac{5}{4}\sin^{2}\hat{i}\right),
\]
and $e_{f}$, known as \textit{frozen eccentricity}, can be expressed
as \cite{cook1966perturbations}:

\begin{equation}
e_{f}=k^{-1}\hat{a}_{0}^{-3/2}\sum_{n=1}^{N}\frac{J_{2n+1}}{\hat{a}_{0}^{2n+1}}\frac{n}{\left(2n+1\right)\left(n+1\right)}P_{2n+1}^{1}(0)P_{2n+1}^{1}(\cos\hat{i})=-\frac{J_{3}}{2J_{2}}\frac{\sin\hat{i}}{\hat{a}}+o\left(J_{3}/J_{2}\right),\label{eq:e_frozen}
\end{equation}
with $P_{n}^{1}$ indicating the associated Legendre function of order
one and degree $n$. 

The solution of Eqs. (\ref{eq:ecc_vector_sec=000026LP}) is:

\begin{equation}
\left\{ \begin{array}{c}
\xi\left(\tau\right)=e_{p}\cos\left(k\tau+\alpha\right),\\
\\
\eta\left(\tau\right)=e_{p}\sin\left(k\tau+\alpha\right)+e_{f},
\end{array}\right.\label{eq:Cook_ecc_diagram}
\end{equation}
where:

\begin{equation}
e_{p}=\sqrt{\left(\hat{e}_{0}\sin\hat{\omega}_{0}-e_{f}\right)^{2}+\hat{e}_{0}^{2}\cos^{2}\hat{\omega}_{0}},\label{eq:e_proper}
\end{equation}

\[
\sin\alpha=\frac{\hat{e}_{0}\sin\hat{\omega}_{0}-e_{f}}{e_{p}},\qquad\cos\alpha=\frac{\hat{e}_{0}\cos\hat{\omega}_{0}}{e_{p}}.
\]

Eqs.(\ref{eq:Cook_ecc_diagram}) corresponds to a circle of radius
$e_{p}$, which is a constant today known as the \textit{proper eccentricity},
and center $\left(0,e_{f}\right)$ in the $\xi-\eta$ plane. By selecting
as initial conditions $\hat{\omega}_{0}=\pi/2$ and $\hat{e}_{0}=e_{f}$
the circle reduces to a point and both $\hat{\omega}$ and $\hat{e}$
remain constant, implying that their long-periodic oscillations have
been eliminated and yielding what is known as a \textit{frozen orbit}.
Note that long-periodic oscillations of the inclination and mean anomaly
are also removed under the frozen orbit conditions as it is evident
from \cite[page 394]{brouwer1959solution}.

\section*{Space Occupancy for the Zonal Problem}

One remarkable feature of frozen orbits is that they have a constant
altitude for a given latitude. This is a consequence of the fact that
the long-periodic variations in the magnitude and direction of the
eccentricity vector are (within the validity of the averaging approximation)
identically zero.

That feature can be shown mathematically by writing the orbital radius
as:

\[
r=\frac{\left(\hat{a}+a_{sp}\right)\left(1-\left(\hat{e}+e_{sp}\right)^{2}\right)}{1+\left(\hat{e}+e_{sp}\right)\cos\nu},
\]
where $a_{sp}$ and $e_{sp}$ are the short-periodic components of,
respectively, the semi-major axis and eccentricity and $\nu$ is the
osculating true anomaly. 

Since all short-periodic components are small quantities we can write:

\begin{equation}
r=\hat{r}+r_{sp}\approx\frac{\hat{a}\left(1-\hat{e}^{2}\right)}{1+\hat{e}\cos\nu}+\left(\frac{1-\hat{e}^{2}}{1+\hat{e}\cos\nu}a_{sp}-\frac{\hat{a}\left[2\hat{e}+\left(1+\hat{e}^{2}\right)\cos\nu\right]}{\left(1+\hat{e}\cos\nu\right)^{2}}e_{sp}\right),\label{eq:radial_oscillations}
\end{equation}

In the above equation $r_{sp}$ and $\hat{r}$ are, respectively,
the fast- and slow-scale of the orbit radius variation. 

On the other hand, the relation between the orbit latitude, $\phi$,
and true anomaly reads:

\begin{equation}
\frac{\sin\phi}{\sin i}=\sin\left(\nu+\omega\right).\label{eq:phi_nu}
\end{equation}

For a frozen orbit, $\hat{e}$ is a constant and, since $\hat{\omega}$
is also constant and equal to $\pi/2$, both $a_{sp}$ and $e_{sp}$
are periodic functions with $\cos\nu$, $\cos2\nu$ and $\cos3\nu$
terms \cite{kozai1959motion}. This means that both $\hat{r}$ and
$r_{sp}$ are explicit functions of $\nu$. Moreover, under frozen-orbit
conditions and neglecting short-periodic oscillations of $i$ (i.e.,
$i\simeq\hat{i}=\mathrm{const}$) as well as short-periodic oscillations
of $\omega$ (i.e. , $\omega\simeq\hat{\omega}=\pi/2$) the true anomaly
$\nu$ is, following Eq. (\ref{eq:phi_nu}), an explicit function
of $\phi$: 

\[
\nu\approx\cos^{-1}\left(\frac{\sin\phi}{\sin\hat{i}}\right).
\]

This proves that for a frozen orbit the terms $\hat{r}$ and $r_{sp}$
in Eq.(\ref{eq:radial_oscillations}) are explicit functions of $\phi$
and the SOR is zero.

When frozen conditions are not met the term $\hat{r}$ is no longer
an explicit function of $\nu$ owing to the long-periodic variations
of $\hat{e}$. Likewise $\nu$ is no longer an explicit function of
$\phi$ owing to the long-periodic variations of $\hat{\omega}$.
Neglecting the contribution of $r_{sp}$ compared to $\hat{r}$, the
SOR corresponds to the maximum ``mean'' apoapsis minus the minimum
``mean'' periapsis, and, accounting for Cook's solution (Eq.(\ref{eq:Cook_ecc_diagram})):

\[
\mathrm{SOR}=\left(\Delta r\right)_{\max}\approx\hat{a}_{0}\left(\hat{e}_{\max}-\hat{e}_{\min}\right)=2\hat{a}e_{p},
\]
showing that space occupancy in the zonal problem is fundamentally
related to the \textit{proper eccentricity, $e_{p}$,} of the orbit.
Note that when $\hat{e}_{0}>>e_{f}$ one has $e_{p}\simeq\hat{e}_{0}$
as it is evident from Eq. (\ref{eq:e_proper}).

\section*{Frozen Orbit Dynamics and Geometry}

In order to fully characterize space occupancy we will now obtain
simple relations characterizing the geometry of frozen orbits. In
order to do that one needs to view frozen orbits in osculating elements
space using the mean-to-osculating orbital elements conversion formulas
(\cite{kozai1959motion,lyddane1963small}) reported, for convenience,
in Appendix I.

\subsection*{Maximum-Latitude Conditions}

Owing to the axial symmetry of the zonal problem and their periodic
nature, frozen orbits are axially symmetric. Therefore, at the maximum
latitude ($\omega+\nu=\pi/2$) the satellite must be either at periapsis
($\omega=\pi/2$, $M=\nu=0$) or apoapsis ($\omega=3\pi/2$, $M=\nu=\pi$)
of its osculating orbit. Consequently, the computation of the frozen
orbit initial conditions is very convenient when referring to the
maximum latitude point.

Following Kozai's \cite{kozai1959motion}, the osculating eccentricity
can be written as a sum of a mean and a short-periodic term:

\[
e=\hat{e}+e_{sp},
\]
where the short-periodic component $e_{sp}$ is dominated by the $J_{2}$
perturbation and obeys Eq. (\ref{eq:e_sp}) given in Appendix I.

In frozen orbit conditions $\hat{e}=e_{f}$ and $\hat{\omega}=\pi/2$.
In addition, the mean true anomaly, here denoted with $\hat{\nu}$,
must be zero at the maximum latitude point for symmetry. From Eq.
(\ref{eq:e_sp}) the short-periodic part of the eccentricity at maximum
latitude (subindex ``$N$'' as in ``North'' ) yields, after neglecting
second order terms in $e_{f}$ and $J_{2}$:

\[
e_{sp,N}\approx\frac{J_{2}}{2\hat{a}^{2}}\left(7\cos^{2}\hat{i}-4\right),
\]

By setting the previous expression to $-e_{f}$ and solving for $\hat{i}$
one obtains the inclination value at which the maximum-latitude osculating
orbit becomes circular:

\begin{equation}
e_{N}=e_{sp,N}+e_{f}=0\;\mathrm{for\;\mathit{\begin{cases}
\hat{i}=\hat{i}^{*},\\
\hat{i}=\pi-\hat{i}^{*},
\end{cases}}}\label{eq:e_N}
\end{equation}
with:

\[
\hat{i}^{*}\approx\cos^{-1}\left(\sqrt{\frac{2}{7}\left(2-\frac{\hat{a}^{2}e_{f}}{J_{2}}-\frac{15e_{f}}{4}\right)}\right).
\]

For the LEO case with altitudes between 400 and 2000 km, $\hat{i}^{*}$
oscillates between $\sim41{}^{\circ}$ and $\sim66^{\circ}$ depending
mainly on $\hat{i}$ .

By considering the short-periodic part of the argument of periapsis
(see Eq.(\ref{eq:omega_sp_tilde}) in Appendix I):

\[
\tilde{\omega}_{sp,N}\approx-\mathrm{atan2}\left(0,e_{sp,N}\right),
\]
so that the maximum-latitude osculating argument of periapsis yields:

\begin{equation}
\omega_{N}=\tilde{\omega}_{sp,N}+\pi/2=\left\{ \begin{array}{ccc}
\pi/2 & \mathrm{for} & \hat{i}^{*}\lesssim\hat{i}\lesssim\pi-\hat{i}^{*},\\
\\
3\pi/2 & \mathrm{} & \mathrm{otherwise},
\end{array}\right.\label{eq:omega_N}
\end{equation}

which means that the maximum-latitude point corresponds to osculating
apoapsis when $\hat{i}^{*}\lesssim\hat{i}\lesssim\pi-\hat{i}^{*}$
and to osculating periapsis otherwise (see Table \ref{tab:TAB1}).
Consequently, the maximum-latitude osculating true anomaly reads:

\begin{equation}
\nu_{N}=\pi/2-\omega_{sp,N}=\left\{ \begin{array}{ccc}
0 & \mathrm{for} & \hat{i}^{*}\lesssim\hat{i}\lesssim\pi-\hat{i}^{*},\\
\\
\pi & \mathrm{} & \mathrm{otherwise.}
\end{array}\right.\label{eq:nu_N}
\end{equation}

\begin{doublespace}
\noindent 
\begin{table}[htbp]
\caption{Frozen orbits mean anomaly and argument of periapsis at maximum latitude}

\smallskip{}

\centering{}\label{tab:TAB1}%
\begin{tabular}{cccccc}
\hline 
\noalign{\vskip\doublerulesep}
inclination range & orbit & $M$ & $\omega$ & $\hat{M}$ & $\hat{\omega}$\tabularnewline[\doublerulesep]
\hline 
\noalign{\vskip\doublerulesep}
\noalign{\vskip\doublerulesep}
\multirow{1}{*}{$\hat{i}\lesssim\hat{i}^{*}$ or $\hat{i}\gtrsim\pi-\hat{i}^{*}$} & periapsis & $0^{\circ}$ & $90^{\circ}$ & $0^{\circ}$ & $90^{\circ}$\tabularnewline[\doublerulesep]
\noalign{\vskip\doublerulesep}
\noalign{\vskip\doublerulesep}
\multirow{1}{*}{$\hat{i}^{*}\lesssim\hat{i}\lesssim\pi-\hat{i}^{*}$} & apoapsis & $180^{\circ}$ & $270^{\circ}$ & $0^{\circ}$ & $90^{\circ}$\tabularnewline[\doublerulesep]
\hline 
\noalign{\vskip\doublerulesep}
\end{tabular}
\end{table}

\end{doublespace}

Similarly, following the formulas reported in Appendix I, we can obtain
compact expressions for the maximum-latitude osculating semi-major
axis and inclination as:

\begin{equation}
i_{N}\approx\hat{i}-\frac{3J_{2}}{8\hat{a}^{2}}\sin2\hat{i},\label{eq:i_N}
\end{equation}

\begin{equation}
a_{N}\approx\hat{a}-\frac{3J_{2}}{2\hat{a}}\sin^{2}\hat{i}.\label{eq:a_N}
\end{equation}

Eqs. (\ref{eq:e_N}-\ref{eq:a_N}) can be employed to obtain frozen
orbit initial conditions at maximum latitude in terms of osculating
orbital elements and starting from a set of desired mean orbital elements.

\subsection*{Frozen-Orbit Polar Equation}

Given the smallness of the eccentricity for a frozen orbit, the orbit
radius obeys, to first order in $e$:

\begin{equation}
r\simeq a\left(1-e\cos M\right).\label{eq:r_frozen_orbit}
\end{equation}
The osculating semi-major axis can be split into a mean and short-period
part (Eq.(\ref{eq:a_sp}) in Appendix I) leading to:

\begin{equation}
a=\hat{a}+\dfrac{J_{2}}{2\hat{a}}\left[\left(2-3\kappa\right)\left(\dfrac{\hat{a}^{3}}{\hat{r}^{3}}-\dfrac{1}{\lambda^{3}}\right)+\dfrac{3\hat{a}^{3}}{\hat{r}^{3}}\kappa\,\cos\left(2\hat{\nu}+2\hat{\omega}\right)\right],\label{eq:a_osc}
\end{equation}
where
\[
\lambda=\sqrt{1-\hat{e}^{2}},\qquad\kappa=\sin^{2}\hat{i}.
\]

In addition, following Lyddane's expansion (Eq.(\ref{eq:ecosM_exp})
in Appendix I), one has:

\begin{equation}
e\cos M\simeq\left(\hat{e}+e_{sp}\right)\cos\hat{M}-\hat{e}M_{sp}\sin\hat{M},\label{eq:e_cosM}
\end{equation}
where the expressions of $e_{sp}$ (Eq. (\ref{eq:e_sp})) and $\hat{e}M_{sp}$
(Eq. (\ref{eq:eM_sp})) are also reported in Appendix I.

In the frozen-orbit condition, one has $\hat{\omega}=\pi/2$, $\hat{e}=e_{f}$
and the ``mean'' mean anomaly can be related to the argument of
latitude $\theta$ neglecting second order terms in $e_{f}$:

\[
\hat{M}\simeq\hat{\nu}\approx\theta-\pi/2
\]

After substituting Eqs.(\ref{eq:a_osc})-(\ref{eq:e_cosM}) into Eq.
(\ref{eq:r_frozen_orbit}), taking into account the preceding relations
and expanding in Taylor series for small $J_{2}$ and $e_{f}$ one
obtains the \textit{frozen-orbit polar equation}:

\[
r\left(\theta\right)\simeq\hat{a}\left(1-e_{f}\sin\theta\right)+\dfrac{J_{2}}{4\hat{a}}\left[\left(9+\cos2\theta\right)\kappa-6\right],
\]
which represents an ellipse whose center is offset along a direction
belonging to the orbital plane and orthogonal to the line of nodes.
The maximum- and minimum-latitude orbit radii yield, respectively:

\[
r_{N}\simeq\hat{a}\left(1-e_{f}\right)+\dfrac{J_{2}\left(4\kappa-3\right)}{2\hat{a}}=\hat{a}+\frac{J_{3}}{2J_{2}}\sin\hat{i}-\frac{J_{2}}{2\hat{a}}\left(3-4\sin^{2}\hat{i}\right)+o\left(J_{3}/J_{2}\right),
\]

\[
r_{S}\simeq\hat{a}\left(1+e_{f}\right)+\dfrac{J_{2}\left(4\kappa-3\right)}{2\hat{a}}=\hat{a}-\frac{J_{3}}{2J_{2}}\sin\hat{i}-\frac{J_{2}}{2\hat{a}}\left(3-4\sin^{2}\hat{i}\right)+o\left(J_{3}/J_{2}\right),
\]

The offset orthogonal to the line of node:

\[
\Delta=r_{N}-r_{S}=-2\hat{a}e_{f}=\frac{J_{3}}{J_{2}}\sin\hat{i}+o\left(J_{3}/J_{2}\right),
\]
is negative (i.e., southward) for the Earth case ($J_{3}<0$). 

Given the smallness of $\Delta$ the orbital radius at node crossing
can be computed as:

\[
r_{eq}\approx r\left(\theta=0\right)\simeq\hat{a}+\dfrac{J_{2}\left(5\kappa-3\right)}{2\hat{a}}=\hat{a}-\frac{J_{2}}{2\hat{a}}\left(3-5\sin^{2}\hat{i}\right)+o\left(J_{3}/J_{2}\right),
\]
and the ellipse flattening yields:

\[
f=\frac{\left(r_{N}+r_{S}\right)/2-r_{eq}}{\left(r_{N}+r_{S}\right)/2}\simeq\frac{J_{2}\sin^{2}\hat{i}}{2\hat{a}}.
\]

It can be easily verified that for the Earth case ($J_{2}\simeq1.08\times10^{-3},J_{3}\simeq-2.54\times10^{-5}$),
for any value of $\hat{i}$:

\[
r_{N}<r_{eq}<r_{S}.
\]

Finally, the nodal (draconitic) period can be evaluated, denoting
with $\tau$ the dimensionless time, according to:

\[
T_{\Omega}\simeq2\pi\left(\frac{d\hat{M}}{d\tau}+\frac{d\hat{\omega}}{d\tau}\right)^{-1},
\]
where the rate of the (secular) evolution of the mean anomaly and
argument of pericenter are, respectively \cite{kozai1959motion}:

\[
\frac{d\hat{M}}{d\tau}=\hat{n}+\frac{3J_{2}}{2\hat{a}^{2}\left(1-\hat{e}^{2}\right)^{3/2}}\left(1-\frac{3}{2}\sin^{2}\hat{i}\right),
\]

\[
\frac{d\hat{\omega}}{d\tau}=\frac{3J_{2}}{2\hat{a}^{2}\left(1-\hat{e}^{2}\right)^{2}}\left(2-\frac{5}{2}\sin^{2}\hat{i}\right).
\]

\subsection*{Shape of the Space Occupancy Region }

\begin{figure}[!t]
\centerline{\includegraphics[clip,width=12cm]{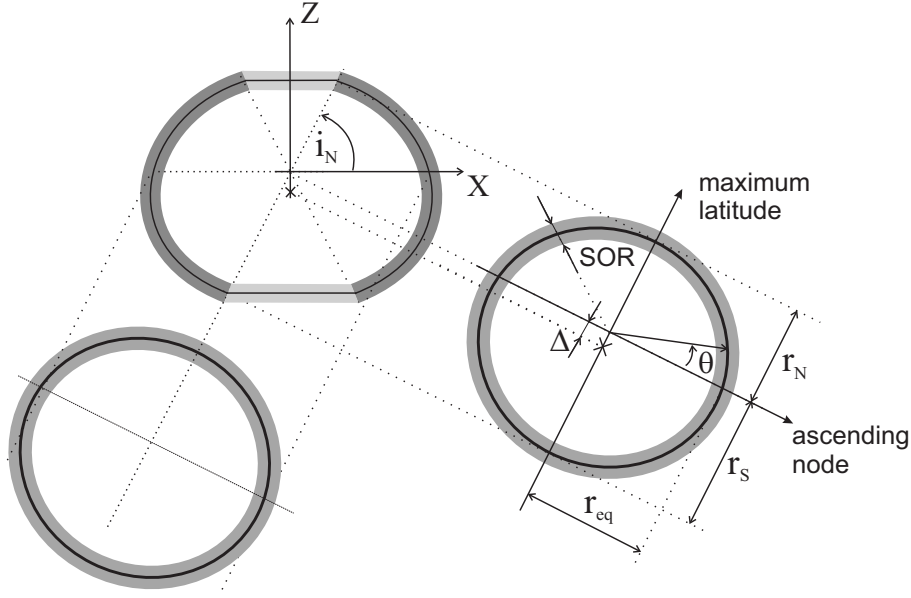}}

\caption{\label{fig:fig1}Geometric relation between the SOV/SOR/SOA and the
frozen orbit trajectory (southward offset and flattening have been
exaggerated for clarity)}
\end{figure}

Based on the considerations of the previous sections we can now characterize
the shape of the space occupancy region as in Figure \ref{fig:fig1}.
With respect to its osculating orbital plane, the orbital motion is
contained inside an annulus, the space occupancy area, whose backbone
is an offset ellipse corresponding to a frozen orbit and whose thickness,
the space occupancy range, is constant and proportional to the orbit
proper eccentricity (Eq.(\ref{eq:e_proper})). As the orbit precesses
around the polar axis $Z$ the orbital motion sweeps a barrel-shaped
3-dimensional region, the space occupancy volume.

In the zonal problem, the SOR is approximately constant and the SOA
and SOV have fixed shape. If the SOR is known, the latter two quantities
can be computed, after neglecting the flattening of the frozen orbit
shape, as:

\[
\mathrm{SOA}\mathrm{\approx2\pi\hat{a}\times SOR},
\]

\[
\mathrm{SOV}\mathrm{\approx4\pi\hat{a}^{2}\sin\hat{i}_{0}\times SOR}.
\]

The preceding expressions highlight the impact of the mean altitude
and inclination, in addition to the SOR, when measuring the occupied
area and orbital volume of a space object.

When time-dependent orbital perturbations are included, on the other
hand, the SOR fluctuate in time as we will show in the next section.
If the cumulative or fixed-timespan SOR is known, the corresponding
SOA and SOV can still be computed with reasonable approximation using
the preceding formulas and taking the average value of the mean semi-major
axis over the SOR computation timespan $\left[t_{0},t_{0}+\Delta t\right]$.

\begin{doublespace}

\end{doublespace}

\section*{Minimum Space Occupancy (MiSO) Orbits}

Let us now consider a much more realistic orbit dynamics model that
includes tesseral harmonics, lunisolar third-body perturbations, solar
radiation pressure, and atmospheric drag. For the results obtained
in this article, the solar radiation pressure perturbation is computed
employing a cannonball model with a reflectivity coefficient $C_{R}=1.2$
and an area to mass ratio of $0.01\:\mathrm{m}^{2}/\mathrm{kg}$,
atmospheric drag is calculated with the same area-to-mass ratio, a
drag coefficient $C_{D}=2.2$, and a simplified static atmospheric
model taken from Vallado \cite[page 564]{vallado2001fundamentals}.
The position of Sun and Moon have been computed using JPL ephemerides.
Finally, we have considered a $23\times23$ geopotential model with
tesseral harmonic coefficients taken from the GRIM5-S1 model \cite{biancale2000new}.

It is clear that zero-occupancy, perfectly frozen orbits cease to
exist in this perturbation environment. The fundamental question is
then how small space occupancy can be made by choosing optimized initial
conditions leading to what we call here \textit{minimum space occupancy
}(MiSO) orbits. The answer to this question can have profound implications
on the design of future mega-constellation of satellites, which could
be organized by stacking non-overlapping space occupancy regions corresponding
to each orbital plane one on top of another by a judicious selection
of the minimum altitude of each plane.

The computation of MiSO initial conditions for the numerical cases
considered in this article has been done numerically using an adaptive
grid-search algorithm to converge to a minimum-occupancy solution
starting from frozen-orbit conditions obtained from the previously
described analytical development. It is important to underline that
each individual point in the grid-search process is a high-fidelity
propagation whose timespan is the one associated to the current SOR
definition (i.e. 100 days) and includes an accurate computation of
the SOR starting from the propagated state vector. This is a very
demanding process in terms of CPU time (the computation of MiSO initial
conditions for an individual constellation plane can take a few hours
with an Intel Core processor i7-4790@3.6GHz) where the use of a very
efficient orbit propagator is paramount. All numerical propagations
were performed using the THALASSA orbit propagator \cite{Amato2019NonaveragedRF},
\cite{amato2018thalassa}.

All MiSO orbits initial conditions derived in this work are reported
in Appendix II for reproducibility purposes.

\begin{doublespace}
\noindent 
\begin{table}[htbp]
\caption{LEO orbits constellations considered in this study}

\centering{}\label{tab:TAB2}%
\begin{tabular}{ccc}
\hline 
\noalign{\vskip\doublerulesep}
orbit class & $h_{N}$ {[}km{]} & $\hat{i}$ {[}deg{]}\tabularnewline[\doublerulesep]
\hline 
\noalign{\vskip\doublerulesep}
\noalign{\vskip\doublerulesep}
class 1 & 550 & 53\tabularnewline[\doublerulesep]
\noalign{\vskip\doublerulesep}
\noalign{\vskip\doublerulesep}
class 2 & 550 & 87.9\tabularnewline[\doublerulesep]
\noalign{\vskip\doublerulesep}
\noalign{\vskip\doublerulesep}
class 3 & 1168 & 53\tabularnewline[\doublerulesep]
\noalign{\vskip\doublerulesep}
\noalign{\vskip\doublerulesep}
class 4 & 1168 & 87.9\tabularnewline[\doublerulesep]
\noalign{\vskip\doublerulesep}
\noalign{\vskip\doublerulesep}
class 5 & 813 & 98.7\tabularnewline[\doublerulesep]
\noalign{\vskip\doublerulesep}
\end{tabular}
\end{table}

\end{doublespace}

Five classes of nominal LEO orbits are considered (see Table \ref{tab:TAB2},
where $h_{N}$ denotes the altitude at maximum-latitude) in line with
existing and upcoming mega-constellations\footnote{At the time of writing of this article, Oneweb has started launching
mega-constellations satellites at around 430 to 620 km mean altitude
and 87.4 degrees of inclination as well as around 1178 km mean altitude
and 87.9 degrees inclination. Starlink on the other hand has launched
at 340 to 550 km mean altitude (presumably with a target 550-km-altitude
orbit) and 53 degrees inclination. We have added the case of a lower-inclination,
high-altitude constellation for completeness.} and including an example of Sun-synchronous orbit (class 5). Each
class comprises 12 orbits with equal mean inclination $\hat{i}$ and
maximum-latitude altitude $h_{N}$ and distributed on 12 orbital planes
spaced by 30 degrees in longitude of node ($\hat{\varOmega}=0^{\circ},30^{\circ},60^{\circ},...,330^{\circ})$.
In other words, each class corresponds to a $"p=12"$ delta-pattern
constellation (see \cite{walker1971some}) except that the number
of satellites in each orbital plane is not specified here. Regarding
the last point, we note that the computation of MiSO initial conditions
for multiple satellites in the same plane can be done by propagating
forward in time the state of one MiSO satellite by a fraction of the
orbital period without expecting any significant departure from individually-computed
MiSO initial conditions. All initial conditions are referred to 1
January 2020 as initial epoch.

Two main scenarios are considered: a drag-free scenario where the
effect of solar radiation pressure and drag is switched off and a
more realistic scenario where both effects are present.

\subsection*{Drag-free MiSO orbits}

Table \ref{tab:TAB3} displays the drag-free, 100-day SOR for 12 orbital
planes of the five classes of MiSO orbits considered in Table \ref{tab:TAB2}.
The results clearly show that lower altitudes and near polar inclinations
(i.e. class 2) results in a wider space occupancy range. This is mainly
due to the combined effect of tesseral harmonics. The corresponding
figures for unoptimized frozen orbits (i.e. orbits obtained by applying
Eqs. (\ref{eq:e_N}-\ref{eq:a_N})) are reported in Table \ref{tab:TAB4}
for comparison and show that MiSO orbits can provide an SOR reduction
of up to almost 600 m compared to the unoptimized case. 

\begin{doublespace}
\noindent 
\begin{table}[htbp]
\caption{100-day SOR {[}km{]} of MiSO orbits in drag-free conditions}

\centering{}\label{tab:TAB3}%
\begin{tabular}{ccccccccccccc}
\hline 
\noalign{\vskip\doublerulesep}
orbit class & $0^{\circ}$  & $30^{\circ}$ & $60^{\circ}$ & $90^{\circ}$ & $120^{\circ}$ & $150^{\circ}$ & $180^{\circ}$ & $210^{\circ}$ & $240^{\circ}$ & $270^{\circ}$ & $300^{\circ}$ & $330^{\circ}$\tabularnewline[\doublerulesep]
\hline 
\noalign{\vskip\doublerulesep}
\noalign{\vskip\doublerulesep}
class 1 & 503 & 493 & 493 & 497 & 498 & 511 & 512 & 513 & 517 & 512 & 508 & 511\tabularnewline[\doublerulesep]
\noalign{\vskip\doublerulesep}
\noalign{\vskip\doublerulesep}
class 2 & 604 & 673 & 604 & 605 & 625 & 643 & 645 & 650 & 647 & 625 & 598 & 687\tabularnewline[\doublerulesep]
\noalign{\vskip\doublerulesep}
\noalign{\vskip\doublerulesep}
class 3 & 378 & 384 & 389 & 392 & 393 & 404 & 404 & 408 & 397 & 387 & 387 & 381\tabularnewline[\doublerulesep]
\noalign{\vskip\doublerulesep}
\noalign{\vskip\doublerulesep}
class 4 & 296 & 297 & 300 & 298 & 290 & 299 & 308 & 309 & 301 & 295 & 295 & 291\tabularnewline[\doublerulesep]
\noalign{\vskip\doublerulesep}
\noalign{\vskip\doublerulesep}
class 5 & 441 & 425 & 462 & 466 & 461 & 471 & 469 & 461 & 448 & 447 & 473 & 445\tabularnewline[\doublerulesep]
\noalign{\vskip\doublerulesep}
\end{tabular}
\end{table}

\noindent 
\begin{table}[htbp]
\caption{100-day SOR {[}km{]} of unoptimized frozen orbits in drag-free conditions}

\centering{}\label{tab:TAB4}%
\begin{tabular}{ccccccccccccc}
\hline 
\noalign{\vskip\doublerulesep}
orbit class & $0^{\circ}$ & $30^{\circ}$ & $60^{\circ}$ & $90^{\circ}$ & $120^{\circ}$ & $150^{\circ}$ & $180^{\circ}$ & $210^{\circ}$ & $240^{\circ}$ & $270^{\circ}$ & $300^{\circ}$ & $330^{\circ}$\tabularnewline[\doublerulesep]
\hline 
\noalign{\vskip\doublerulesep}
\noalign{\vskip\doublerulesep}
class 1 & 575 & 658 & 726 & 754 & 747 & 603 & 631 & 901 & 700 & 860 & 775 & 801\tabularnewline[\doublerulesep]
\noalign{\vskip\doublerulesep}
\noalign{\vskip\doublerulesep}
class 2 & 889 & 1183 & 1040 & 1114 & 983 & 781 & 818 & 868 & 933 & 887 & 906 & 1272\tabularnewline[\doublerulesep]
\noalign{\vskip\doublerulesep}
\noalign{\vskip\doublerulesep}
class 3 & 460 & 505 & 539 & 549 & 560 & 484 & 477 & 603 & 539 & 637 & 578 & 584\tabularnewline[\doublerulesep]
\noalign{\vskip\doublerulesep}
\noalign{\vskip\doublerulesep}
class 4 & 495 & 434 & 520 & 470 & 453 & 426 & 354 & 570 & 482 & 337 & 467 & 586\tabularnewline[\doublerulesep]
\noalign{\vskip\doublerulesep}
\noalign{\vskip\doublerulesep}
class 5 & 535 & 508 & 847 & 876 & 693 & 664 & 668 & 536 & 613 & 771 & 812 & 776\tabularnewline[\doublerulesep]
\noalign{\vskip\doublerulesep}
\end{tabular}
\end{table}

\end{doublespace}

\begin{figure}[!t]
\centerline{\includegraphics[clip,width=8cm]{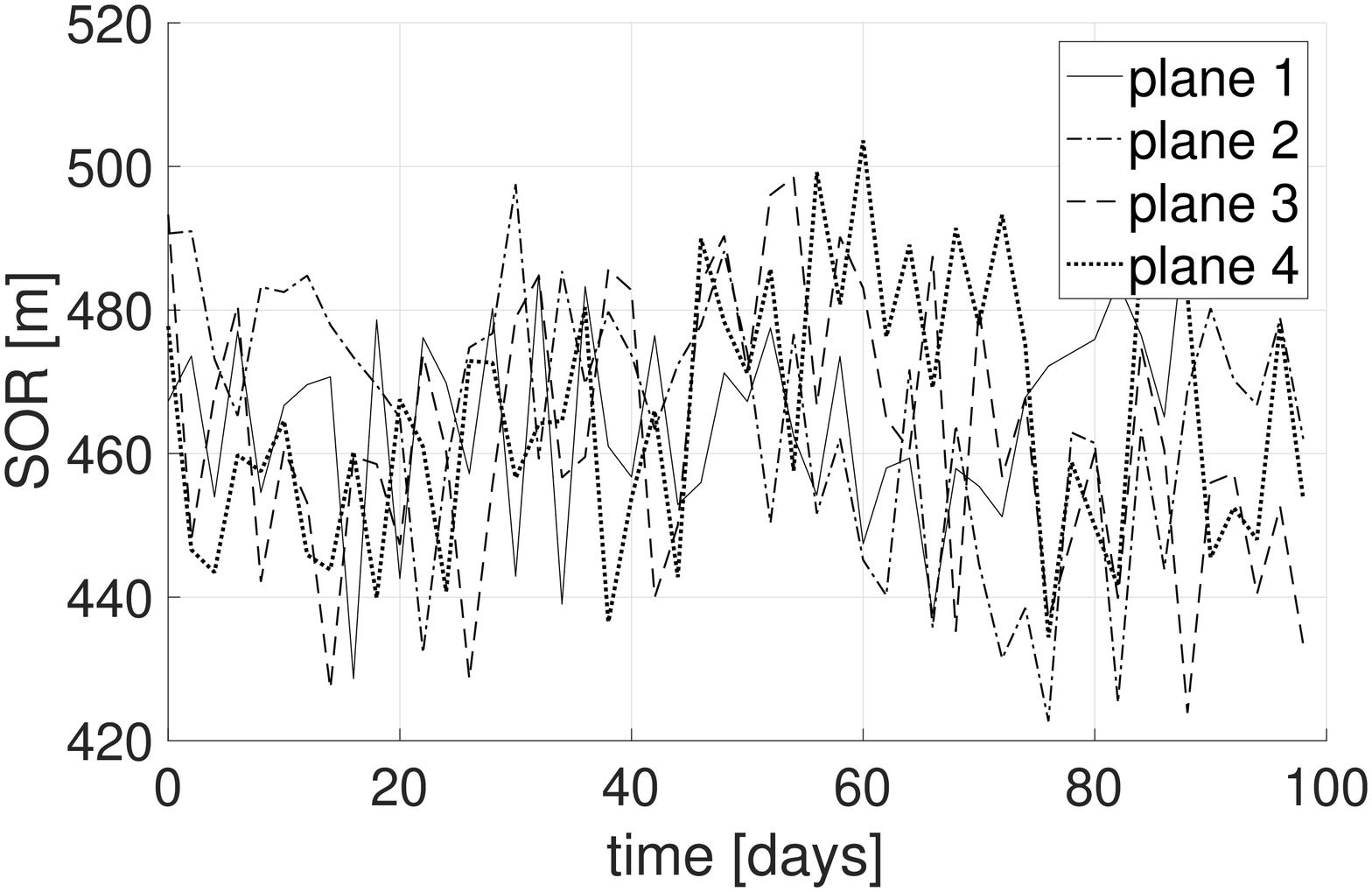}\includegraphics[clip,width=8cm]{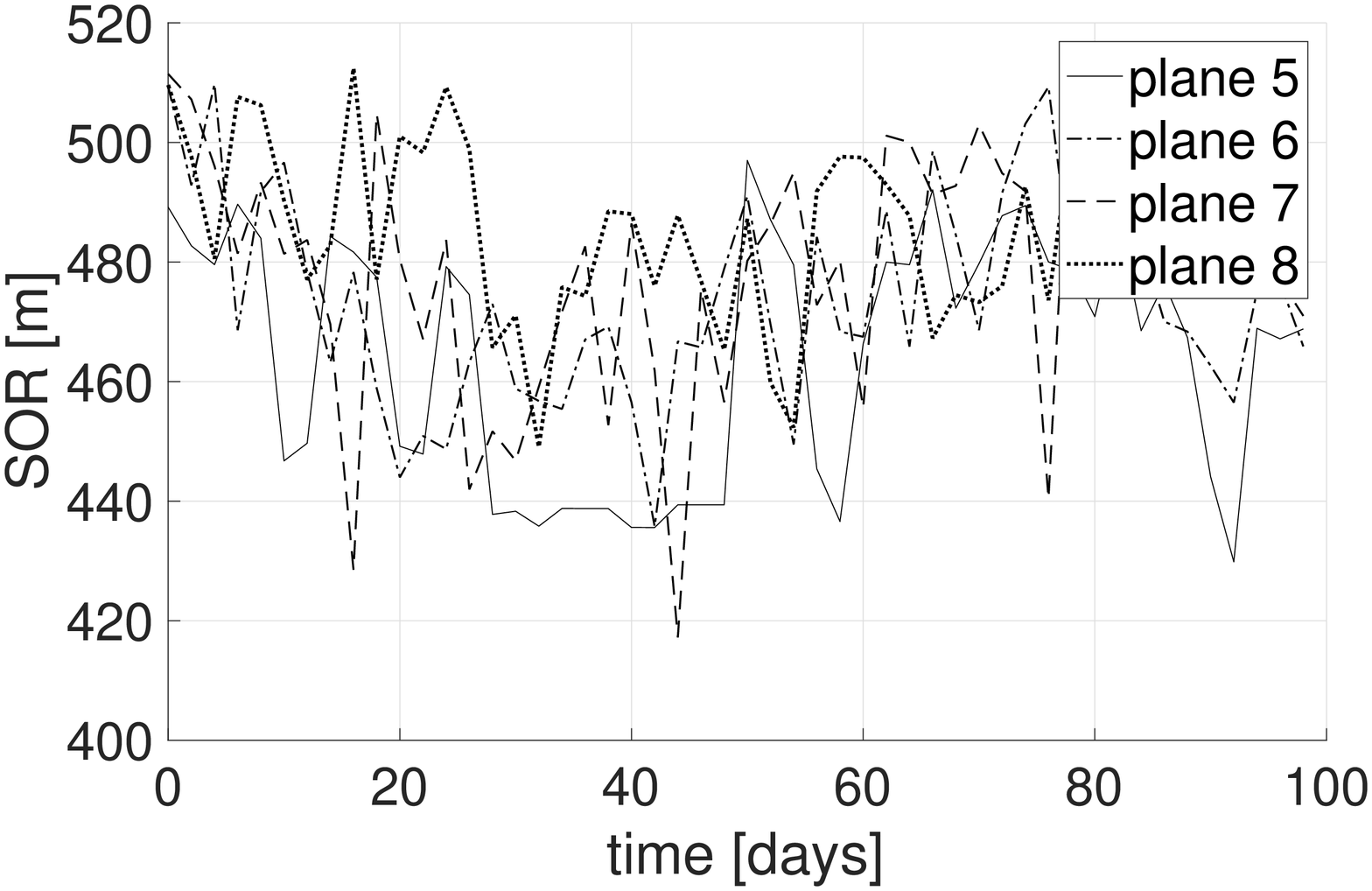}}

\centerline{\includegraphics[clip,width=8cm]{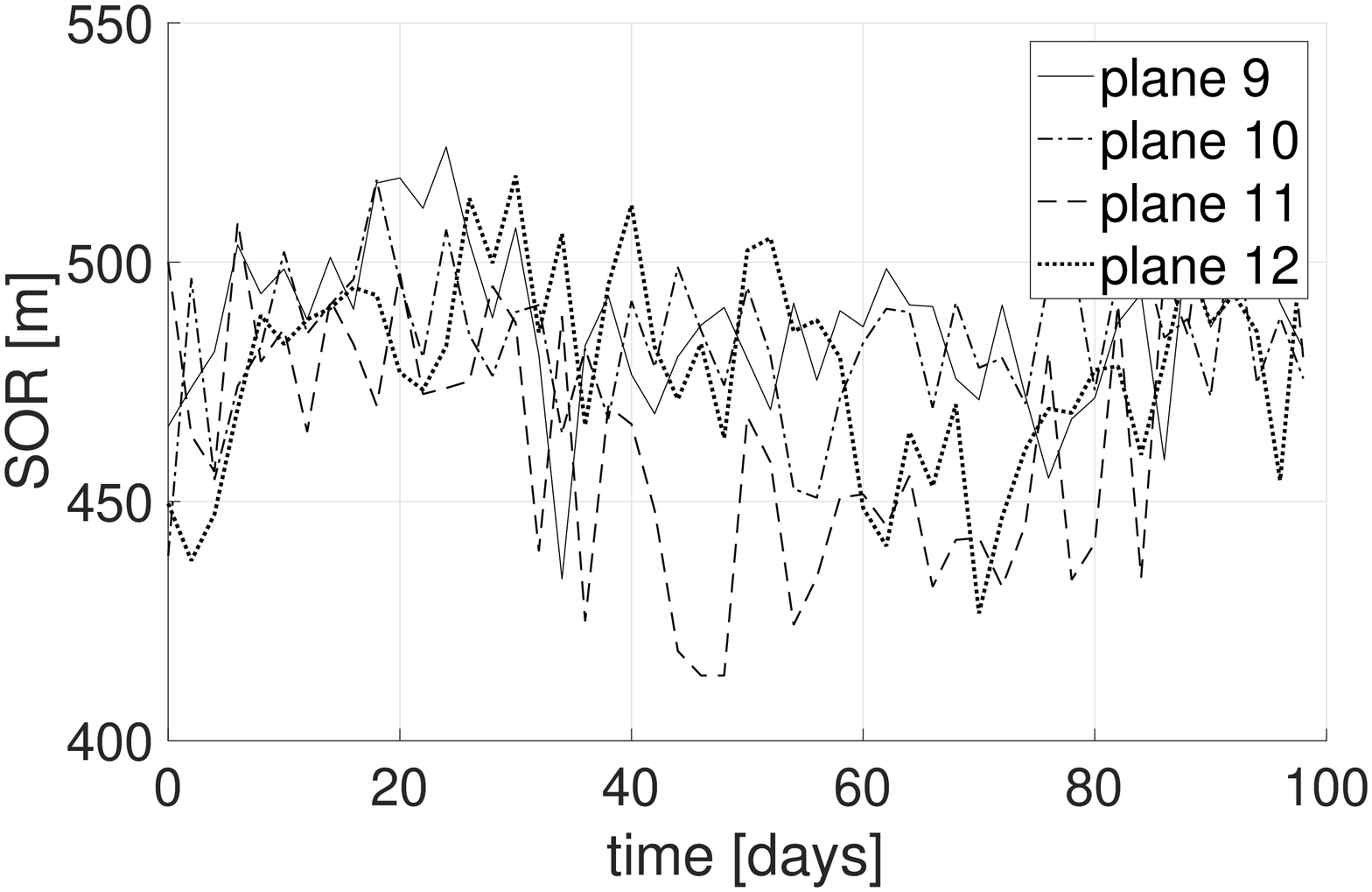}}

\caption{\label{fig:fig2}SOR evolution for class 1 MiSO orbits in drag-free
conditions}
\end{figure}

\begin{figure}[!t]
\centerline{\includegraphics[clip,width=8cm]{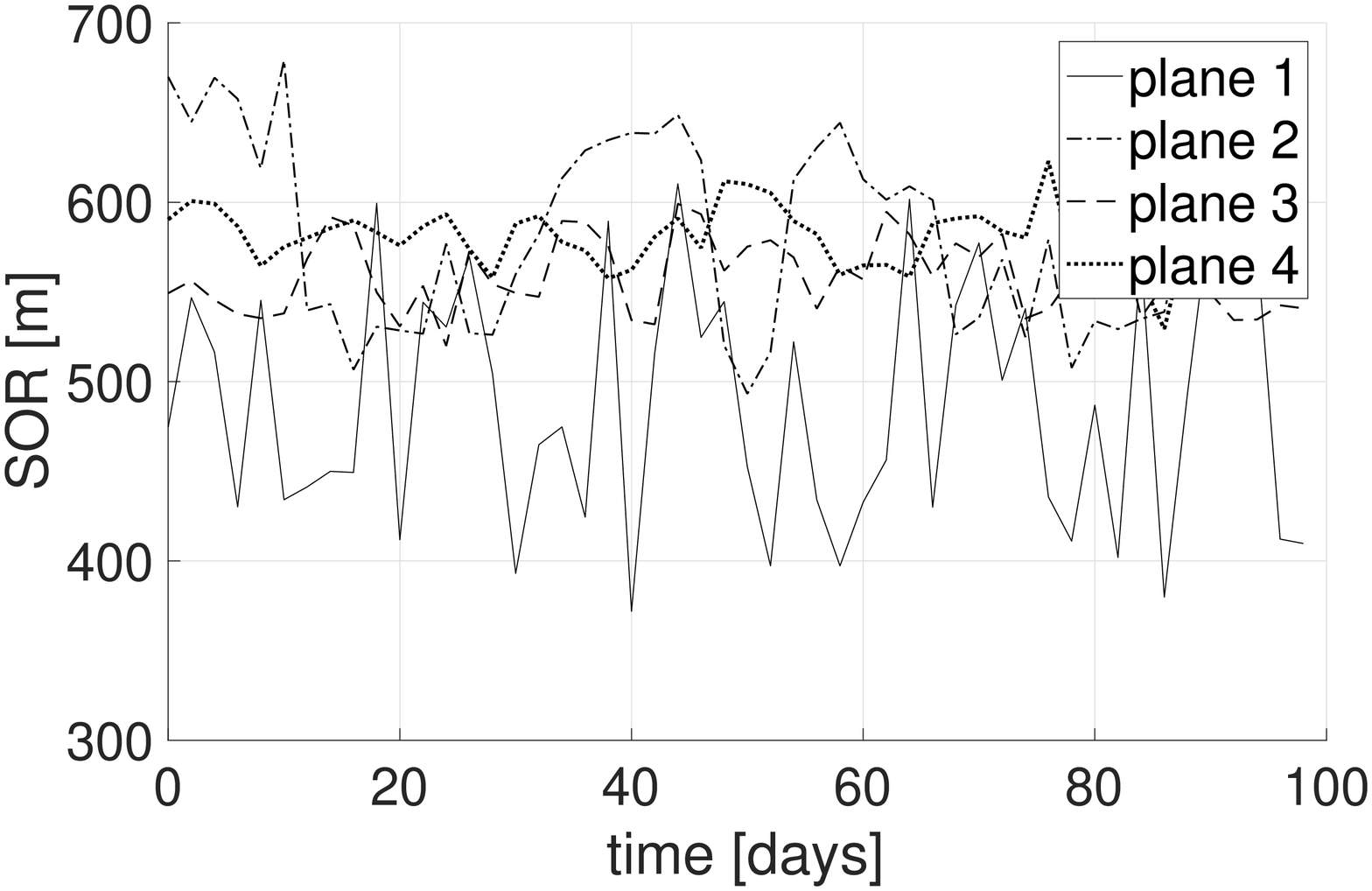}\includegraphics[clip,width=8cm]{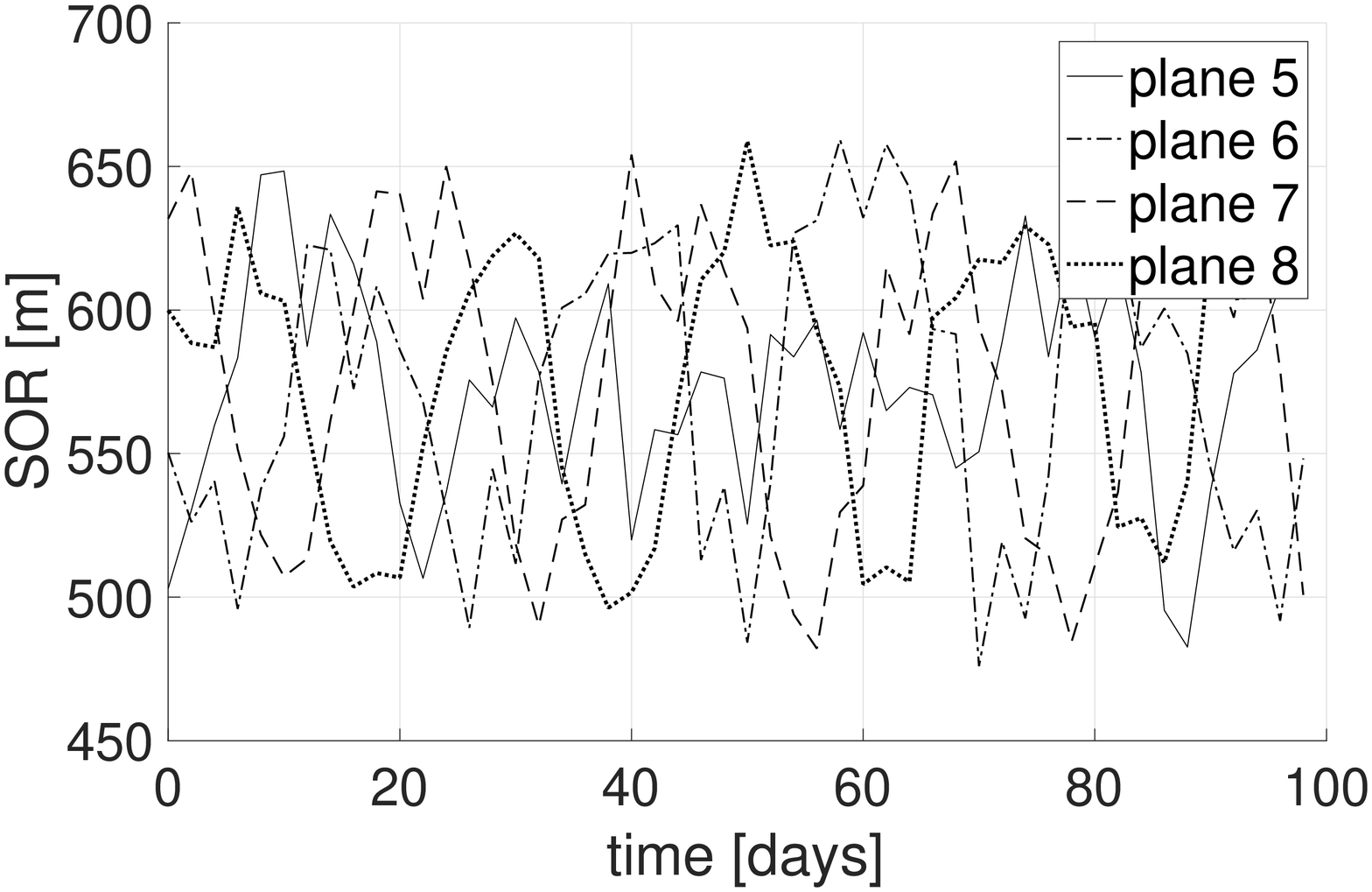}}

\centerline{\includegraphics[clip,width=8cm]{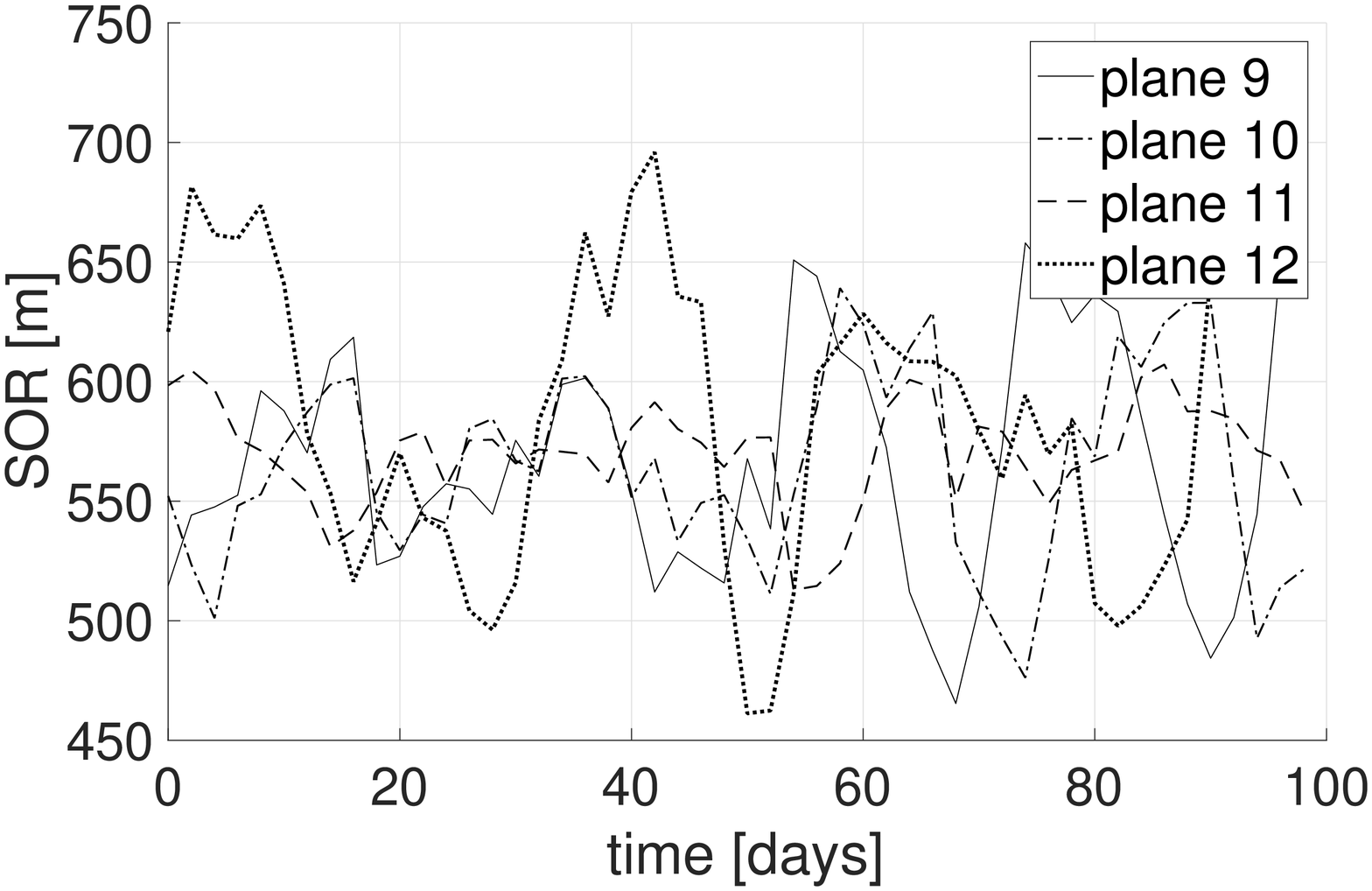}}

\caption{\label{fig:fig3}SOR evolution for class 2 MiSO orbits in drag-free
conditions }
\end{figure}

\begin{figure}[!t]
\centerline{\includegraphics[clip,width=8cm]{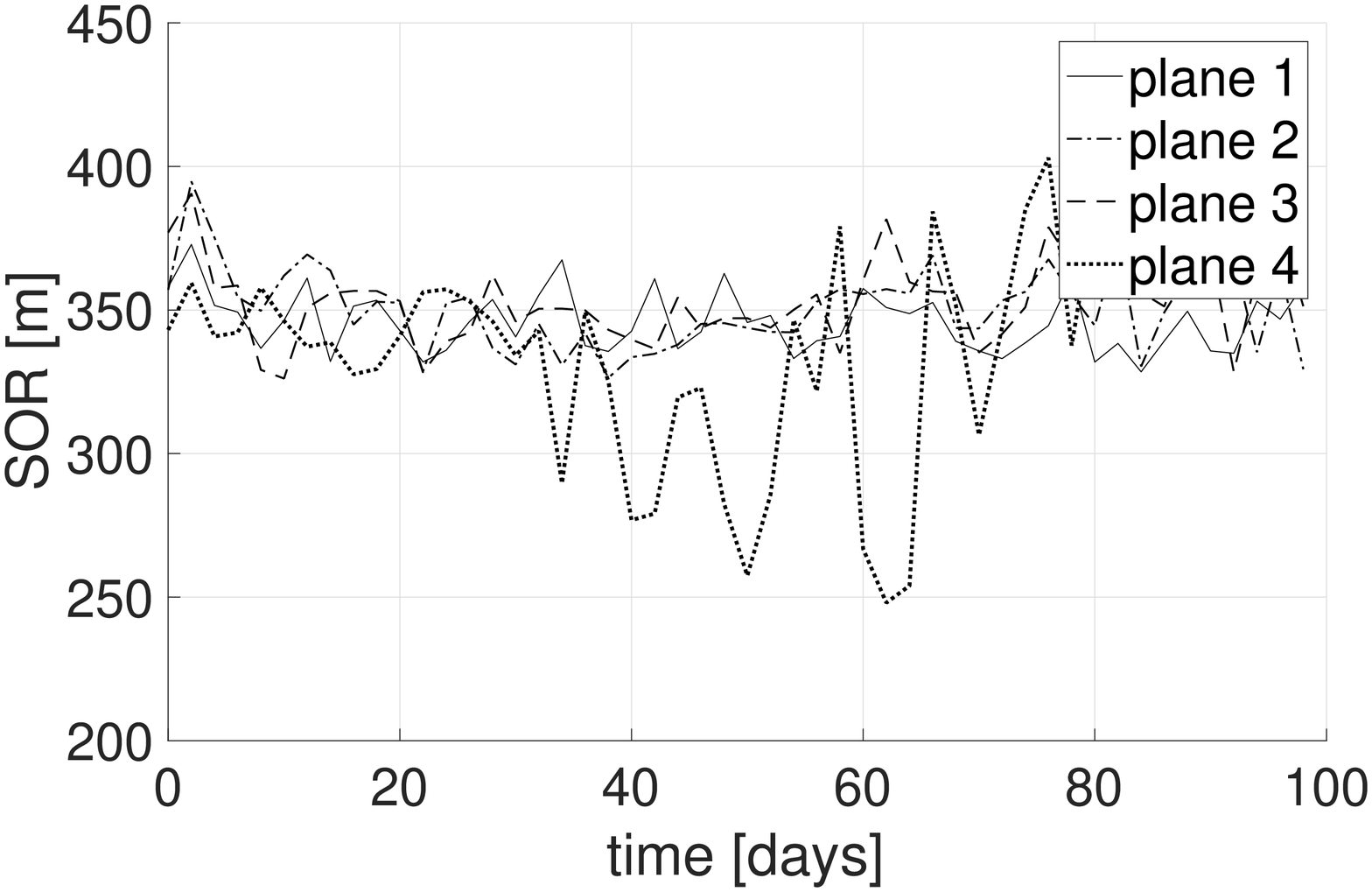}\includegraphics[clip,width=8cm]{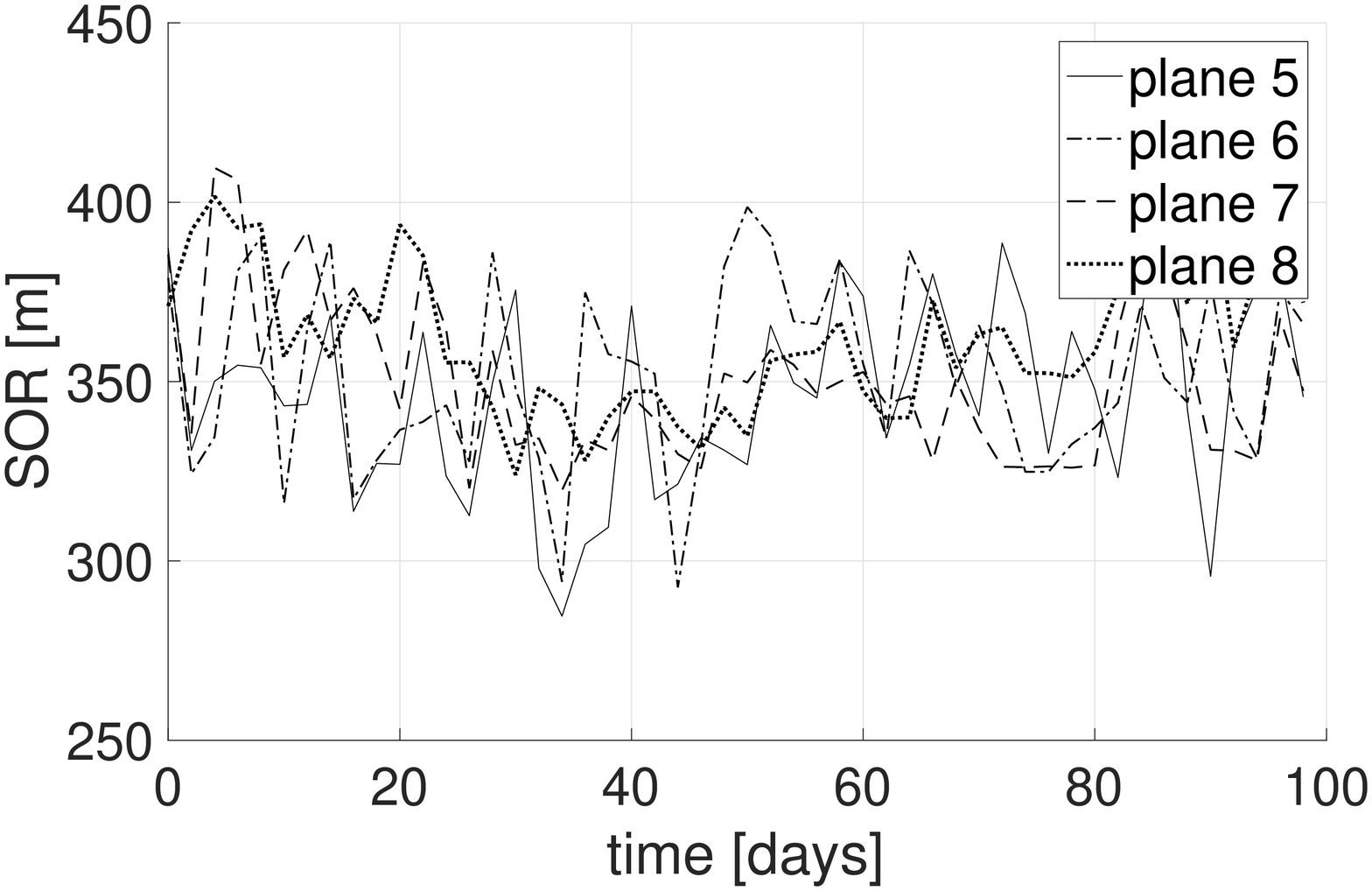}}

\centerline{\includegraphics[clip,width=8cm]{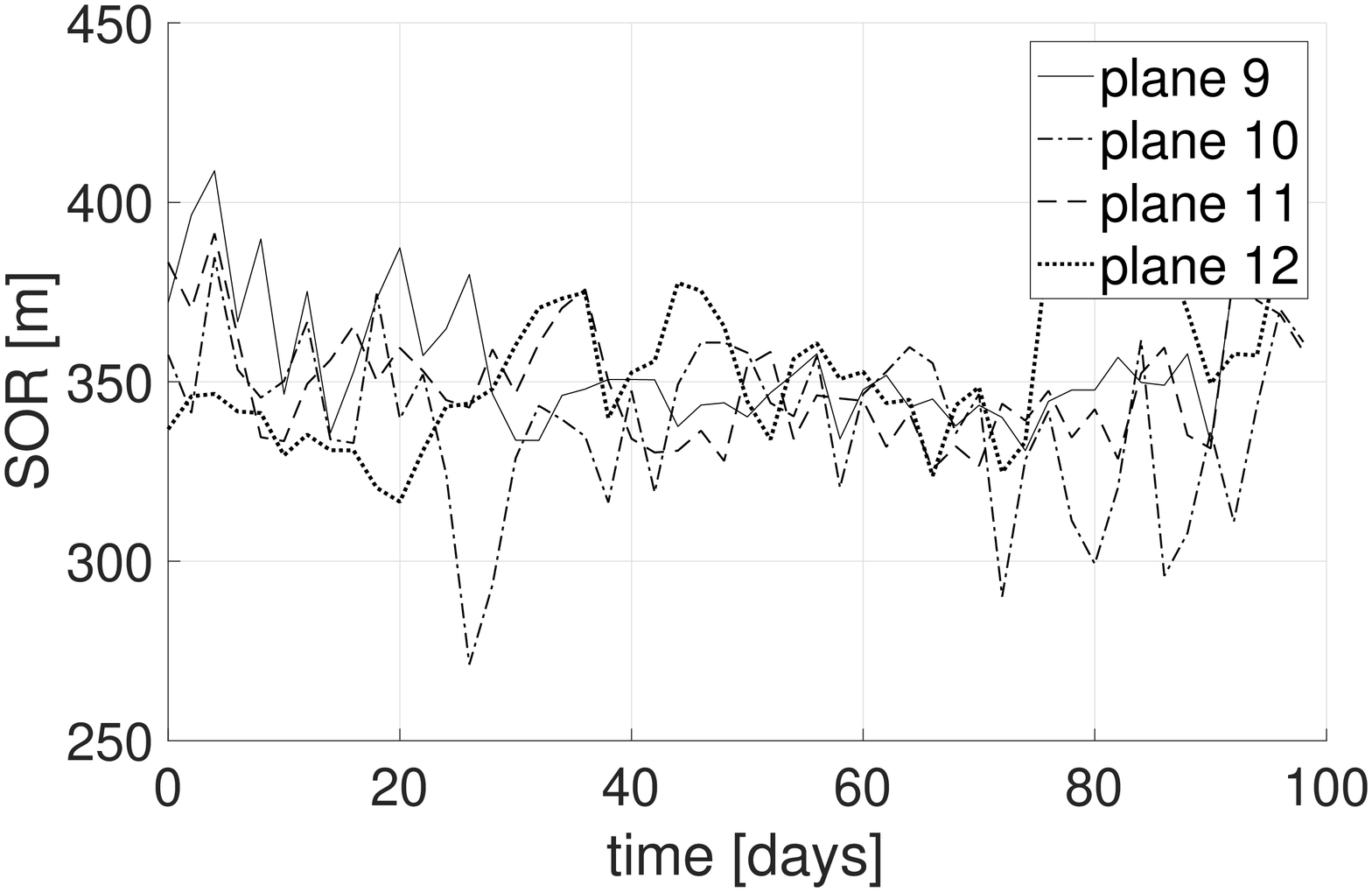}}

\caption{\label{fig:fig4}SOR evolution for class 3 MiSO orbits in drag-free
conditions }
\end{figure}

\begin{figure}[!t]
\centerline{\includegraphics[clip,width=8cm]{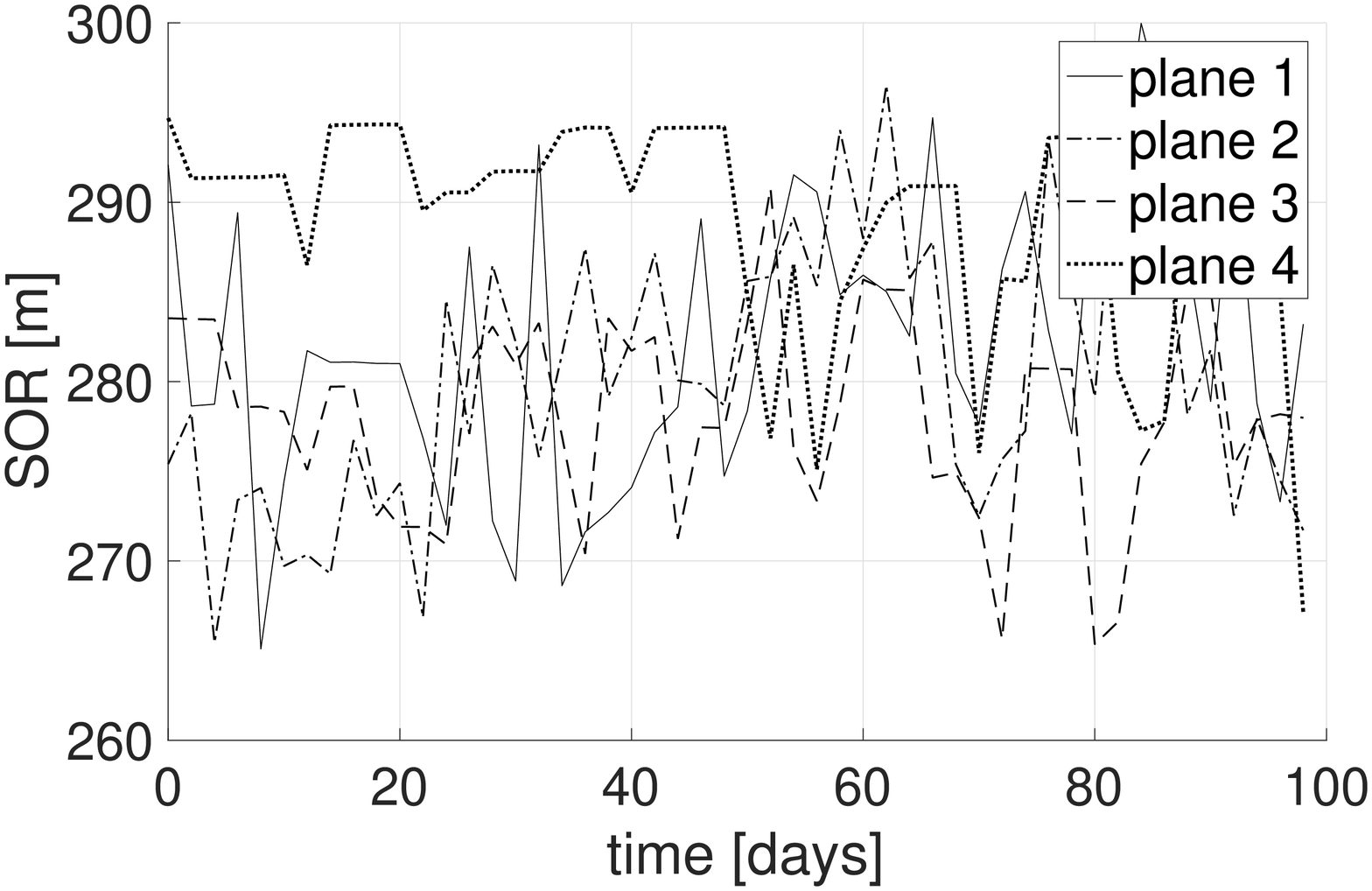}\includegraphics[clip,width=8cm]{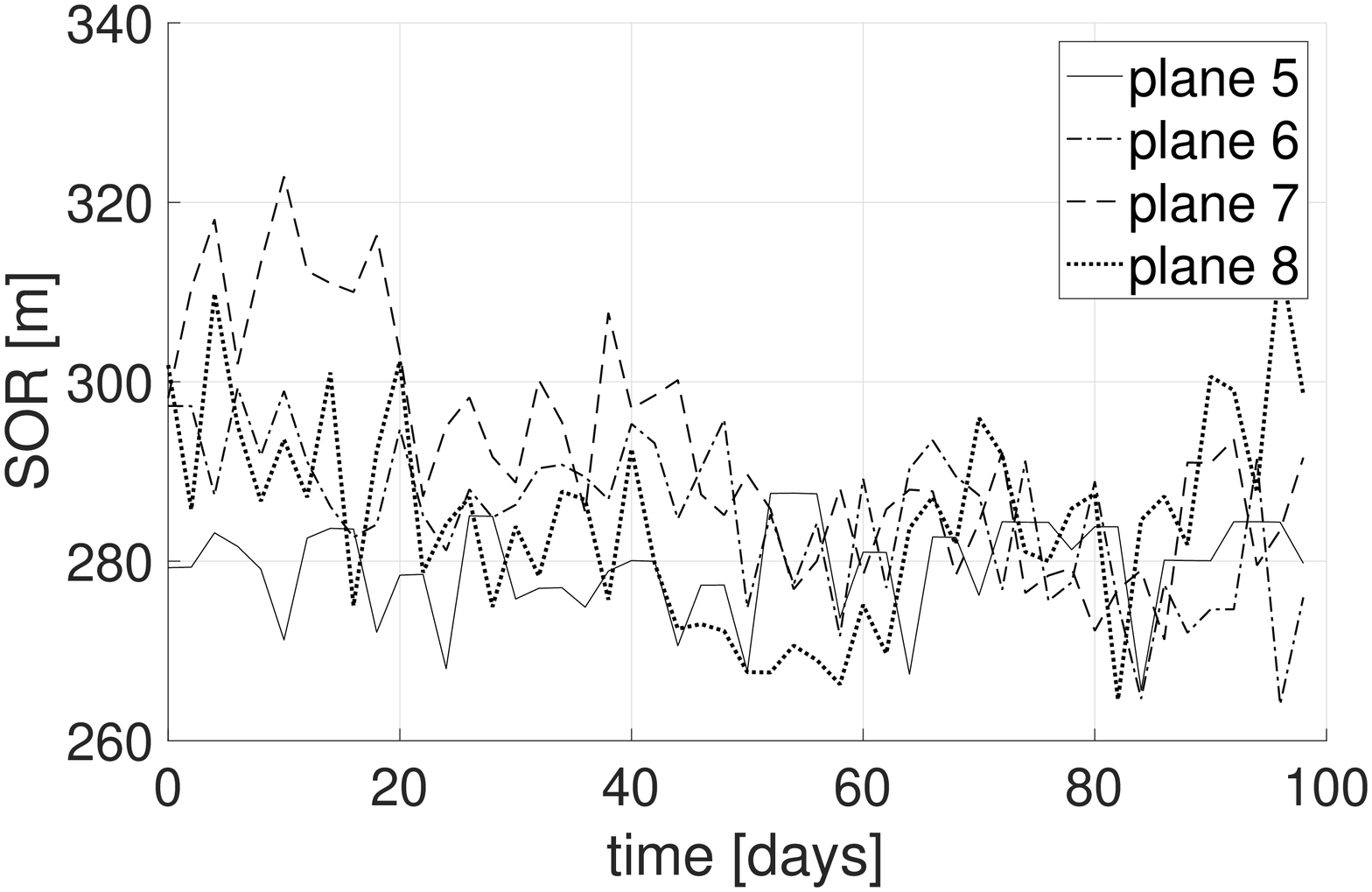}}

\centerline{\includegraphics[clip,width=8cm]{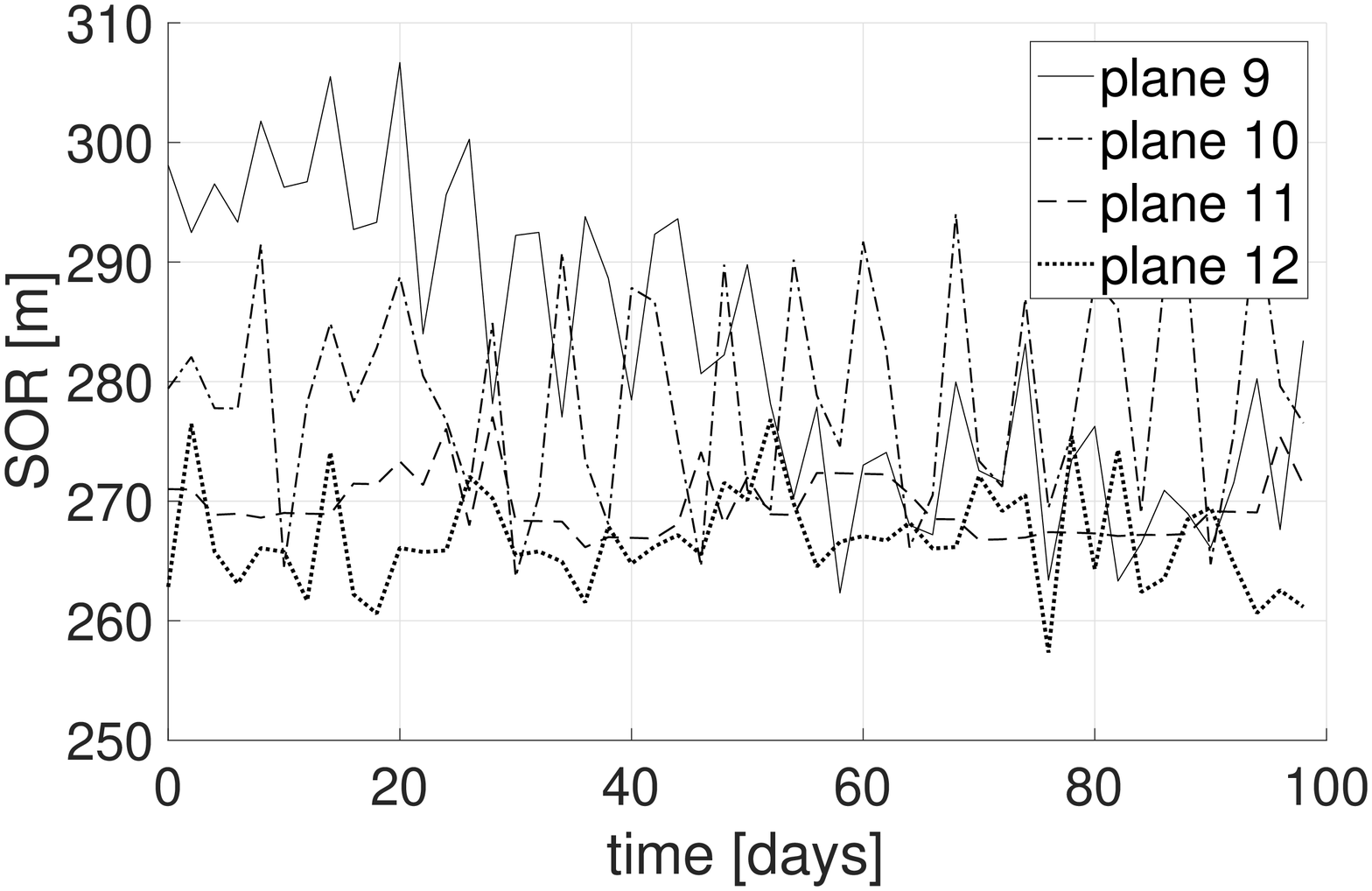}}

\caption{\label{fig:fig5}SOR evolution for class 4 MiSO orbits in drag-free
conditions }
\end{figure}

\begin{figure}[!t]
\centerline{\includegraphics[clip,width=8cm]{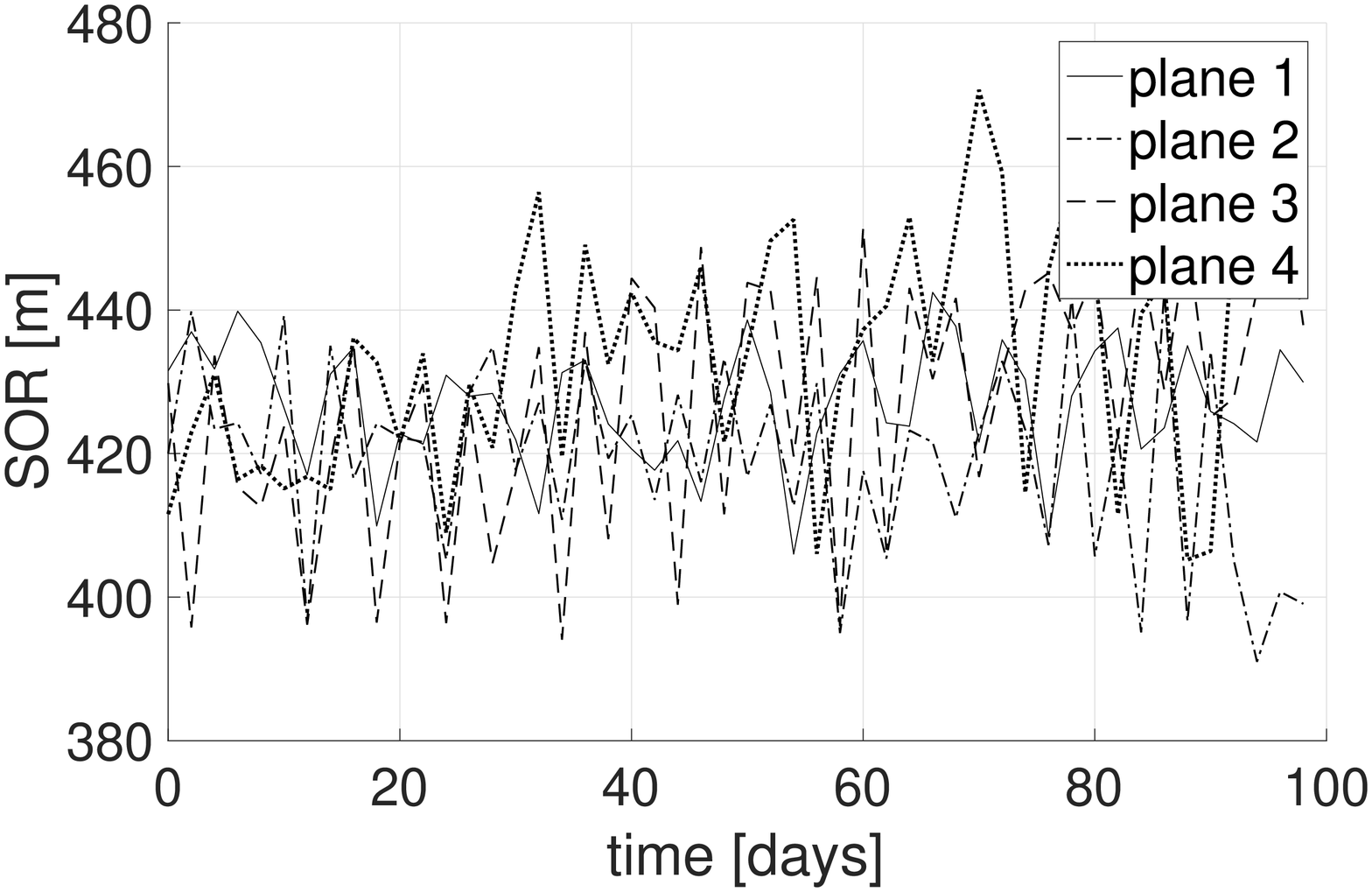}\includegraphics[clip,width=8cm]{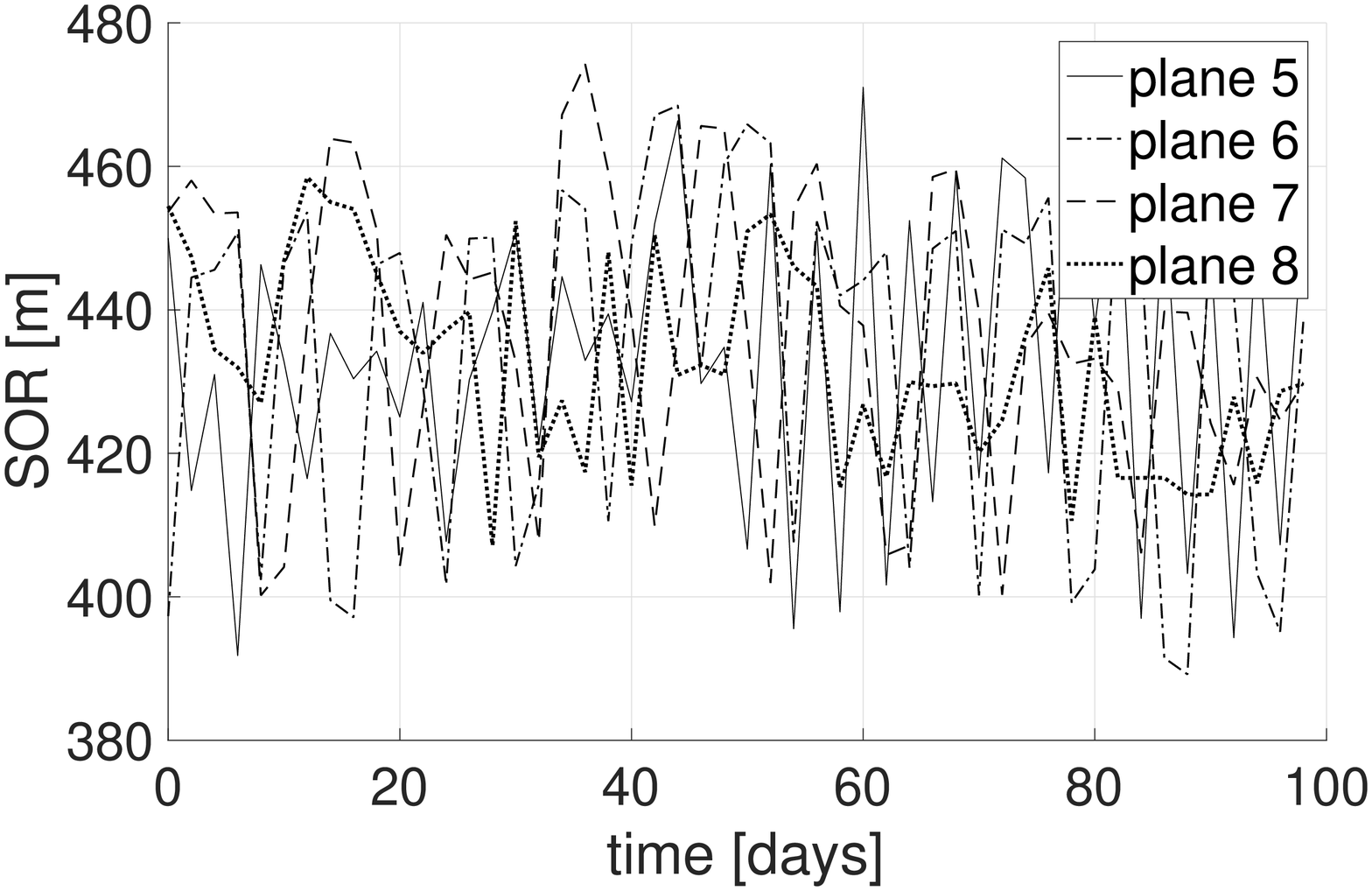}}

\centerline{\includegraphics[clip,width=8cm]{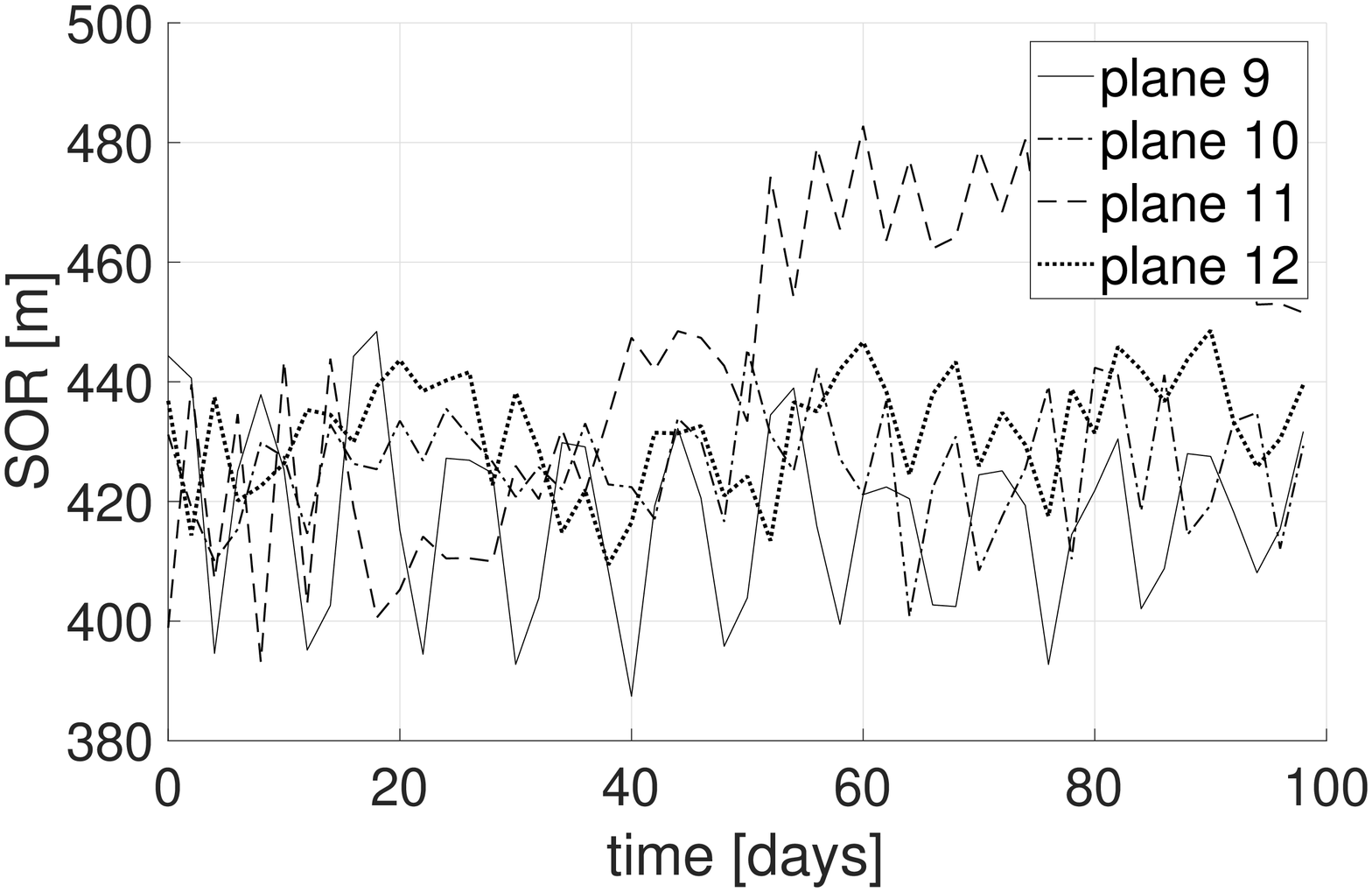}}

\caption{\label{fig:fig6}SOR evolution for class 5 MiSO orbits in drag-free
conditions}
\end{figure}

Figures (\ref{fig:fig2})-(\ref{fig:fig6}) show the evolution of
the \textit{10-day fixed-timespan} SOR function over a period of 100
days for the 12 planes of the five classes of orbits. The size of
the space occupancy region appears to fluctuate without experiencing
any significant secular increase, which implies that the different
gravitational perturbations do not have a significant long-term deteriorating
effect on drag-free MiSO orbits.

\begin{doublespace}

\end{doublespace}

\begin{figure}[!t]
\centerline{\includegraphics[clip,width=8cm]{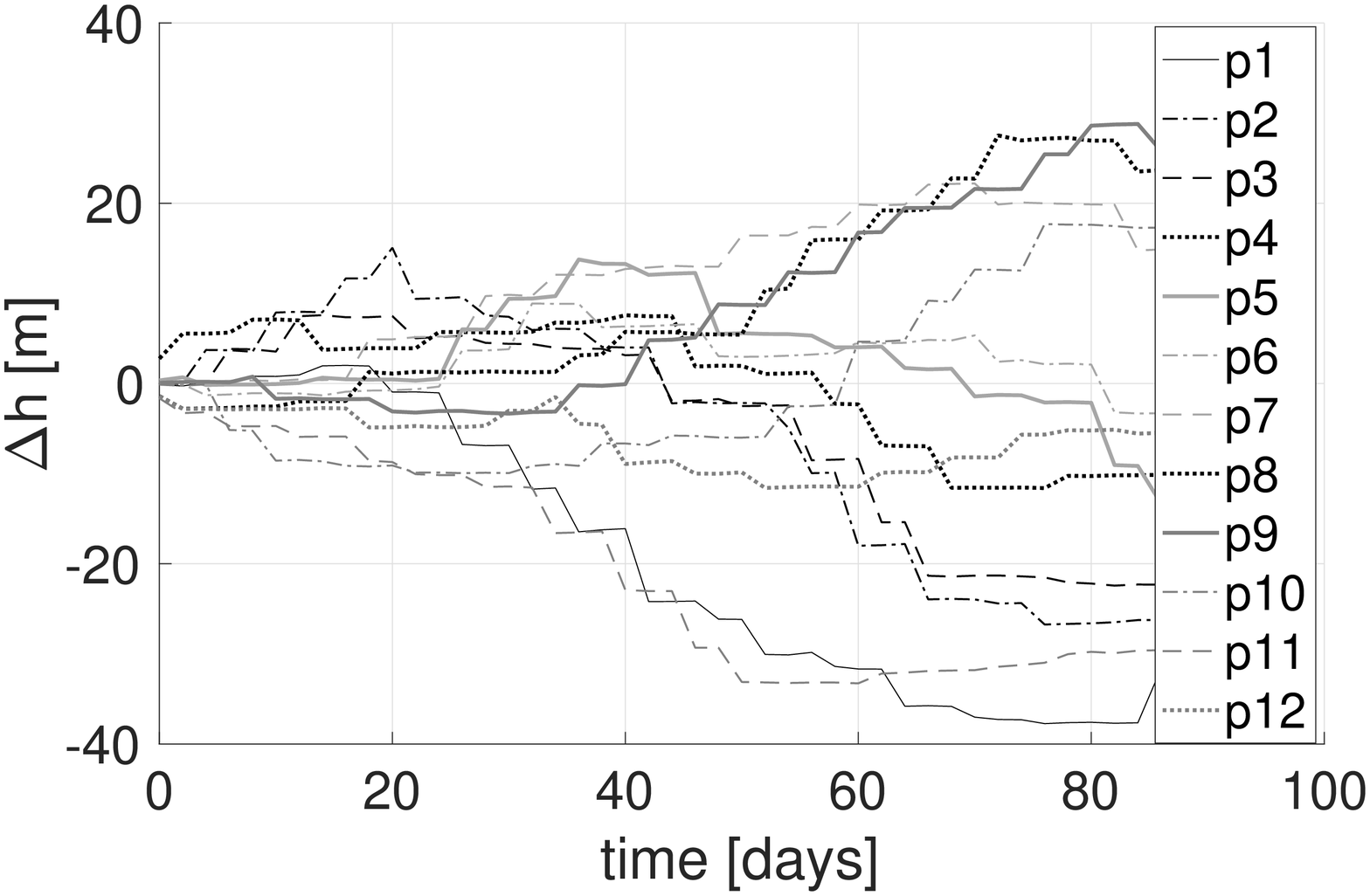}\includegraphics[clip,width=8cm]{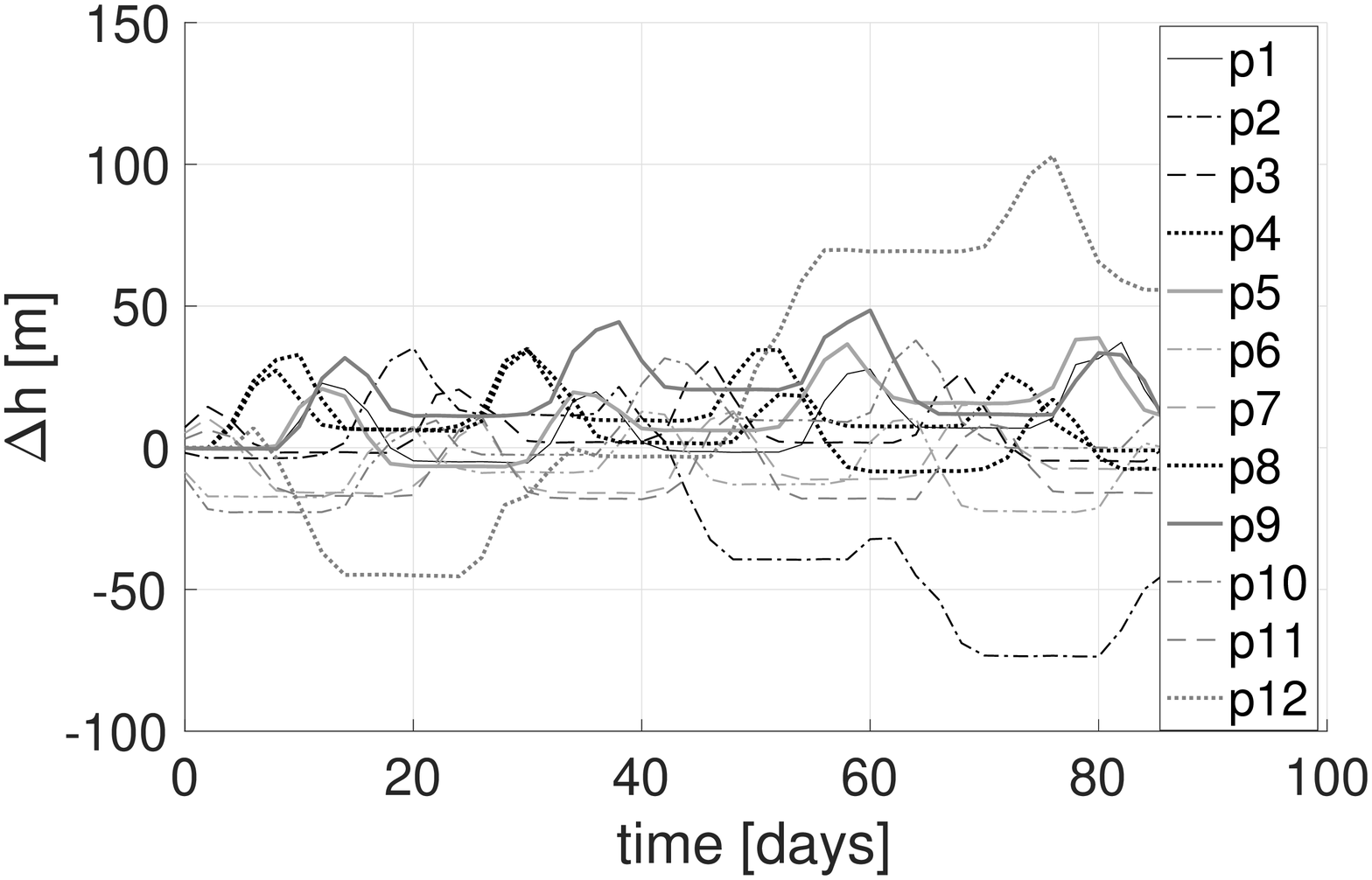}}

\caption{\label{fig:dh_drag_free_class1_2}Evolution of the minimum altitude
for all orbital planes of class 1 (left) and class 2 (right) MiSO
orbits}
\end{figure}

\begin{figure}[!t]
\centerline{\includegraphics[clip,width=8cm]{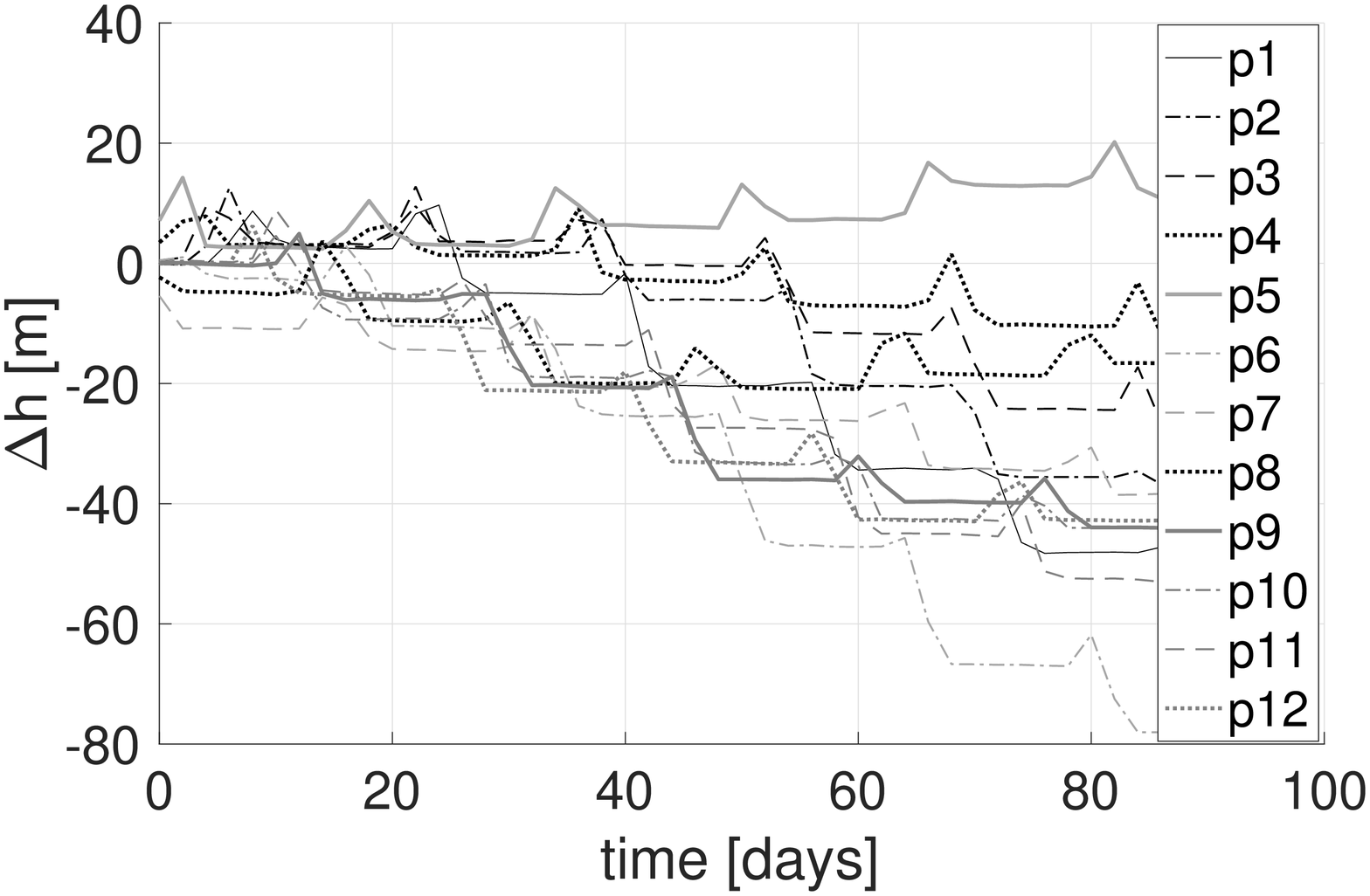}\includegraphics[clip,width=8cm]{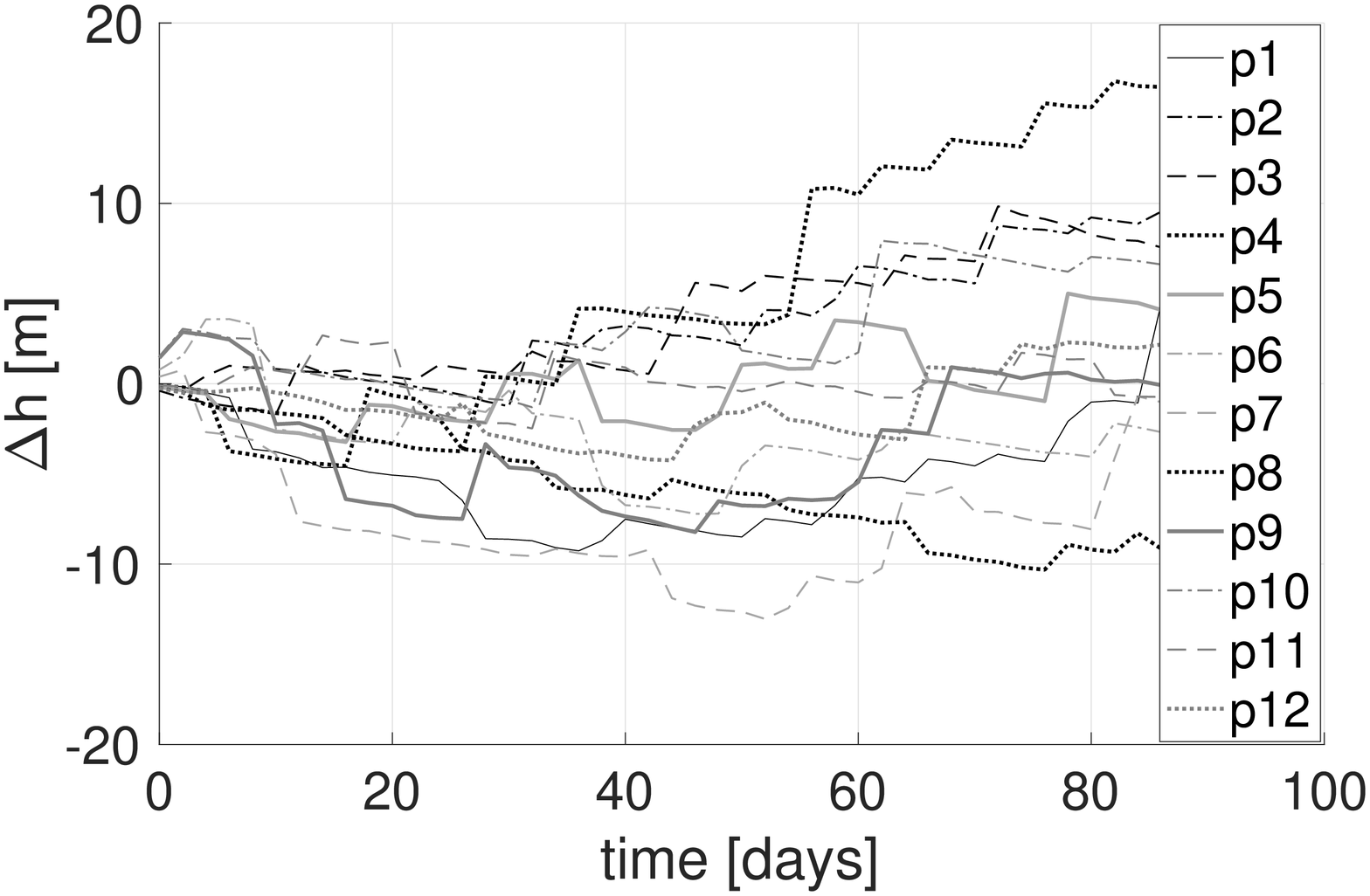}}

\caption{\label{fig:dh_drag_free_class3_4}Evolution of minimum altitude for
all orbital planes of class 3 (left) and class 4 (right) MiSO orbits}
\end{figure}

\begin{figure}[!t]
\centerline{\includegraphics[clip,width=8cm]{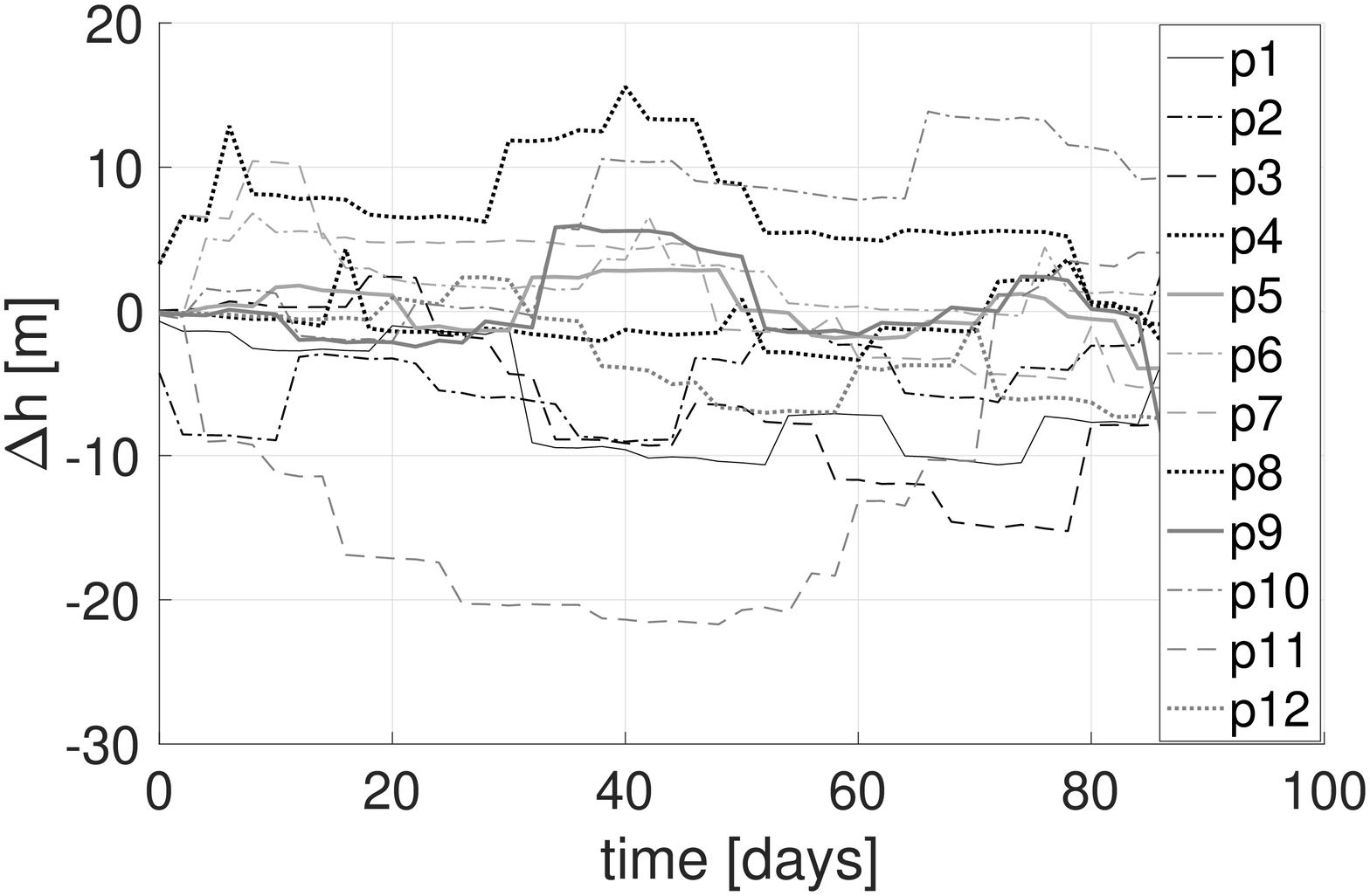}}

\caption{\label{fig:dh_drag_free_class5}Evolution of minimum altitude for
all orbital planes of class 5 MiSO orbits}
\end{figure}

To conclude the analysis we plot the variation of the minimum altitude
in time for the different orbital planes. Such variation, measured
with respect to initial minimum altitude (i.e. computed during the
first 10 days), is displayed in Figures (\ref{fig:dh_drag_free_class1_2})-(\ref{fig:dh_drag_free_class5}),
where the minimum altitude function is computed in a similar way as
the fixed-timespan SOR (i.e. over a moving 10-day time interval).
Altitude fluctuations are contained below 50 meters for all cases
with the exception of class 2, which experiences 100-m-wide altitude
fluctuations in two of its planes, and class 3, which experiences
a 80-m-wide altitude fluctuations in one of its planes. 

\begin{doublespace}

\end{doublespace}

\subsection*{Impact of SRP and drag }

As one can expect from the available frozen-orbit literature (see
in particular Shapiro \cite{shapiro1995phase}), solar radiation pressure
and drag have a major impact on the minimum achievable space occupancy
and its evolution. Even if the effect of these perturbations can be
compensated by correction maneuvers it is extremely important to be
able to delay the need to perform such maneuvers as much as possible
by including these perturbations in the MiSO orbit design process.
As an example, correction maneuvers for the frozen-orbit based Sentinel-3
mission can be as frequent as every two weeks \cite{taboada2019sentinel}.

Table \ref{tab:TAB5} displays the 100-day SOR for 12 orbital planes
of the five classes of MiSO orbits previously considered but with
both atmospheric drag and solar radiation pressure active. The corresponding
results for unoptimized frozen orbits are reported in Table \ref{tab:TAB6}
showing the benefit of MiSO orbits in terms of SOR reduction (up to
almost 400 m). As expected, the minimum space occupancy of lower altitude
orbits (class 1,2) is considerably higher compared to their drag-free
counterpart mainly because of drag-induced altitude decay. For higher-altitude
orbits (class 3,4,5) non-gravitational perturbations (mainly SRP)
also result in an increased SOR, but to a much lesser extent. We must
stress here the importance of including both types of perturbations
in the process of MiSO initial conditions generation as adopting initial
conditions of a drag-free MiSO orbit would result in a much bigger
SOR in this scenario.

\begin{figure}[!t]
\centerline{\includegraphics[clip,width=8cm]{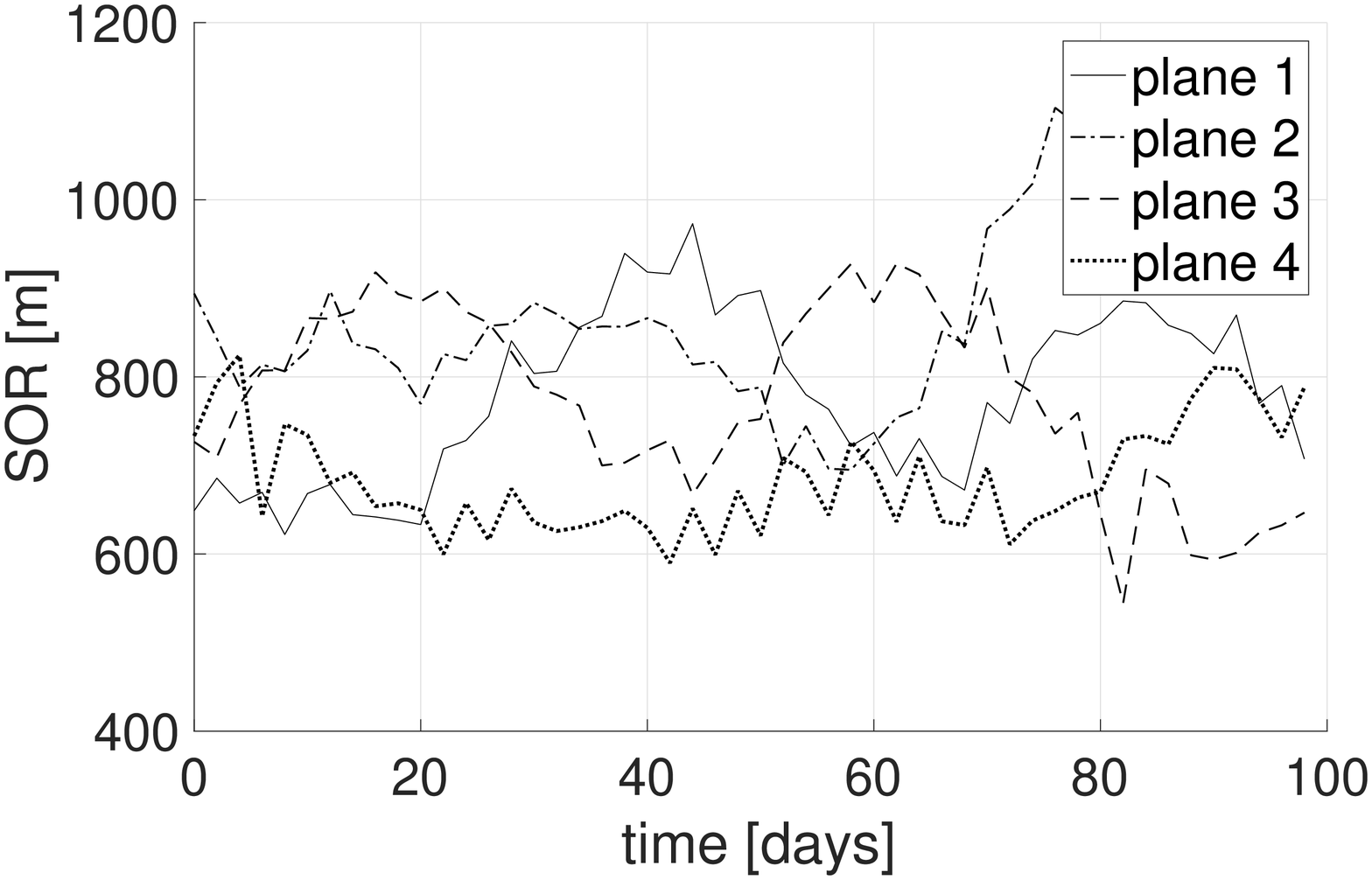}\includegraphics[clip,width=8cm]{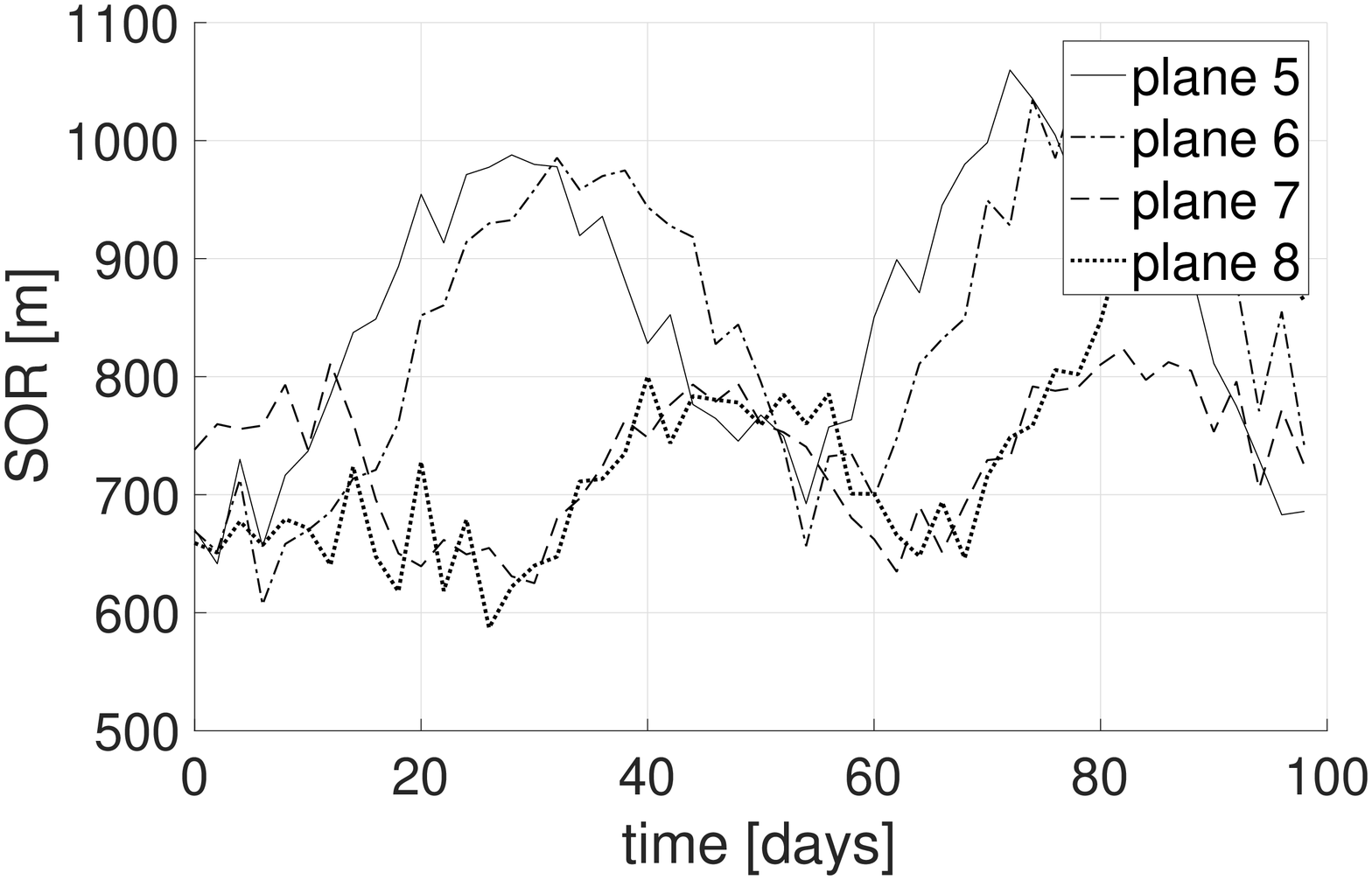}}

\centerline{\includegraphics[clip,width=8cm]{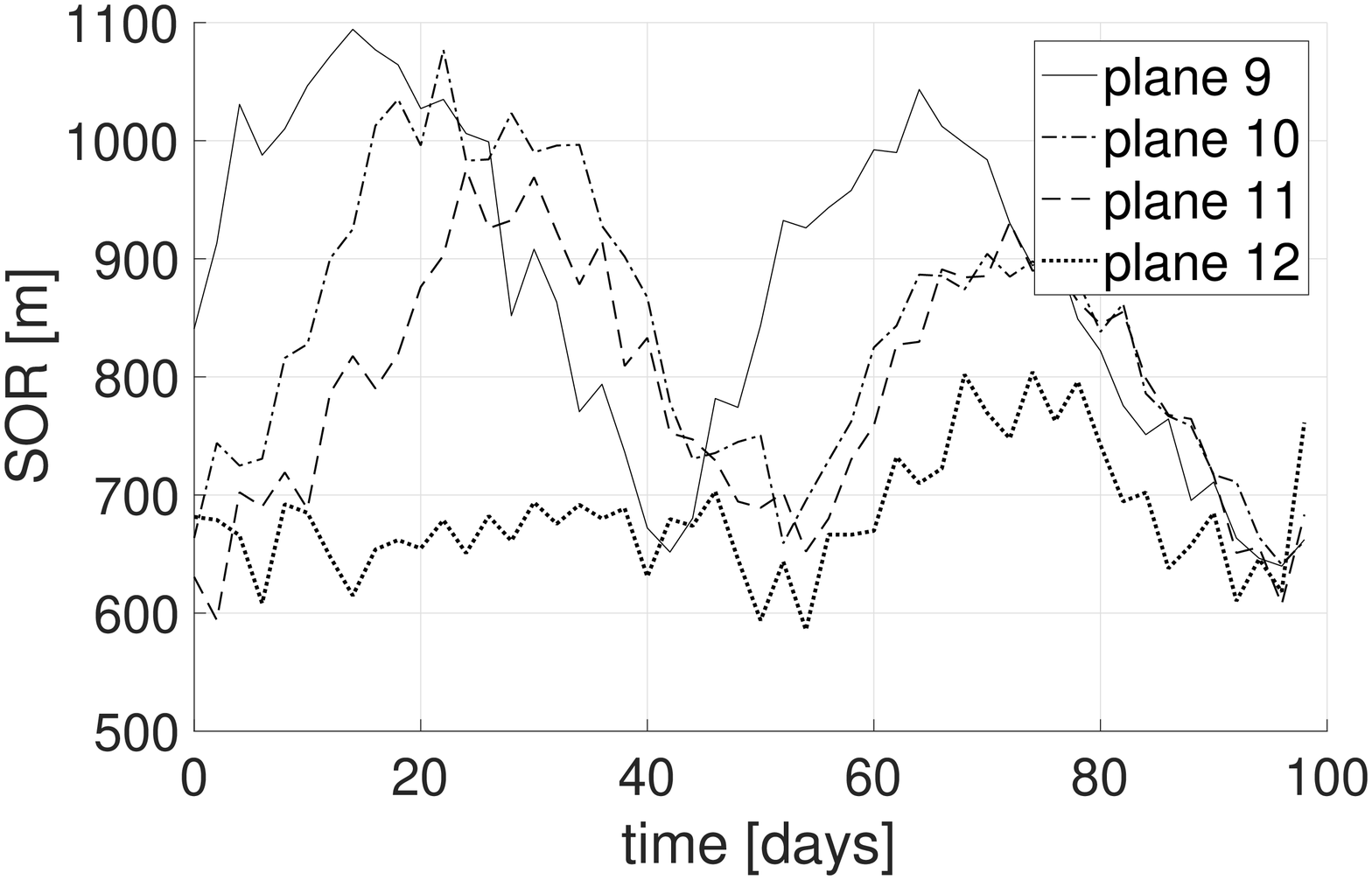}}

\caption{\label{fig:fig7}SOR evolution for class 1 MiSO orbits considering
non-gravitational perturbations}
\end{figure}

\begin{figure}[!t]
\centerline{\includegraphics[clip,width=8cm]{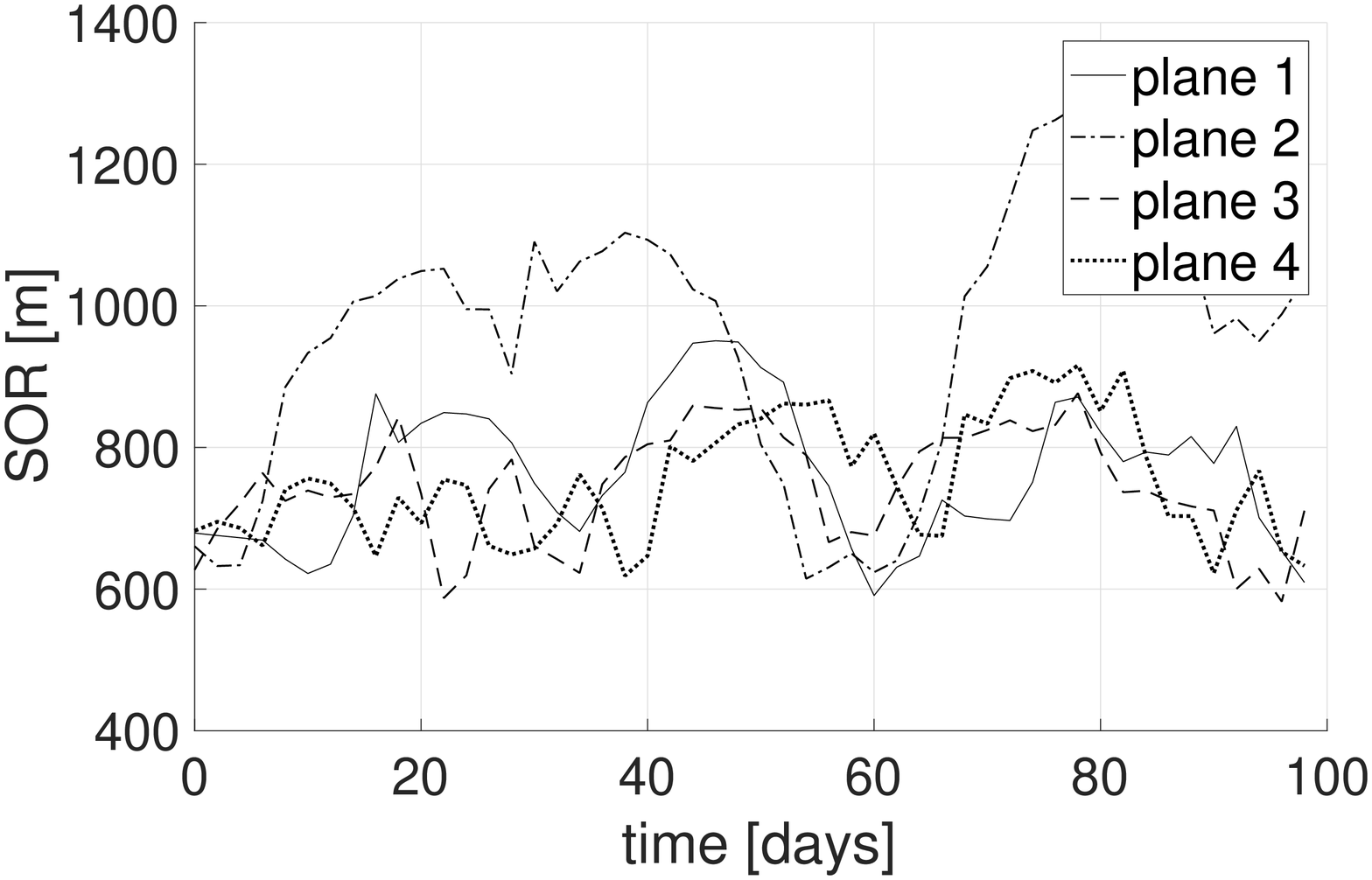}\includegraphics[clip,width=8cm]{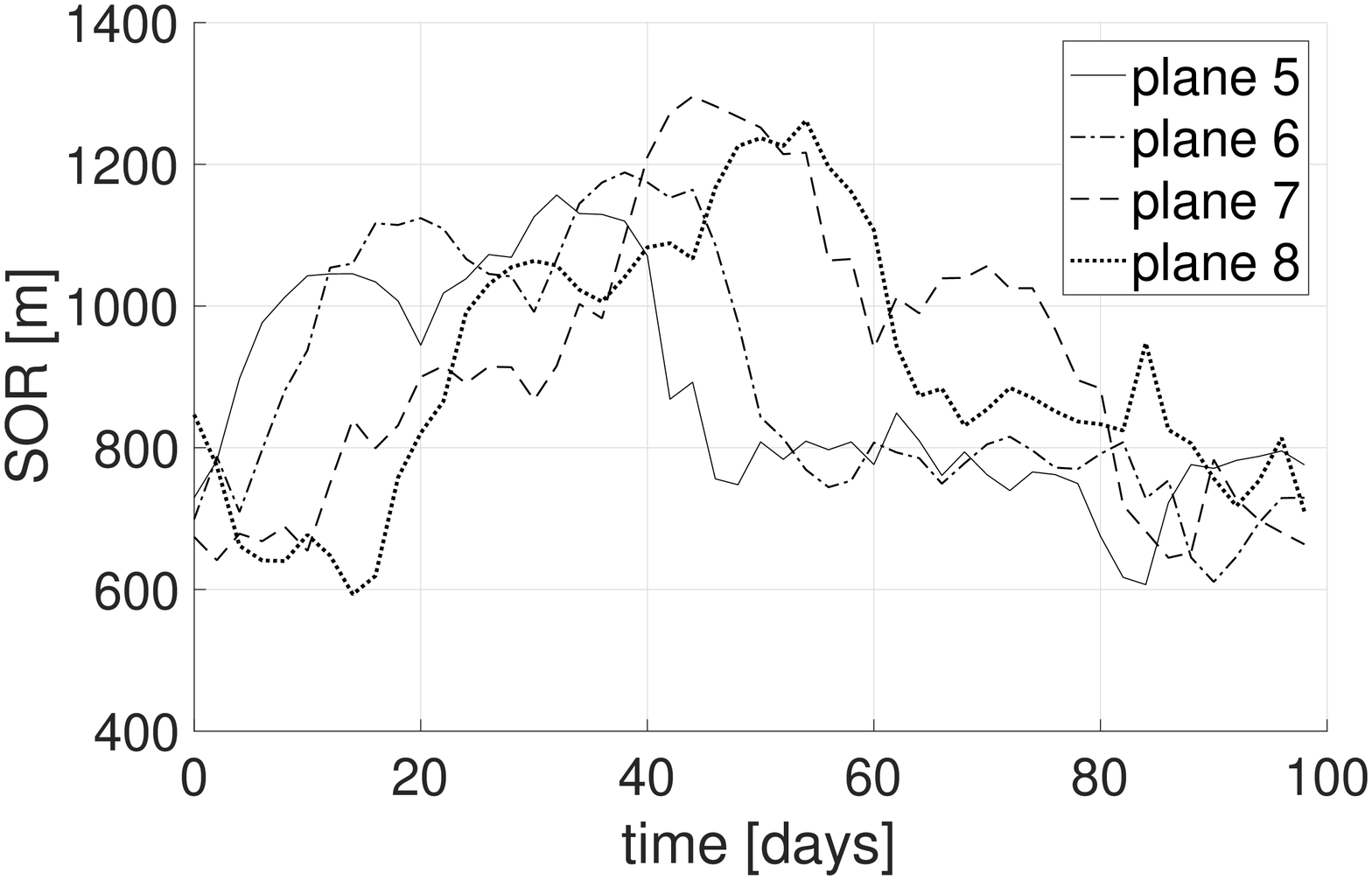}}

\centerline{\includegraphics[clip,width=8cm]{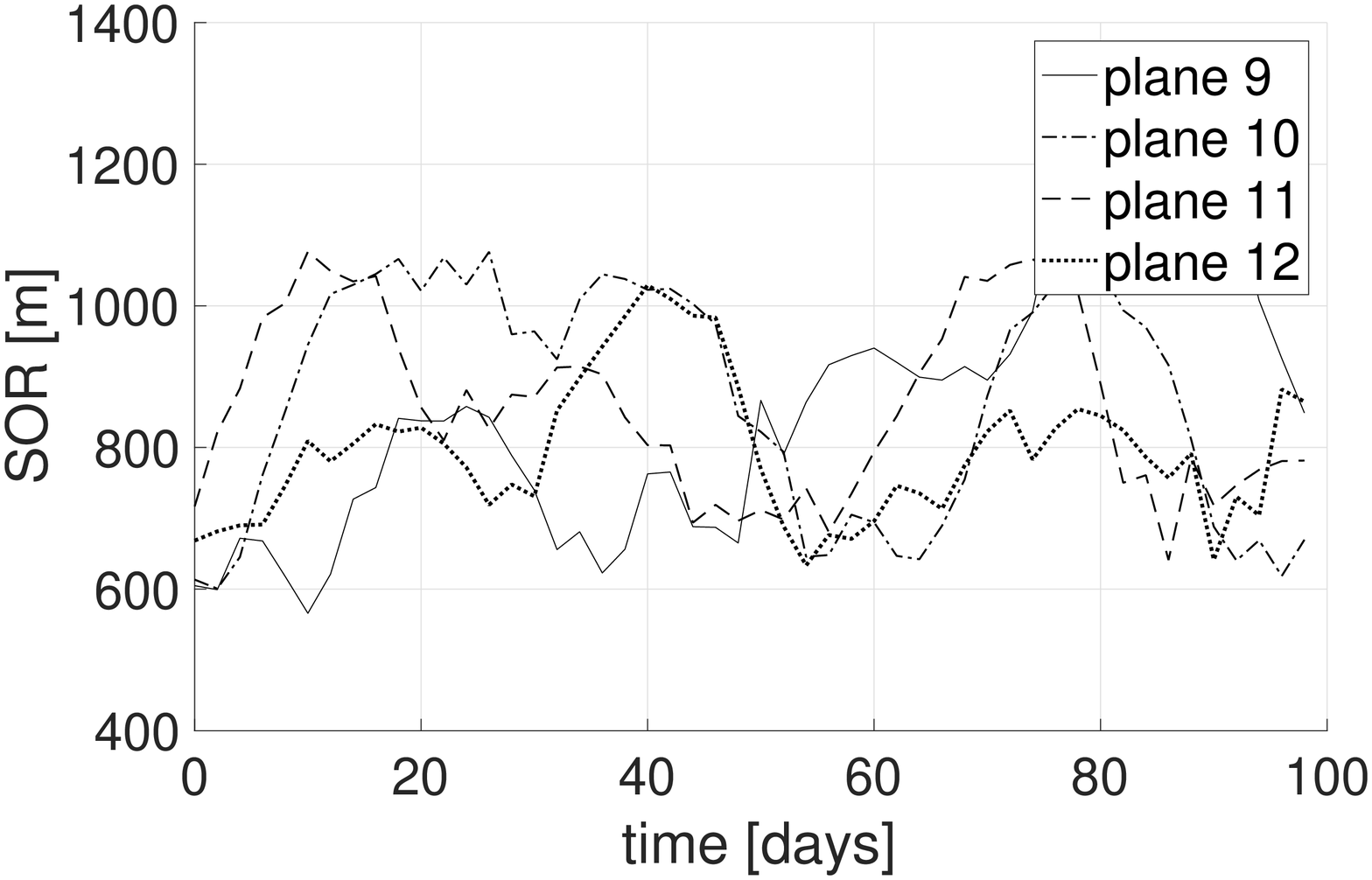}}

\caption{\label{fig:fig8}SOR evolution for class 2 MiSO orbits considering
non-gravitational perturbations}
\end{figure}

\begin{figure}[!t]
\centerline{\includegraphics[clip,width=8cm]{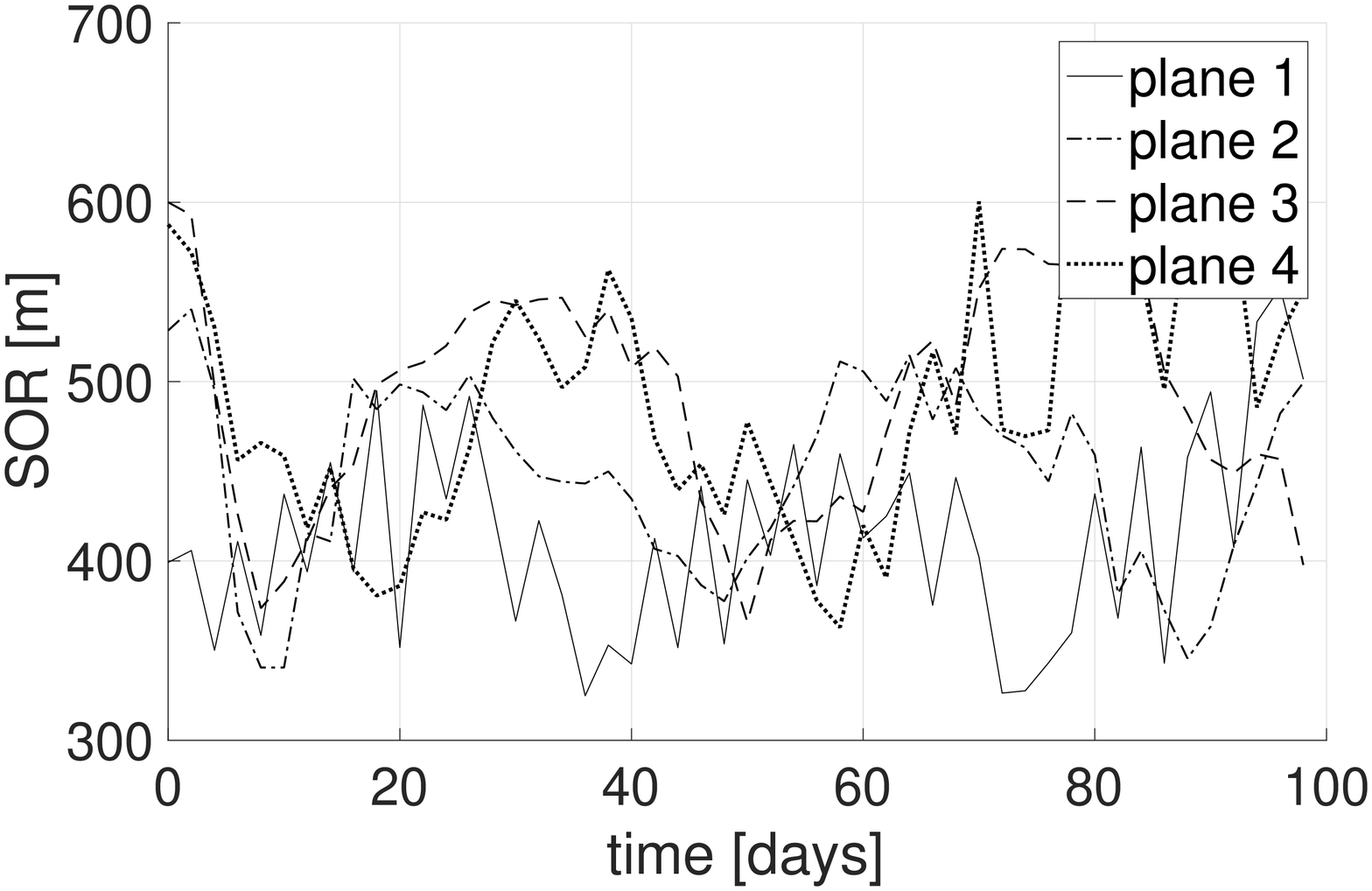}\includegraphics[clip,width=8cm]{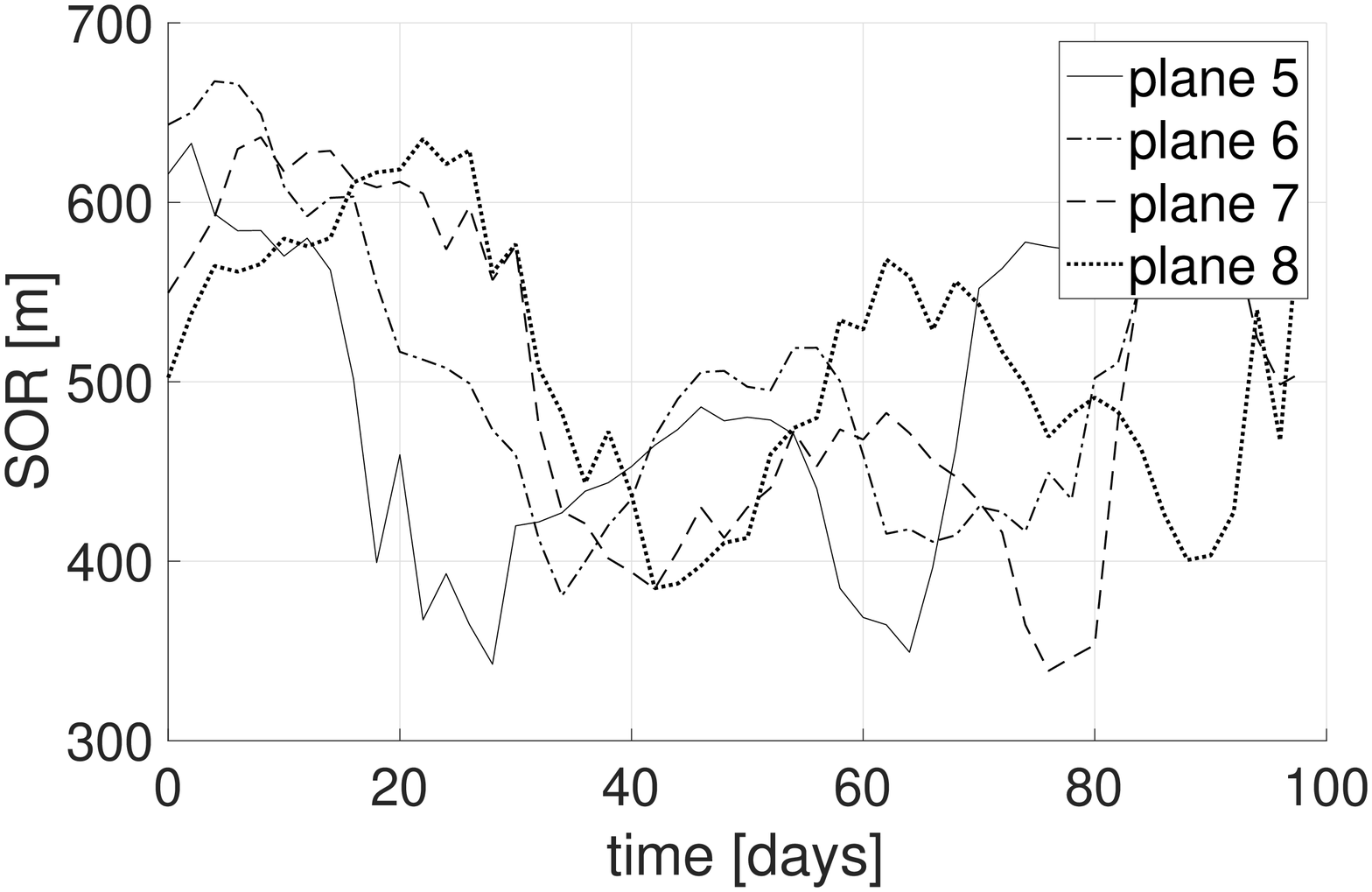}}

\centerline{\includegraphics[clip,width=8cm]{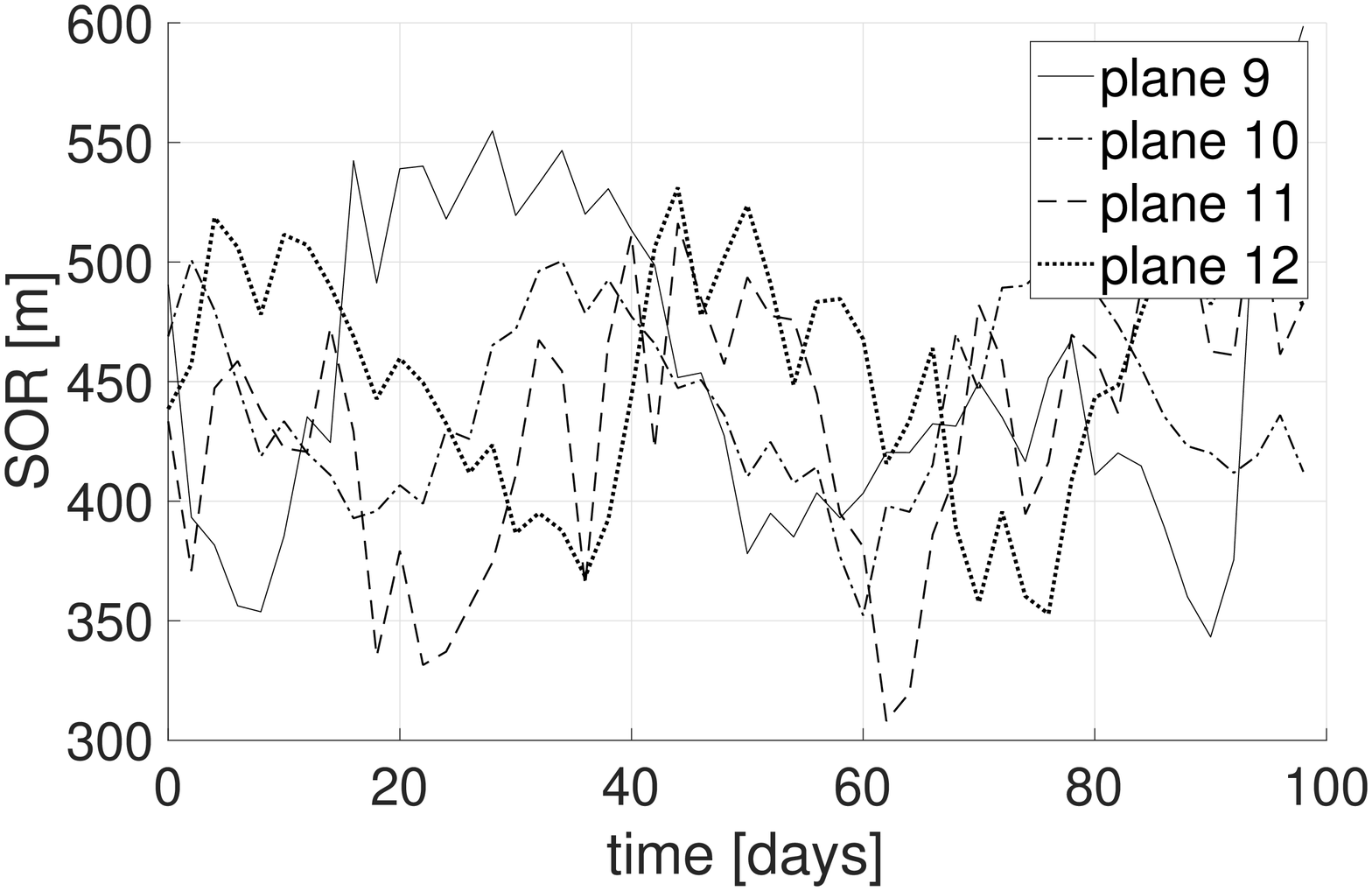}}

\caption{\label{fig:fig9}SOR evolution for class 3 MiSO orbits considering
non-gravitational perturbations}
\end{figure}

\begin{figure}[!t]
\centerline{\includegraphics[clip,width=8cm]{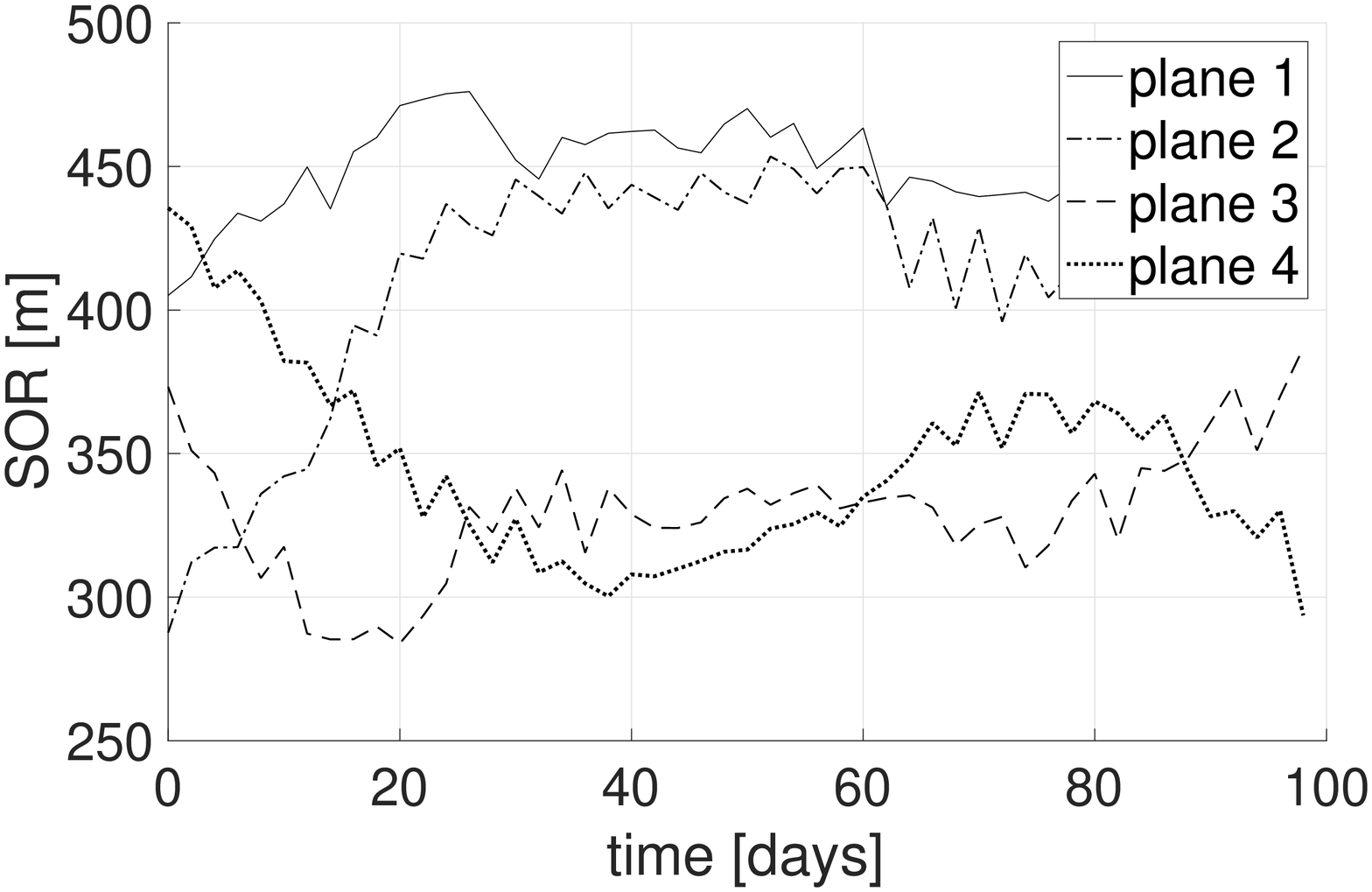}\includegraphics[clip,width=8cm]{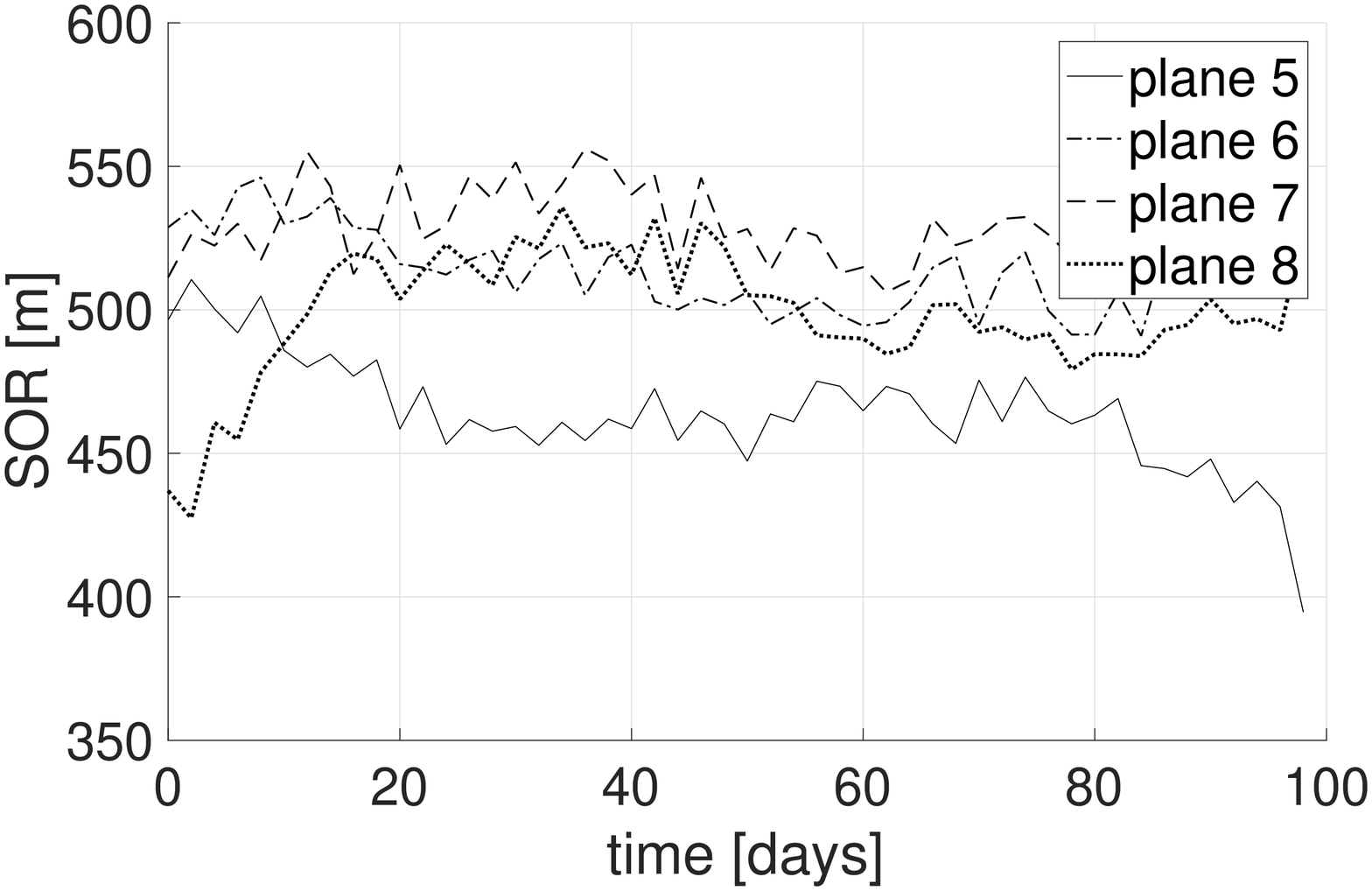}}

\centerline{\includegraphics[clip,width=8cm]{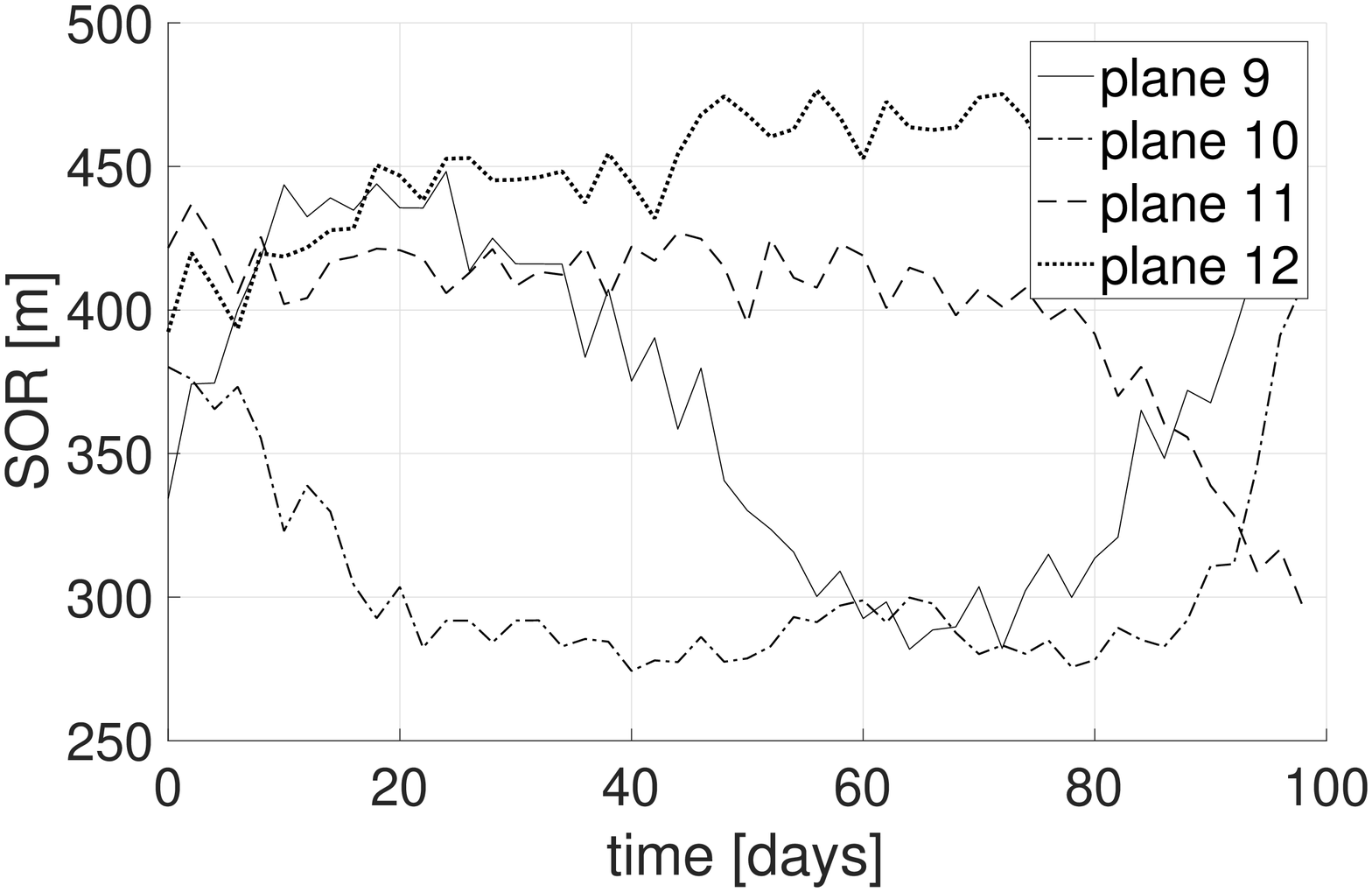}}

\caption{\label{fig:fig10}SOR evolution for class 4 MiSO orbits considering
non-gravitational perturbations}
\end{figure}

\begin{figure}[!t]
\centerline{\includegraphics[clip,width=8cm]{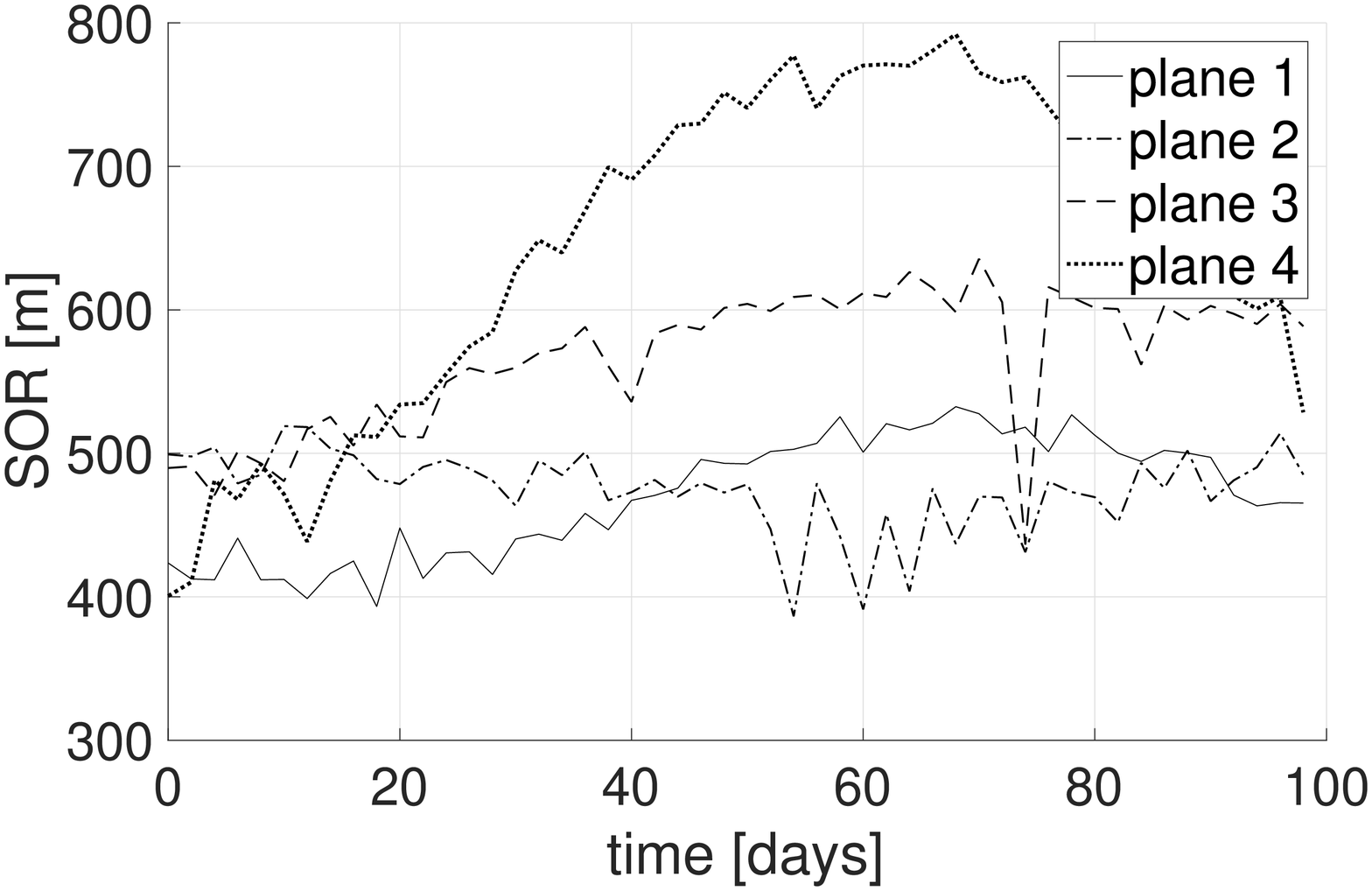}\includegraphics[clip,width=8cm]{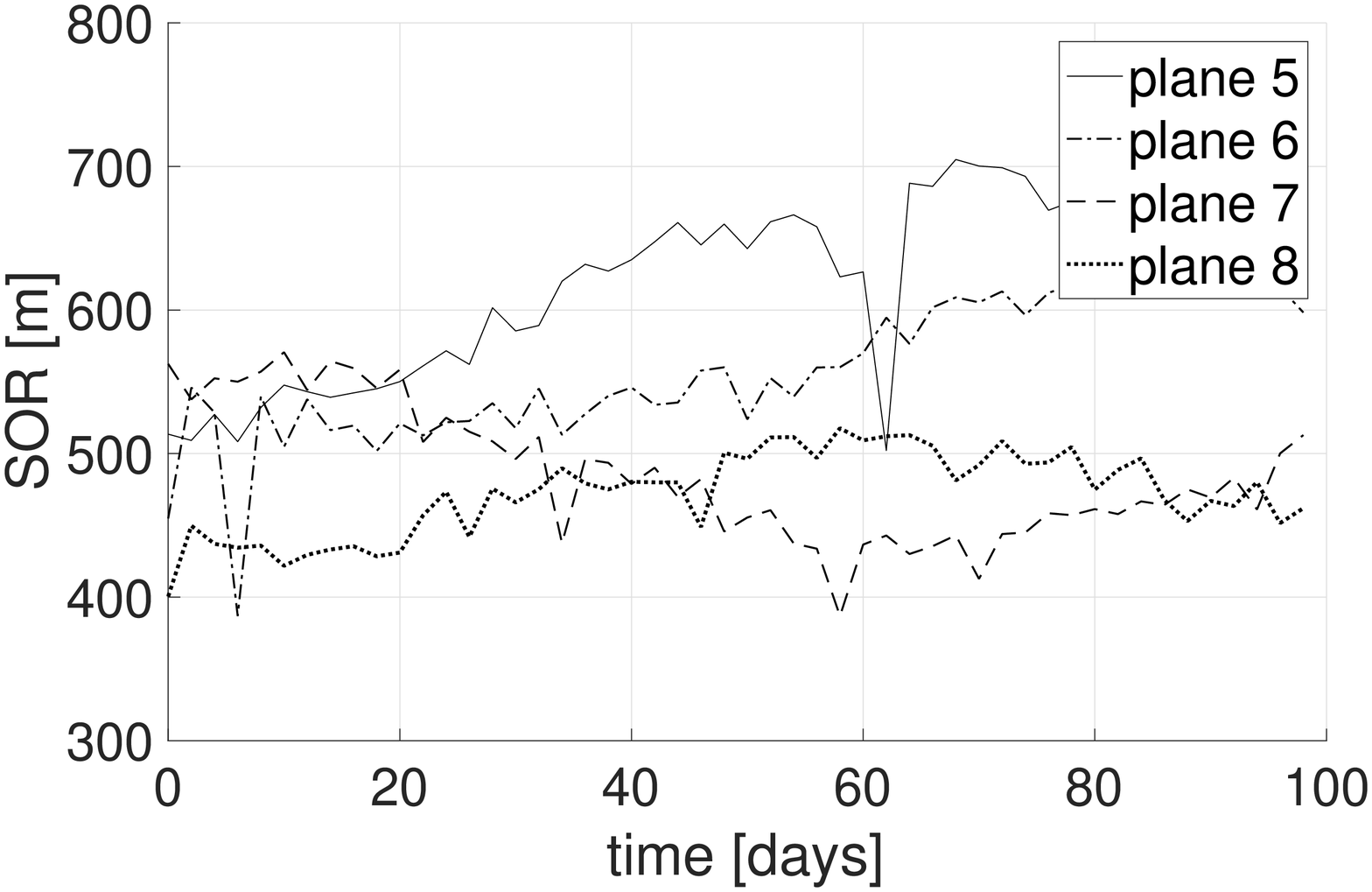}}

\centerline{\includegraphics[clip,width=8cm]{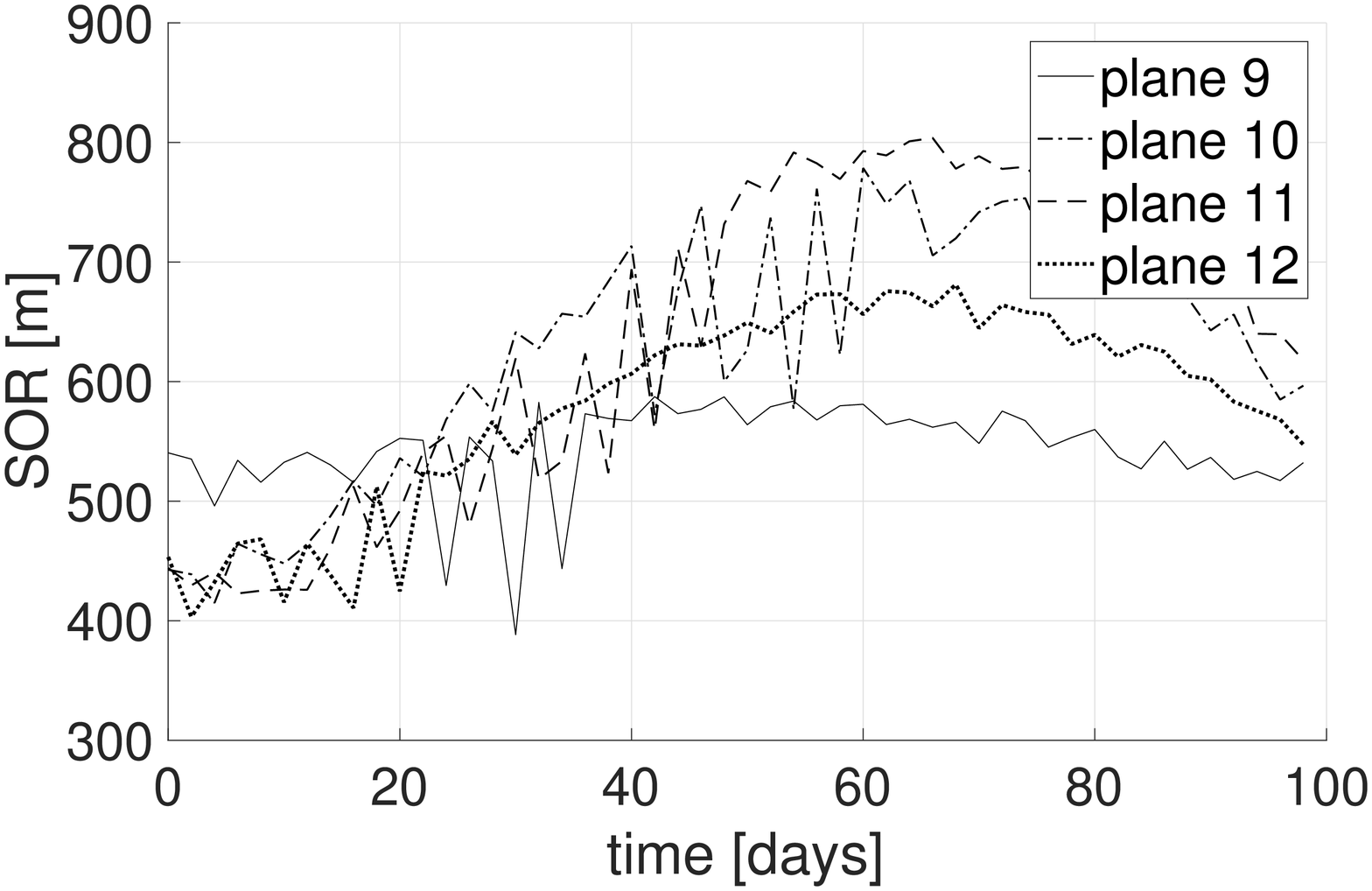}}

\caption{\label{fig:fig11}SOR evolution for class 5 MiSO orbits considering
non-gravitational perturbations}
\end{figure}

\begin{figure}[!t]
\centerline{\includegraphics[clip,width=8cm]{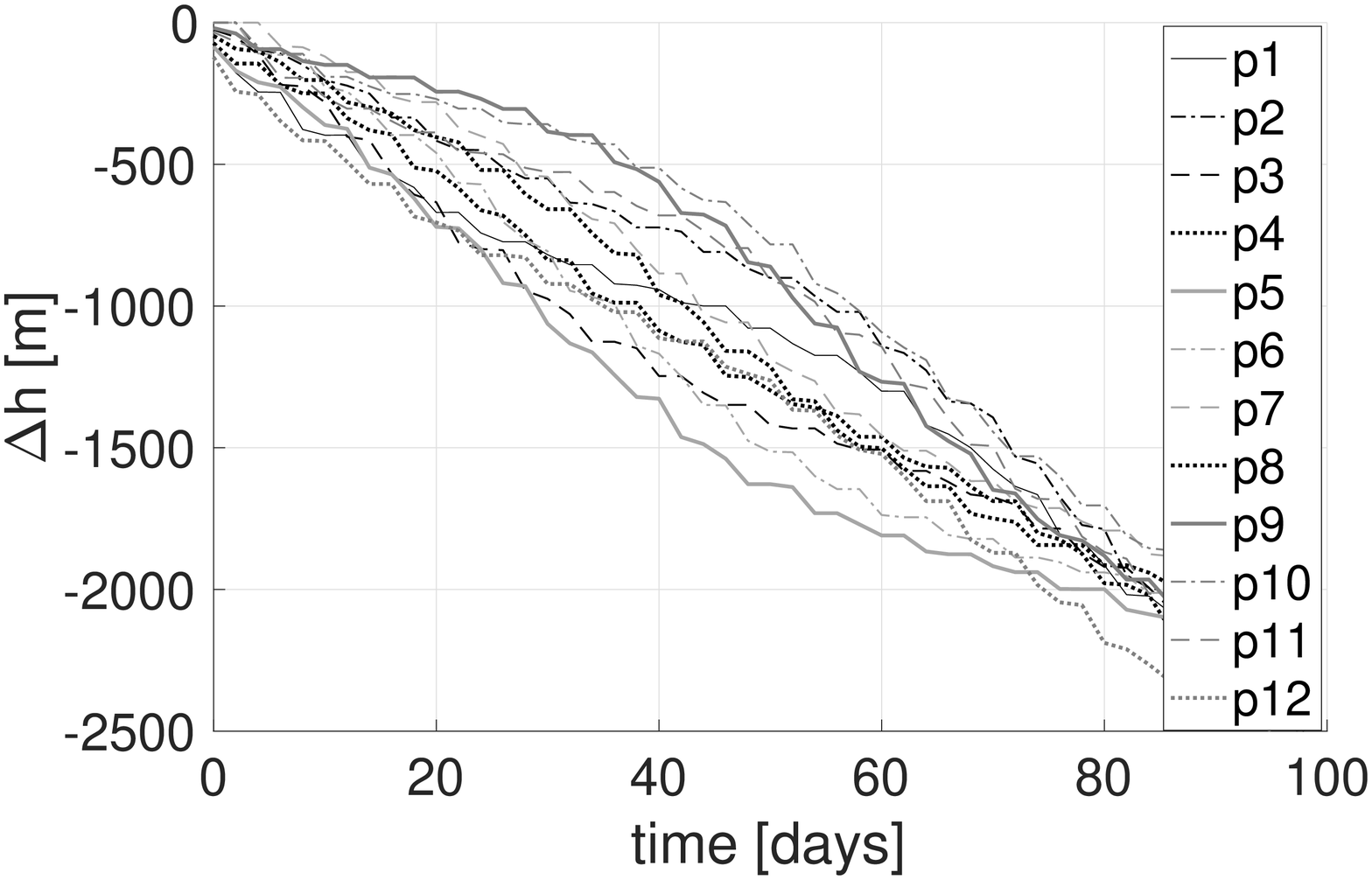}\includegraphics[clip,width=8cm]{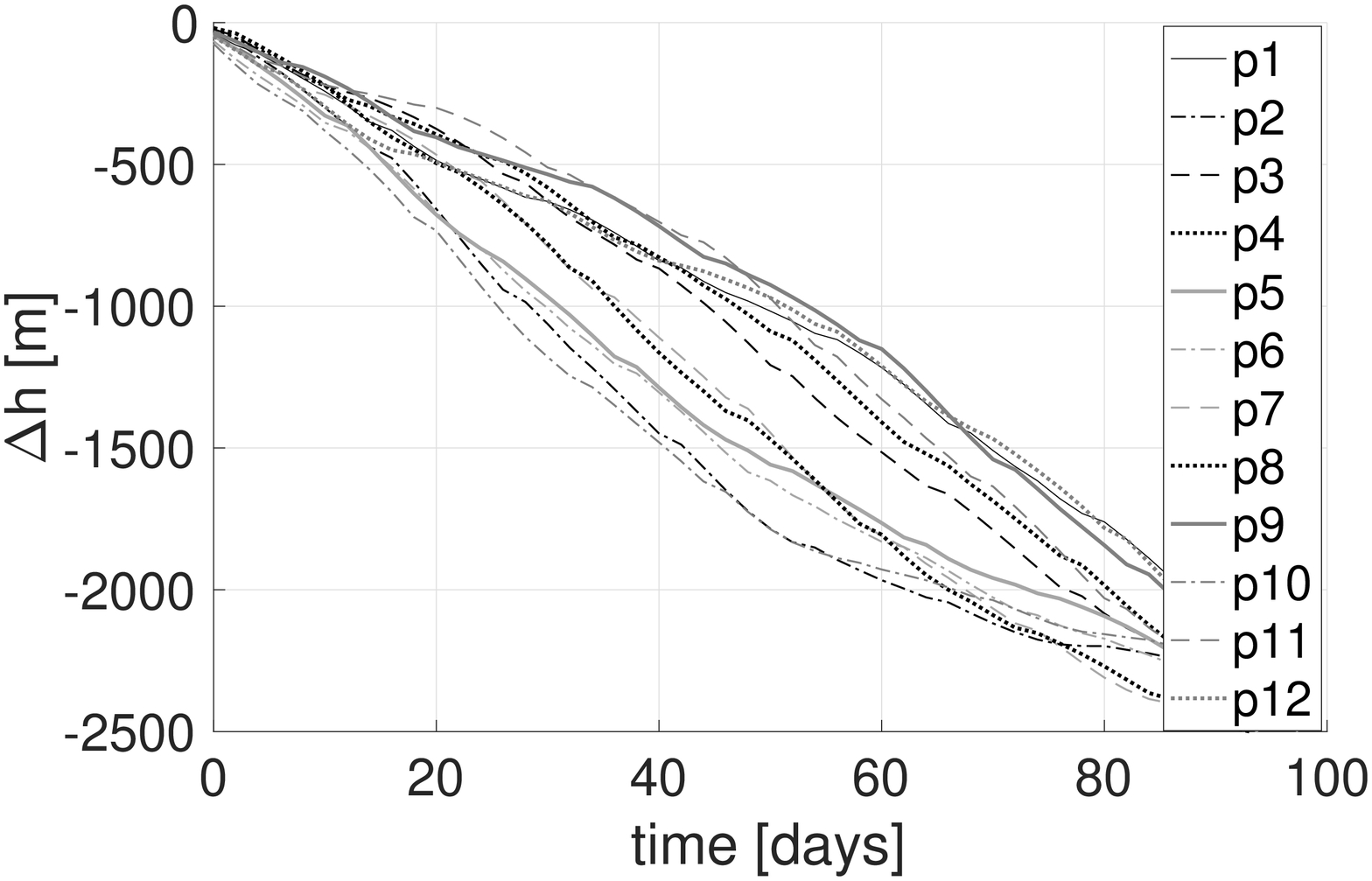}}

\caption{\label{fig:fig13}Evolution of the minimum altitude for all orbital
planes of class 1 (left) and class 2 (right) MiSO orbits }
\end{figure}

\begin{figure}[!t]
\centerline{\includegraphics[clip,width=8cm]{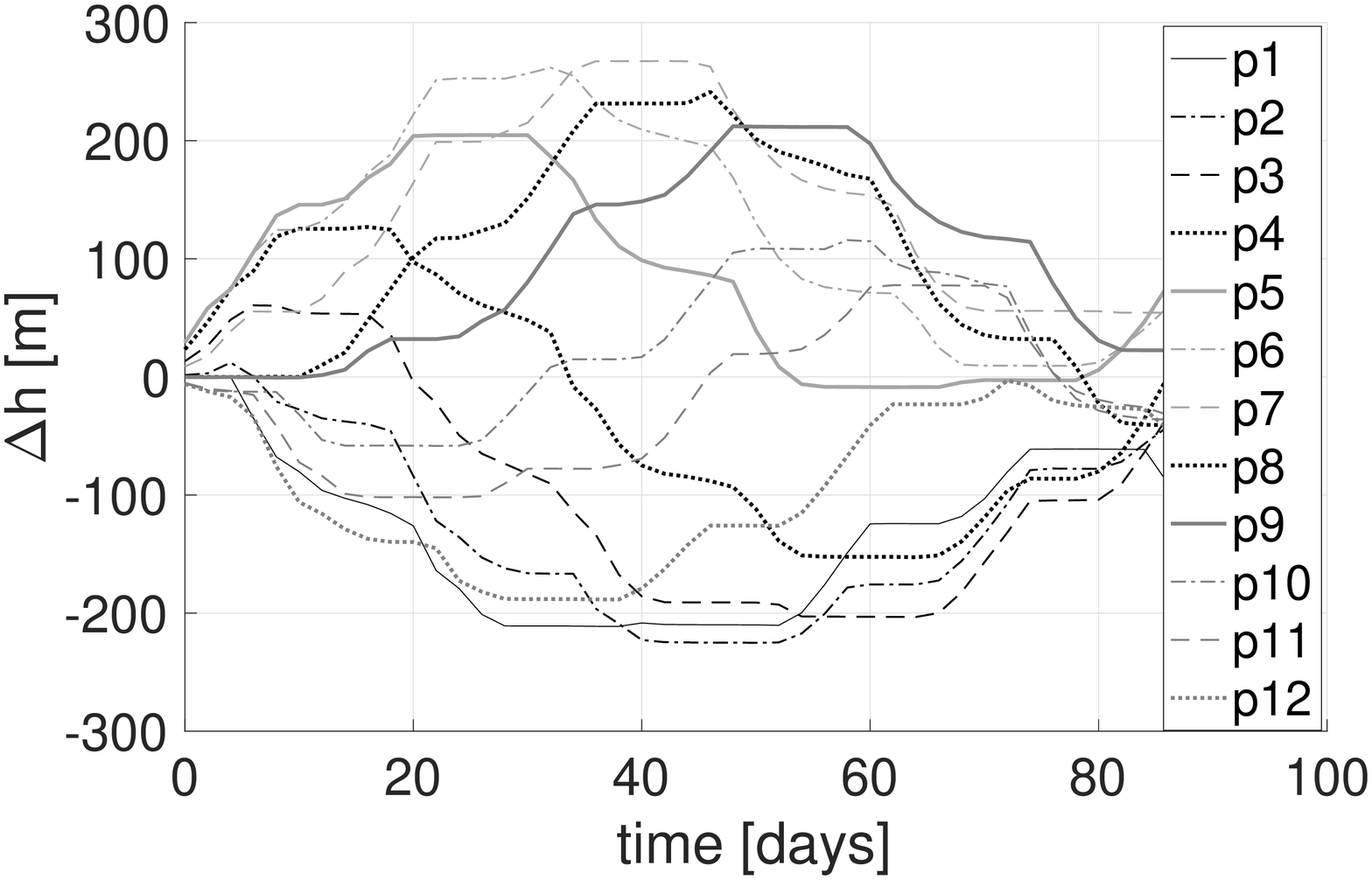}\includegraphics[clip,width=8cm]{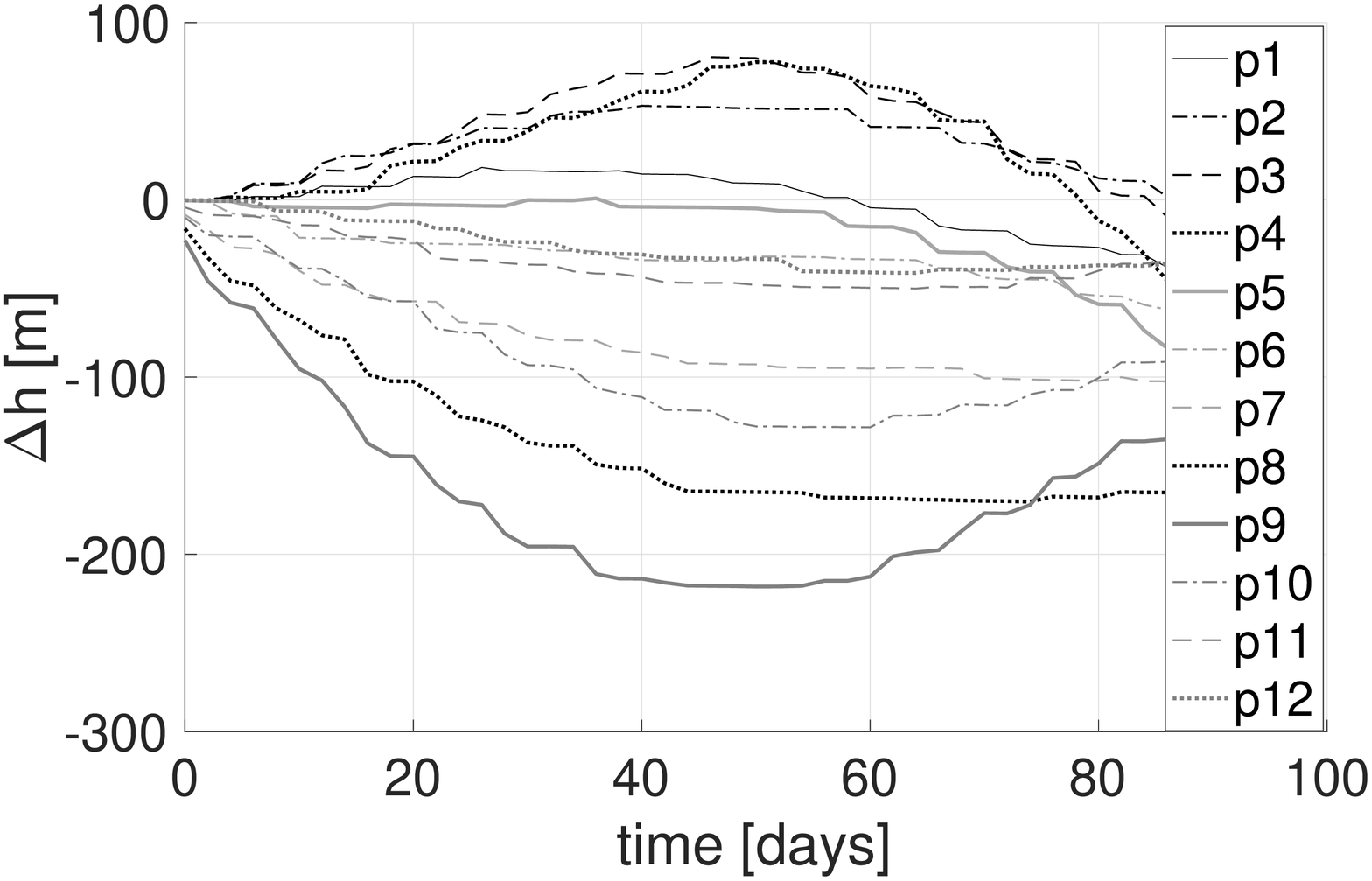}}

\caption{\label{fig:fig14}Evolution of minimum altitude for all orbital planes
of class 3 (left) and class 4 (right) MiSO orbits }
\end{figure}

\begin{figure}[!t]
\centerline{\includegraphics[clip,width=8cm]{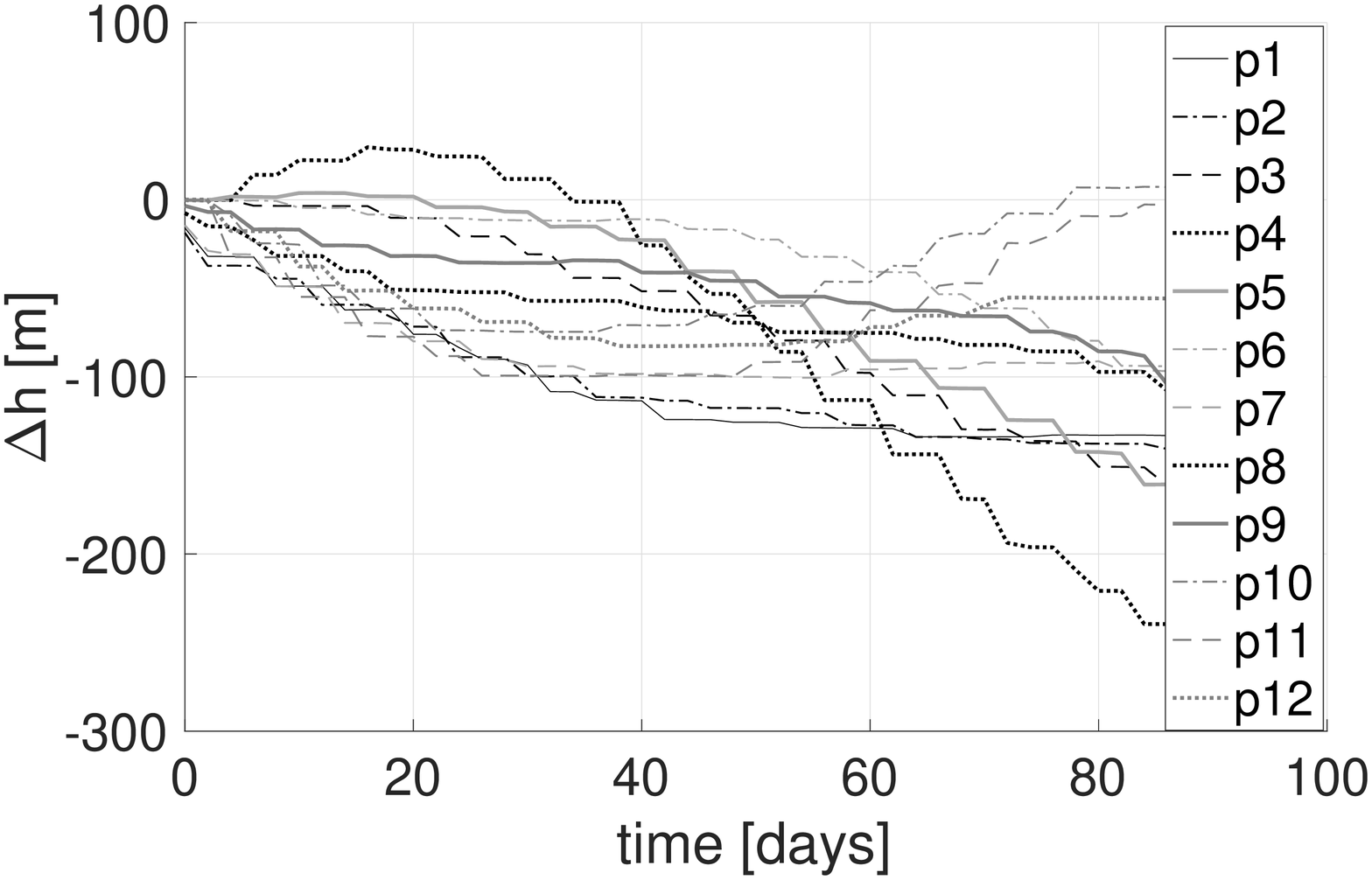}}

\caption{\label{fig:fig15}Evolution of minimum altitude for all orbital planes
of class 5 MiSO orbits}
\end{figure}

Figures (\ref{fig:fig7})-(\ref{fig:fig11}) plot the 100-day SOR
for the 12 orbital planes of the five classes of orbits while the
evolution of the minimum altitude variation in time for the different
orbital planes of each orbit class is displayed in Figures (\ref{fig:fig13})-(\ref{fig:fig15}). 

What clearly emerges from these plots is that the action of both types
of non-gravitational perturbations tends to shift the mean altitude
of the whole space occupancy region without significantly changing
its size. In other words, both perturbations do not appear to be able
to disrupt the frozen-like character of these orbits, at least over
the 100 days time-scale considered here. This is a considerable merit
of the MiSO orbit design concept. Regarding the specific influence
of SRP it appears to have a stronger influence on the SOR of Sun-synchronous
(class 5, Figure \ref{fig:fig11}) and lower inclination orbits (class
3 rather than 4, as it is evident in Figure \ref{fig:fig9} and \ref{fig:fig10})
although for the case of lower altitude orbits this tends to be masked
by the dominant effect of atmospheric drag (Figure \ref{fig:fig7}). 

Regarding the time evolution of the minimum altitude, lower altitude
MiSO orbits (Figure \ref{fig:fig13}) tend to exhibit a uniform secular
decay superposed to an oscillating behavior while for higher altitude
the behavior is predominantly oscillatory (Figure \ref{fig:fig14},\ref{fig:fig15}). 

A possible design strategy for a mega-constellations with non-overlapping
$P$ planes is the following. After sorting the different constellation
planes by ascending minimum altitude (over the desired time-span,
e.g., 100 days) the nominal maximum-latitude altitude of each plane
can be set to:

\[
h_{N,p+1}=h_{N,p}+\mathrm{SOR_{\mathrm{\mathit{p}}}}\qquad p=0..P.
\]

In the preceding equation, $\mathrm{SOR_{\mathrm{\mathit{p}}}}$ is
the space occupancy range of the $P-$plane orbit accumulated over
the total time-span and does take into account altitude oscillations.

This design process can be refined iteratively by recomputing the
new SOR of each plane after the addition of the required altitude
offset. The effectiveness of this approach has been demonstrated in
a recent paper currently under review \cite{reiland2020assessing}.

\begin{doublespace}
\noindent 
\begin{table}[htbp]
\caption{100-day SOR {[}km{]} of MiSO orbits including non-gravitational perturbations}

\centering{}\label{tab:TAB5}%
\begin{tabular}{ccccccccccccc}
\hline 
\noalign{\vskip\doublerulesep}
orbit class & $0^{\circ}$ & $30^{\circ}$ & $60^{\circ}$ & $90^{\circ}$ & $120^{\circ}$ & $150^{\circ}$ & $180^{\circ}$ & $210^{\circ}$ & $240^{\circ}$ & $270^{\circ}$ & $300^{\circ}$ & $330^{\circ}$\tabularnewline[\doublerulesep]
\hline 
\noalign{\vskip\doublerulesep}
\noalign{\vskip\doublerulesep}
class 1 & 2997 & 3134 & 2952 & 3073 & 2958 & 3056 & 3072 & 3049 & 3004 & 2933 & 2956 & 2996\tabularnewline[\doublerulesep]
\noalign{\vskip\doublerulesep}
\noalign{\vskip\doublerulesep}
class 2 & 2894 & 3210 & 3049 & 2983 & 2966 & 3100 & 3034 & 3239 & 3036 & 2925 & 3022 & 3183\tabularnewline[\doublerulesep]
\noalign{\vskip\doublerulesep}
\noalign{\vskip\doublerulesep}
class 3 & 525 & 539 & 600 & 616 & 669 & 659 & 650 & 640 & 576 & 522 & 512 & 515\tabularnewline[\doublerulesep]
\noalign{\vskip\doublerulesep}
\noalign{\vskip\doublerulesep}
class 4 & 474 & 452 & 404 & 436 & 502 & 541 & 563 & 537 & 472 & 400 & 432 & 463\tabularnewline[\doublerulesep]
\noalign{\vskip\doublerulesep}
\noalign{\vskip\doublerulesep}
class 5 & 542 & 609 & 687 & 814 & 746 & 679 & 602 & 532 & 611 & 766 & 797 & 674\tabularnewline[\doublerulesep]
\noalign{\vskip\doublerulesep}
\end{tabular}
\end{table}

\noindent 
\begin{table}[htbp]
\caption{100-day SOR {[}km{]} of unoptimized frozen orbits including non-gravitational
perturbations}

\centering{}\label{tab:TAB6}%
\begin{tabular}{ccccccccccccc}
\hline 
\noalign{\vskip\doublerulesep}
orbit class & $0^{\circ}$ & $30^{\circ}$ & $60^{\circ}$ & $90^{\circ}$ & $120^{\circ}$ & $150^{\circ}$ & $180^{\circ}$ & $210^{\circ}$ & $240^{\circ}$ & $270^{\circ}$ & $300^{\circ}$ & $330^{\circ}$\tabularnewline[\doublerulesep]
\hline 
\noalign{\vskip\doublerulesep}
\noalign{\vskip\doublerulesep}
class 1 & 3143 & 3186 & 3166 & 3180 & 3156 & 3109 & 3210 & 3335 & 3206 & 3174 & 3178 & 3291\tabularnewline[\doublerulesep]
\noalign{\vskip\doublerulesep}
\noalign{\vskip\doublerulesep}
class 2 & 3275 & 3622 & 3148 & 3454 & 3207 & 3356 & 3270 & 3337 & 3286 & 3108 & 3515 & 3692\tabularnewline[\doublerulesep]
\noalign{\vskip\doublerulesep}
\noalign{\vskip\doublerulesep}
class 3 & 673 & 707 & 711 & 732 & 764 & 908 & 1002 & 1009 & 632 & 649 & 682 & 593\tabularnewline[\doublerulesep]
\noalign{\vskip\doublerulesep}
\noalign{\vskip\doublerulesep}
class 4 & 677 & 575 & 718 & 658 & 558 & 704 & 867 & 883 & 837 & 676 & 718 & 799\tabularnewline[\doublerulesep]
\noalign{\vskip\doublerulesep}
\noalign{\vskip\doublerulesep}
class 5 & 592 & 754 & 1063 & 1186 & 1055 & 1019 & 946 & 533 & 808 & 1102 & 1132 & 1011\tabularnewline[\doublerulesep]
\noalign{\vskip\doublerulesep}
\end{tabular}
\end{table}

\end{doublespace}

\begin{doublespace}

\end{doublespace}

\begin{doublespace}

\end{doublespace}

\section*{Conclusions}

The concept of space occupancy and minimum space occupancy (MiSO)
orbits are promising tools to quantify and mitigate the risk of space
debris accumulation in LEO as well as to minimize the frequency of
collision avoidance maneuvers, especially in light of upcoming LEO
mega-constellations of satellites. MiSO orbits can be seen as a generalization
of frozen orbits beyond the zonal problem, where tesseral harmonics,
third-body effects, and non-gravitational perturbations make it impossible
to achieve constant altitude at equal latitude, i.e. ``zero occupancy''
conditions. In the zonal problem, frozen orbits can be conveniently
characterized in osculating element space leading to a newly derived
frozen orbit polar equation and providing a first guess solution for
the computation of MiSO orbits when additional perturbations are included.
We have numerically obtained initial conditions leading to MiSO orbits
in five different scenarios in LEO and studied their behavior in time.
For higher altitude orbits ($h\sim$1200 km), and with a standard
area-to-mass ratio, a SOR of less than 700 m over 100 days is achievable
and can be reduced below 600 m for near polar orbits (thanks to a
reduced negative influence of SRP). A slightly higher SOR, 814 m in
the worst case, was obtained for Sun-synchronous MiSO orbits. Lower
altitude orbits ($h\sim$500 km) are characterized by a wider SOR
(around 3 km in 100 days for the cases considered in this article)
mainly due to atmospheric drag decay (clearly inflating the space
occupancy region inwards), but also due to a stronger tesseral harmonics
effect. In all cases, non-gravitational perturbations have a detrimental
effect on the size of the space occupancy region but do not appear
to be capable of completely disrupting the frozen-like character of
the orbit, at least over a timescale of several months. It is important
to add that we did not perform a detailed investigation of the behavior
of MiSO orbits near to the critical inclination where bifurcations
between orbit families and instability arise\cite{coffey1986critical};
we leave this task for a future study.

These results suggest that the lay-out of large constellations of
satellites could be effectively optimized by having all satellites
flying in MiSO orbits and with an incremental stacking of non-intersecting
constellation planes. The effectiveness of this solution will be further
investigated.
\begin{doublespace}

\section*{Appendix I: Kozai-Lyddane Conversion Formulas}
\end{doublespace}

\begin{doublespace}
\noindent Following Kozai \cite{kozai1959motion} (or equivalently,
Brouwer \cite{brouwer1959solution}) , the ($J_{2}$-dominated) short-periodic
terms for the orbital elements of the zonal problem, after indicating
with $\hat{}$ the mean ( = secular + long-periodic) component of
each element, are as follows:
\end{doublespace}

\noindent semi-major axis:

\begin{equation}
a_{sp}=\dfrac{J_{2}}{2\hat{a}}\left[\left(2-3\kappa\right)\left(\dfrac{\hat{a}^{3}}{\hat{r}^{3}}-\dfrac{1}{\eta^{3}}\right)+\dfrac{3\kappa\hat{a}^{3}}{\hat{r}^{3}}\,c_{2,2}\right].\label{eq:a_sp}
\end{equation}

eccentricity:

\begin{equation}
\begin{array}{c}
e_{sp}=\dfrac{3J_{2}\lambda^{2}}{4\hat{e}\hat{a}^{2}}\left[\left(2-3\kappa\right)\left(\dfrac{\hat{a}^{3}}{\hat{r}^{3}}-\dfrac{1}{\lambda^{3}}\right)+\dfrac{3\kappa\hat{a}^{3}}{\hat{r}^{3}}\,c_{2,2}\right]\\
\\
-\dfrac{3J_{2}\kappa}{4\hat{e}\hat{a}^{2}\lambda^{2}}\left[c_{2,2}+\hat{e}c_{1,2}+\dfrac{\hat{e}}{3}c_{3,2}\right]-\dfrac{J_{2}\kappa\hat{e}\left(2\lambda+1\right)\cos\left(2\hat{\omega}\right)}{4\hat{a}^{2}\lambda^{2}\left(\lambda+1\right)^{2}}.
\end{array}\label{eq:e_sp}
\end{equation}

inclination:

\begin{equation}
\begin{array}{c}
\begin{array}{c}
i_{sp}=\dfrac{J_{2}}{8\hat{a}^{2}\lambda^{4}}\sin2\hat{i}\left[3c_{2,2}+3\hat{e}c_{1,2}+\hat{e}c_{3,2}\right]\end{array}\\
\\
-\dfrac{J_{2}\sin2\hat{i}\left(2\lambda^{2}-\lambda-1\right)\cos\left(2\hat{\omega}\right)}{8\hat{a}^{2}\lambda^{2}\left(\lambda+1\right)}.
\end{array}\label{eq:i_sp}
\end{equation}

longitude of the ascending node:

\begin{equation}
\begin{array}{c}
\begin{array}{c}
\Omega_{sp}=-\dfrac{3J_{2}\sqrt{1-\kappa}}{2a^{2}\lambda^{4}}\left[\hat{\nu}-\hat{M}+\hat{e}\,s_{1,0}-\dfrac{1}{2}\left(s_{2,2}+\hat{e}\,s_{1,2}+\dfrac{\hat{e}\,s_{3,2}}{3}\right)\right]\end{array}\\
\\
-\dfrac{J_{2}\sqrt{1-\kappa}\left(2\lambda^{2}-\lambda-1\right)\sin\left(2\hat{\omega}\right)}{4\hat{a}^{2}\lambda^{4}\left(\lambda+1\right)}.
\end{array}\label{eq:raan_sp}
\end{equation}

argument of pericenter:

\begin{equation}
\begin{array}{c}
\omega_{sp}=\dfrac{3J_{2}}{2\hat{a}^{2}\lambda^{4}}\left\{ \dfrac{4-5\kappa}{2}\left(\hat{\nu}-\hat{M}+\hat{e}\,s_{1,0}\right)+\dfrac{5\kappa-2}{4}\left(s_{2,2}+\hat{e}s_{1,2}+\dfrac{\hat{e}}{3}s_{3,2}\right)\right.\\
\\
+\dfrac{\lambda^{2}}{\hat{e}}\left(\dfrac{2-3\kappa}{2}\left[\left(1-\dfrac{\hat{e}^{2}}{4}\right)s_{1,0}+\dfrac{\hat{e}}{2}s_{2,0}+\dfrac{\hat{e}^{2}}{12}s_{3,0}\right]\right.\\
\\
+\kappa\left[\dfrac{1}{4}\left(1+\dfrac{5}{4}\hat{e}^{2}\right)s_{1,2}-\dfrac{\hat{e}^{2}}{16}s_{1,-2}\right.\\
\\
\left.\left.\left.-\dfrac{7}{12}\left(1-\dfrac{\hat{e}^{2}}{28}\right)s_{3,2}-\dfrac{3}{8}\hat{e}s_{4,2}-\dfrac{\hat{e}^{2}}{16}s_{5,2}\right]\right)\right\} \\
\\
+\dfrac{3J_{2}\sqrt{1-\kappa}}{2\hat{a}^{2}\lambda^{4}}\left[\dfrac{\kappa}{8}+\dfrac{\left(1+2\lambda\right)\left(2\kappa\lambda^{2}-\lambda^{2}-\kappa+1\right)}{6\left(\lambda+1\right)^{2}}\right]\sin\left(2\hat{\omega}\right).
\end{array}\label{eq:omega_sp}
\end{equation}

mean anomaly:
\begin{equation}
\begin{array}{c}
\hat{e}M_{sp}=-\dfrac{3J_{2}}{2\hat{a}^{2}\lambda^{3}}\left\{ \dfrac{2-3\kappa}{2}\left[\left(1-\dfrac{\hat{e}^{2}}{4}\right)s_{1,0}+\dfrac{\hat{e}}{2}s_{2,0}+\dfrac{\hat{e}^{2}}{12}s_{3,0}\right]\right.\\
\\
-\kappa\left[\dfrac{1}{4}\left(1+\dfrac{5}{4}\hat{e}^{2}\right)s_{1,2}-\dfrac{\hat{e}^{2}}{16}s_{1,-2}\right.\\
\\
\left.\left.-\dfrac{7}{12}\left(1-\dfrac{\hat{e}^{2}}{28}\right)s_{3,2}-\dfrac{3}{8}\hat{e}s_{4,2}-\dfrac{\hat{e}^{2}}{16}s_{5,2}\right]\right\} \\
\\
+\hat{e}\dfrac{J_{2}\kappa\left(4\lambda^{3}-\lambda^{2}-18\lambda-9\right)\sin\left(2\hat{\omega}\right)}{16\hat{a}^{2}\lambda^{3}\left(\lambda+1\right)^{2}},
\end{array}\label{eq:eM_sp}
\end{equation}
with:

\[
\lambda=\sqrt{1-\hat{e}^{2}},\qquad\kappa=\sin^{2}\hat{i},
\]

\[
\hat{r}=\frac{\hat{a}\left(1-\hat{e}^{2}\right)}{1+\hat{e}\cos\hat{\nu}},
\]

\[
s_{1,0}=\sin\hat{\nu},\qquad s_{2,0}=\sin2\hat{\nu},\qquad s_{3,0}=\sin3\hat{\nu},
\]

\[
s_{1,2}=\sin\left(\hat{\nu}+2\hat{\omega}\right),\qquad s_{1,-2}=\sin\left(\hat{\nu}-2\hat{\omega}\right),\qquad s_{2,2}=\sin\left(2\hat{\nu}+2\hat{\omega}\right),
\]

\[
s_{3,2}=\sin\left(3\hat{\nu}+2\hat{\omega}\right),\qquad s_{4,2}=\sin\left(4\hat{\nu}+2\hat{\omega}\right),\qquad s_{5,2}=\sin\left(5\hat{\nu}+2\hat{\omega}\right),
\]

\[
c_{1,0}=\cos\hat{\nu},\qquad c_{1,1}=\cos\left(\hat{\nu}+\hat{\omega}\right),\qquad c_{1,2}=\cos\left(\hat{\nu}+2\hat{\omega}\right),
\]

\[
c_{2,2}=\cos\left(2\hat{\nu}+2\hat{\omega}\right),\qquad c_{3,2}=\cos\left(3\hat{\nu}+2\hat{\omega}\right).
\]

With the exception of the semi-major axis, all above expressions may
become numerically unstable near circular and/or equatorial conditions.
Following

Lyddane's method \cite{lyddane1963small}, a numerically stable expression
for the mean anomaly short-periodic can be obtained based on the expansion:

\begin{equation}
\left(\hat{e}+e_{sp}\right)\cos\left(\hat{M}+M_{sp}\right)\simeq\left(\hat{e}+e_{sp}\right)\cos\hat{M}-\hat{e}M_{sp}\sin\hat{M}=\varsigma,\label{eq:ecosM_exp}
\end{equation}

\begin{equation}
\left(\hat{e}+e_{sp}\right)\sin\left(\hat{M}+M_{sp}\right)\simeq\left(\hat{e}+e_{sp}\right)\sin\hat{M}+\hat{e}M_{sp}\sin\hat{M}=\iota,\label{eq:esinM_exp}
\end{equation}
providing numerically stable expressions (denoted with a tilde) for
the mean anomaly and eccentricity short-periodic components as:

\begin{equation}
\tilde{M}_{sp}\simeq\mathrm{atan2}\left(\iota,\varsigma\right)-\hat{M},\label{eq:M_sp_tilde}
\end{equation}

\begin{equation}
\tilde{e}_{sp}=\sqrt{\iota^{2}+\varsigma^{2}}-\hat{e}.\label{eq:e_sp_tilde}
\end{equation}

Similarly, using Lyddane's expansion:

\[
\sin\left(\frac{\hat{i}+i_{sp}}{2}\right)\cos\left(\hat{\Omega}+\Omega_{sp}\right)\simeq\left(\sin\frac{\hat{i}}{2}+\frac{i_{sp}}{2}\cos\frac{\hat{i}}{2}\right)\cos\hat{\Omega}-\sin\frac{\hat{i}}{2}\sin\hat{\Omega}\,\Omega_{sp}=\varrho,
\]

\[
\sin\left(\frac{\hat{i}+i_{sp}}{2}\right)\sin\left(\hat{\Omega}+\Omega_{sp}\right)\simeq\left(\sin\frac{\hat{i}}{2}+\frac{i_{sp}}{2}\cos\frac{\hat{i}}{2}\right)\sin\hat{\Omega}+\sin\frac{\hat{i}}{2}\cos\hat{\Omega}\,\Omega_{sp}=\kappa,
\]
stable expressions for the right ascension of the ascending node and
the inclination are obtained:

\begin{equation}
\tilde{\Omega}_{sp}\simeq\mathrm{atan2}\left(\kappa,\varrho\right)-\hat{\Omega},\label{eq:raan_sp_tilde}
\end{equation}

\begin{equation}
\tilde{i}_{sp}=2\sin^{-\text{1}}\sqrt{\left(\varrho^{2}+\kappa^{2}\right)}.\label{eq:i_sp_tilde}
\end{equation}

The non-singular expression for the argument of periapsis short-periodic
component can be computed as:

\begin{equation}
\tilde{\omega}_{sp}=\ell_{sp}-\tilde{M}_{sp}-\tilde{\Omega}_{sp}.\label{eq:omega_sp_tilde}
\end{equation}
where:

\[
\ell_{sp}=M_{sp}+\omega_{sp}+\Omega_{sp}
\]
is the short-periodic component of the mean longitude and reads (from
Eqs. (\ref{eq:raan_sp}-\ref{eq:eM_sp})):

\begin{equation}
\begin{array}{c}
\ell_{sp}=\dfrac{3J_{2}}{2\hat{a}^{2}\lambda^{4}}\left\{ \left(\dfrac{4-5\kappa}{2}-\sqrt{1-\kappa}\right)\left(\hat{\nu}-\hat{M}+\hat{e}\,s_{1,0}\right)\right.\\
\\
+\left(\dfrac{5\kappa-2}{4}+\dfrac{\sqrt{1-\kappa}}{2}\right)\left(s_{2,2}+es_{1,2}+\dfrac{\hat{e}}{3}s_{3,2}\right)\\
\\
+\dfrac{\hat{e}}{1+\lambda}\left(\dfrac{2-3\kappa}{2}\left[\left(1-\dfrac{\hat{e}^{2}}{4}\right)s_{1,0}+\dfrac{\hat{e}}{2}s_{2,0}+\dfrac{\hat{e}^{2}}{12}s_{3,0}\right]\right.\\
\\
+\kappa\left[\dfrac{1}{4}\left(1+\dfrac{5}{4}\hat{e}^{2}\right)s_{1,2}-\dfrac{\hat{e}^{2}}{16}s_{1,-2}\right.\\
\\
\left.\left.\left.-\dfrac{7}{12}\left(1-\dfrac{\hat{e}^{2}}{28}\right)s_{3,2}-\dfrac{3}{8}\hat{e}s_{4,2}-\dfrac{\hat{e}^{2}}{16}s_{5,2}\right]\right)\right\} \\
\\
+\dfrac{3J_{2}\sqrt{1-\kappa}}{2\hat{a}^{2}\lambda^{4}}\left[\dfrac{\kappa}{8}+\dfrac{\left(1+2\lambda\right)\left(2\kappa\lambda^{2}-\lambda^{2}-\kappa+1\right)}{6\left(\lambda+1\right)^{2}}\right]\sin\left(2\hat{\omega}\right)\\
\\
+\dfrac{J_{2}\kappa\left(4\lambda^{3}-\lambda^{2}-18\lambda-9\right)\sin\left(2\hat{\omega}\right)}{16\hat{a}^{2}\lambda^{3}\left(\lambda+1\right)^{2}}-\dfrac{J_{2}\sqrt{1-\kappa}\left(2\lambda^{2}-\lambda-1\right)\sin\left(2\hat{\omega}\right)}{4\hat{a}^{2}\lambda^{4}\left(\lambda+1\right)}.
\end{array}\label{eq:ell_sp}
\end{equation}

\begin{doublespace}

\section*{Appendix II: MiSO orbit initial conditions}
\end{doublespace}
\begin{itemize}
\begin{doublespace}
\item We report the initial conditions in terms of classical orbital elements
for the five classes of orbits with and without SRP and drag. The
reference epoch is 1 January 2020 (JD=2458849.5).
\end{doublespace}
\end{itemize}

\subsubsection*{class 1, drag-free}

\begin{align*}
\left[
\begin{array}{cccccc} 
\Omega(^{\circ})\ & a(\mathrm{km})\ & e()\ & inc(^{\circ})\ & \omega(^{\circ})\ & M_{0}(^{\circ})\\ 0 & 6932.759064 & 0.0003334517026 & 52.98106159 & 90.54762182 & 359.4527433\\ 30.0 & 6932.601577 & 0.0003222155192 & 52.98106159 & 87.73258259 & 2.265956955\\ 60.0 & 6932.584891 & 0.0003210841031 & 52.98106159 & 87.15531659 & 2.842857835\\ 90.0 & 6932.567921 & 0.0003199118596 & 52.98106159 & 86.80193533 & 3.196020025\\ 120.0 & 6932.550011 & 0.0003186039891 & 52.98106159 & 86.90359531 & 3.094433071\\ 150.0 & 6932.738275 & 0.0003320120194 & 52.98106159 & 88.37426286 & 1.624658026\\ 180.0 & 6932.829753 & 0.0003385453073 & 52.98106159 & 89.56849536 & 0.4312125478\\ 210.0 & 6933.077243 & 0.0003563831289 & 52.98106159 & 90.0 & 4.042782711\cdot 10^{-12}\\ 240.0 & 6932.583903 & 0.0003209417217 & 52.98106159 & 92.27641993 & 357.7250405\\ 270.0 & 6932.505325 & 0.0003155593633 & 52.98106159 & 94.13133918 & 355.8712653\\ 300.0 & 6932.641127 & 0.0003253686865 & 52.98106159 & 94.15663717 & 355.8460647\\ 330.0 & 6932.921076 & 0.0003453388649 & 52.98106159 & 92.85652342 & 357.1454482 
\end{array}
\right]
\end{align*}

\medskip{}

\medskip{}

\subsubsection*{class 2, drag-free}

\begin{align*}
\left[\begin{array}{cccccc} \Omega(^{\circ})\ & a(\mathrm{km})\ & e()\ & inc(^{\circ})\ & \omega(^{\circ})\ & M_{0}(^{\circ})\\ 0 & 6928.137945 & 0.0004628998797 & 87.89855475 & 270.0 & 180.0\\ 30.0 & 6928.142547 & 0.0004686537004 & 87.89855475 & 266.9618906 & 183.0409567\\ 60.0 & 6928.138892 & 0.000454128096 & 87.89855475 & 268.5534594 & 181.4478548\\ 90.0 & 6928.14189 & 0.0004596507278 & 87.89855475 & 272.859197 & 177.1381748\\ 120.0 & 6928.140508 & 0.0004709049191 & 87.89855475 & 272.2478274 & 177.7500554\\ 150.0 & 6928.138081 & 0.0004832807396 & 87.89855475 & 269.9244934 & 180.0755796\\ 180.0 & 6928.138124 & 0.0004883770586 & 87.89855475 & 269.850563 & 180.149583\\ 210.0 & 6928.138271 & 0.0004667646813 & 87.89855475 & 269.2181866 & 180.7825434\\ 240.0 & 6928.139876 & 0.0004759041062 & 87.89855475 & 268.0827124 & 181.9191128\\ 270.0 & 6928.139195 & 0.0004719882596 & 87.89855475 & 268.4535424 & 181.5479178\\ 300.0 & 6928.138817 & 0.0004617530503 & 87.89855475 & 271.3435928 & 178.6551661\\ 330.0 & 6928.142847 & 0.0004577776823 & 87.89855475 & 273.1961256 & 176.8009487 \end{array}\right]
\end{align*}

\subsubsection*{class 3, drag-free}

\begin{align*}
\left[\begin{array}{cccccc} \Omega(^{\circ})\ & a(\mathrm{km})\ & e()\ & inc(^{\circ})\ & \omega(^{\circ})\ & M_{0}(^{\circ})\\ 0 & 7551.070081 & 0.0003267389013 & 52.98403631 & 90.28264403 & 359.7175406\\ 30.0 & 7550.941064 & 0.0003182546944 & 52.98403631 & 88.45220381 & 1.546811358\\ 60.0 & 7550.925068 & 0.0003172114877 & 52.98403631 & 88.25294551 & 1.745946553\\ 90.0 & 7550.925471 & 0.0003172647177 & 52.98403631 & 87.96199778 & 2.036709626\\ 120.0 & 7550.942161 & 0.0003183997827 & 52.98403631 & 87.67901118 & 2.319511573\\ 150.0 & 7550.98916 & 0.0003213985795 & 52.98403631 & 89.13795204 & 0.8614939899\\ 180.0 & 7551.086314 & 0.0003278134345 & 52.98403631 & 90.2817177 & 359.718467\\ 210.0 & 7551.232544 & 0.0003375013701 & 52.98403631 & 90.63848614 & 359.3619447\\ 240.0 & 7550.860418 & 0.0003129504061 & 52.98403631 & 91.96767998 & 358.0335511\\ 270.0 & 7550.796865 & 0.0003088343336 & 52.98403631 & 92.79201685 & 357.2097066\\ 300.0 & 7550.910426 & 0.0003163473872 & 52.98403631 & 92.72566777 & 357.2760557\\ 330.0 & 7551.136052 & 0.0003311743495 & 52.98403631 & 91.67338604 & 358.3277219 \end{array}\right]
\end{align*}

\subsubsection*{\medskip{}
}

\medskip{}

\subsubsection*{class 4, drag-free}

\begin{align*}
\left[\begin{array}{cccccc} \Omega(^{\circ})\ & a(\mathrm{km})\ & e()\ & inc(^{\circ})\ & \omega(^{\circ})\ & M_{0}(^{\circ})\\ 0 & 7546.137417 & 0.0003554791211 & 87.89878205 & 269.9134623 & 180.0865992\\ 30.0 & 7546.137812 & 0.0003619697371 & 87.89878205 & 269.0651164 & 180.9355606\\ 60.0 & 7546.137518 & 0.0003544195479 & 87.89878205 & 269.4792143 & 180.521155\\ 90.0 & 7546.13762 & 0.0003576521232 & 87.89878205 & 270.6881109 & 179.3113968\\ 120.0 & 7546.13854 & 0.0003642120497 & 87.89878205 & 271.6049943 & 178.3938364\\ 150.0 & 7546.138352 & 0.000372771327 & 87.89878205 & 271.4030248 & 178.595929\\ 180.0 & 7546.137523 & 0.0003748075505 & 87.89878205 & 269.9179258 & 180.0821358\\ 210.0 & 7546.140531 & 0.0003612565296 & 87.89878205 & 267.2740604 & 182.7279089\\ 240.0 & 7546.139481 & 0.0003686287757 & 87.89878205 & 267.8297766 & 182.1718234\\ 270.0 & 7546.137505 & 0.0003672941004 & 87.89878205 & 270.2512612 & 179.7485542\\ 300.0 & 7546.138211 & 0.0003598764869 & 87.89878205 & 271.367811 & 178.6312043\\ 330.0 & 7546.138634 & 0.0003534942316 & 87.89878205 & 271.7407412 & 178.258028 \end{array}\right]
\end{align*}

\subsubsection*{\medskip{}
}

\medskip{}

\subsubsection*{class 5, drag-free}

\begin{align*}
\left[\begin{array}{cccccc} \Omega(^{\circ})\ & a(\mathrm{km})\ & e()\ & inc(^{\circ})\ & \omega(^{\circ})\ & M_{0}(^{\circ})\\ 0 & 7191.138473 & 0.0004861880768 & 98.7382975 & 270.7663379 & 179.2329167\\ 30.0 & 7191.138377 & 0.0004873559266 & 98.7382975 & 269.3745053 & 180.6261046\\ 60.0 & 7191.13945 & 0.0004662437432 & 98.7382975 & 268.3289104 & 181.6726482\\ 90.0 & 7191.141517 & 0.0004700743544 & 98.7382975 & 267.4051815 & 182.5972581\\ 120.0 & 7191.139764 & 0.000482823931 & 98.7382975 & 268.2459627 & 181.7557314\\ 150.0 & 7191.139363 & 0.0004993041537 & 98.7382975 & 268.5753276 & 181.4260955\\ 180.0 & 7191.138775 & 0.0005039469162 & 98.7382975 & 269.0590254 & 180.9419233\\ 210.0 & 7191.138153 & 0.0004837814401 & 98.7382975 & 269.8599766 & 180.140159\\ 240.0 & 7191.138789 & 0.0004803262336 & 98.7382975 & 271.1283193 & 178.8705965\\ 270.0 & 7191.142094 & 0.000486690954 & 98.7382975 & 272.715144 & 177.2822131\\ 300.0 & 7191.142097 & 0.0004843290287 & 98.7382975 & 272.7283982 & 177.268959\\ 330.0 & 7191.141897 & 0.000484301249 & 98.7382975 & 272.6585409 & 177.338884 \end{array}\right]
\end{align*}

\subsubsection*{\medskip{}
}

\medskip{}

\subsubsection*{class 1, with drag and SRP}

\begin{align*}
\left[\begin{array}{cccccc} \Omega(^{\circ})\ & a(\mathrm{km})\ & e()\ & inc(^{\circ})\ & \omega(^{\circ})\ & M_{0}(^{\circ})\\ 0 & 6932.847704 & 0.0003398589364 & 52.98106159 & 89.03282984 & 0.966512953\\ 30.0 & 6932.847373 & 0.0003398111148 & 52.98106159 & 89.89252657 & 0.1074004103\\ 60.0 & 6932.374999 & 0.0003061144251 & 52.98106159 & 86.0599343 & 3.937655933\\ 90.0 & 6932.688387 & 0.000328359523 & 52.98106159 & 89.33265961 & 0.6669022514\\ 120.0 & 6932.373341 & 0.0003058754377 & 52.98106159 & 86.77463005 & 3.223398318\\ 150.0 & 6932.635623 & 0.0003245754356 & 52.98106159 & 88.87475421 & 1.124515558\\ 180.0 & 6933.006515 & 0.0003512844274 & 52.98106159 & 89.89603617 & 0.1038908093\\ 210.0 & 6933.165975 & 0.000362802605 & 52.98106159 & 89.09398865 & 0.9053541528\\ 240.0 & 6932.958716 & 0.0003482162884 & 52.98106159 & 93.77840363 & 356.2242252\\ 270.0 & 6932.850391 & 0.000340246043 & 52.98106159 & 92.89930889 & 357.1026627\\ 300.0 & 6932.849824 & 0.0003401643041 & 52.98106159 & 92.61337609 & 357.3884008\\ 330.0 & 6932.795681 & 0.0003361819257 & 52.98106159 & 91.95578046 & 358.0455339 \end{array}\right]
\end{align*}

\subsubsection*{\medskip{}
}

\medskip{}

\subsubsection*{class 2, with drag and SRP}

\begin{align*}
\left[\begin{array}{cccccc} \Omega(^{\circ})\ & a(\mathrm{km})\ & e()\ & inc(^{\circ})\ & \omega(^{\circ})\ & M_{0}(^{\circ})\\ 0 & 6928.139579 & 0.000455924017 & 87.89855475 & 268.1321102 & 181.8695933\\ 30.0 & 6928.140254 & 0.0005039540111 & 87.89855475 & 268.0445949 & 181.9573763\\ 60.0 & 6928.138861 & 0.0004464878217 & 87.89855475 & 268.5286952 & 181.4726189\\ 90.0 & 6928.138877 & 0.000450307952 & 87.89855475 & 271.4588177 & 178.5398682\\ 120.0 & 6928.139415 & 0.0004961982704 & 87.89855475 & 271.5445355 & 178.4539314\\ 150.0 & 6928.138288 & 0.0005108813245 & 87.89855475 & 269.8095296 & 180.1906651\\ 180.0 & 6928.140232 & 0.0004963159694 & 87.89855475 & 271.9855173 & 178.0125115\\ 210.0 & 6928.139991 & 0.0004818593136 & 87.89855475 & 271.9440925 & 178.0540336\\ 240.0 & 6928.141251 & 0.0004671942162 & 87.89855475 & 267.4215977 & 182.5808115\\ 270.0 & 6928.13852 & 0.0005151562307 & 87.89855475 & 270.6375138 & 179.3618291\\ 300.0 & 6928.141947 & 0.0004392966006 & 87.89855475 & 272.9918218 & 177.0055499\\ 330.0 & 6928.140676 & 0.0004492946643 & 87.89855475 & 272.4373195 & 177.5604903 \end{array}\right]
\end{align*}

\subsubsection*{\medskip{}
}

\medskip{}

\subsubsection*{class 3, with drag and SRP}

\begin{align*}
\left[\begin{array}{cccccc} \Omega(^{\circ})\ & a(\mathrm{km})\ & e()\ & inc(^{\circ})\ & \omega(^{\circ})\ & M_{0}(^{\circ})\\ 0 & 7550.845248 & 0.0003118971737 & 52.98403631 & 91.34898058 & 358.6518606\\ 30.0 & 7550.754644 & 0.0003058712851 & 52.98403631 & 90.80515351 & 359.1953389\\ 60.0 & 7550.794622 & 0.0003085376134 & 52.98403631 & 91.21951891 & 358.7812334\\ 90.0 & 7550.929887 & 0.0003174913716 & 52.98403631 & 91.19590232 & 358.8048568\\ 120.0 & 7551.133363 & 0.0003309378911 & 52.98403631 & 90.69249227 & 359.3079659\\ 150.0 & 7551.346447 & 0.0003450581501 & 52.98403631 & 91.03096422 & 358.969747\\ 180.0 & 7551.492367 & 0.0003547038822 & 52.98403631 & 90.75218097 & 359.2483525\\ 210.0 & 7551.549974 & 0.0003585090176 & 52.98403631 & 90.5247479 & 359.4756282\\ 240.0 & 7551.109944 & 0.00032938964 & 52.98403631 & 89.25232889 & 0.7471786939\\ 270.0 & 7550.842793 & 0.0003116915439 & 52.98403631 & 89.93415873 & 0.06580023575\\ 300.0 & 7550.843054 & 0.0003117261164 & 52.98403631 & 90.85587323 & 359.1446602\\ 330.0 & 7550.956833 & 0.0003192678224 & 52.98403631 & 91.0606653 & 358.9400118 \end{array}\right]
\end{align*}

\subsubsection*{\medskip{}
}

\medskip{}

\subsubsection*{class 4, with drag and SRP}

\begin{align*}
\left[\begin{array}{cccccc} \Omega(^{\circ})\ & a(\mathrm{km})\ & e()\ & inc(^{\circ})\ & \omega(^{\circ})\ & M_{0}(^{\circ})\\ 0 & 7546.138016 & 0.0003562560579 & 87.79872416 & 271.2089743 & 178.7901641\\ 30.0 & 7546.138063 & 0.0003585275642 & 87.79872416 & 268.7605451 & 181.2403439\\ 60.0 & 7546.139345 & 0.0003497552251 & 87.79872416 & 267.7908264 & 182.210719\\ 90.0 & 7546.138529 & 0.0003540585378 & 87.79872416 & 268.3392998 & 181.6618763\\ 120.0 & 7546.137756 & 0.0003688594585 & 87.79872416 & 269.2030605 & 180.7975275\\ 150.0 & 7546.138011 & 0.0003804577162 & 87.79872416 & 268.9757809 & 181.0249987\\ 180.0 & 7546.139199 & 0.0003891986586 & 87.79872416 & 268.1115078 & 181.8899624\\ 210.0 & 7546.141828 & 0.0003856124813 & 87.79872416 & 266.9050342 & 183.0973522\\ 240.0 & 7546.139077 & 0.0003932358682 & 87.79872416 & 268.2091741 & 181.7922345\\ 270.0 & 7546.139725 & 0.0003836650407 & 87.79872416 & 272.2099263 & 177.7883779\\ 300.0 & 7546.142083 & 0.0003684783715 & 87.79872416 & 273.2948053 & 176.7027673\\ 330.0 & 7546.142174 & 0.0003595486915 & 87.79872416 & 273.3862663 & 176.6112994 \end{array}\right]
\end{align*}

\subsubsection*{class 5, with drag and SRP}

\begin{align*}
\left[\begin{array}{cccccc} \Omega(^{\circ})\ & a(\mathrm{km})\ & e()\ & inc(^{\circ})\ & \omega(^{\circ})\ & M_{0}(^{\circ})\\ 0 & 7191.138373 & 0.000493261026 & 98.7382975 & 270.549335 & 179.4501229\\ 30.0 & 7191.139938 & 0.0004946594852 & 98.7382975 & 268.2194503 & 181.7823116\\ 60.0 & 7191.141711 & 0.0004712825164 & 98.7382975 & 267.3398952 & 182.6626121\\ 90.0 & 7191.141996 & 0.0004768634963 & 98.7382975 & 267.2674714 & 182.7351346\\ 120.0 & 7191.142092 & 0.0004890529023 & 98.7382975 & 267.2979821 & 182.7046608\\ 150.0 & 7191.141703 & 0.0005043534467 & 98.7382975 & 267.5144753 & 182.488032\\ 180.0 & 7191.140867 & 0.0005148671179 & 98.7382975 & 267.8944508 & 182.1077177\\ 210.0 & 7191.138209 & 0.0004922321172 & 98.7382975 & 269.8904146 & 180.1096933\\ 240.0 & 7191.140188 & 0.0004864262899 & 98.7382975 & 271.9500414 & 178.0480612\\ 270.0 & 7191.142089 & 0.0004925958671 & 98.7382975 & 272.6825648 & 177.3147924\\ 300.0 & 7191.142092 & 0.0004890529023 & 98.7382975 & 272.7020179 & 177.2953392\\ 330.0 & 7191.142092 & 0.0004890529023 & 98.7382975 & 272.7020179 & 177.2953392 \end{array}\right]
\end{align*}

\section*{Acknowledgments}

A substantial part of this work was conducted during a visit of the
first author to the University of Arizona in the summer of 2018. Also,
funding has been provided by the Spanish Ministry of Economy and Competitiveness
within the framework of the research project ESP2017-87271-P . We
thank Joe Carroll from Tethers Applications Inc. for many fruitful
discussions that motivated the present work.

\begin{doublespace}
\medskip{}

\end{doublespace}

\bibliographystyle{AAS_publication}
\bibliography{bibliography}

\begin{lyxcode}
\end{lyxcode}

\end{document}